\documentclass[12pt,a4paper,openright]{thesis}

\advance\voffset by.7cm

\usepackage{lhcb-definitions}
\usepackage{charmproduction-definitions}
\usepackage{measurement-definitions}
\usepackage{amsmath, amsthm}
\DeclareRobustCommand{\thesismath}[1]{\ensuremath{\maybebmsf{#1}}}
\usepackage{picture}
\usepackage{multirow}
\usepackage{color}
\usepackage{ifpdf}
\usepackage{ctable}
\usepackage{subfigure}
\usepackage{ctable}
\usepackage{textpos}
\usepackage{verbatim}
\usepackage{lineno}
\usepackage{SIunits, chronology}
\usepackage{hepunits,abhepexpt,abhep,hyperref} 
\usepackage{graphics}
\usepackage{graphicx}

\newcommand{\thesistitle}{Conceptualization, implementation, and commissioning of real-time analysis in the High Level Trigger of the LHCb experiment}

\newcommand{\authorname}{Vladimir Vava Gligorov}

\newcommand{\collegename}{LPNHE}

\newcommand{\DPhiltext}{Habilitation a Diriger des Recherches}

\newcommand{\DPhildate}{2018}

\newcommand{\abstext}{
LHCb is a general purpose forward detector located at the Large Hadron Collider (LHC) at CERN. 
Although initially optimized for the study of hadrons containing beauty quarks, the better than
expected performance of the detector hardware and trigger system allowed LHCb to perform precise
measurements of particle properties across a wide range of light hadron species produced at the LHC.
The abundance of these light hadron species, and the large branching ratios of many theoretically
interesting decay modes, have made it mandatory for LHCb to perform a large part of its data analysis
within the experiment's trigger system, that is to say in real-time. This thesis describes the
conceptualization, development, and commissioning of real-time analysis in LHCb, culminating in the
proof-of-concept measurements produced with the first data collected in Run~II of the LHC. It 
also describes mistakes made in these first real-time analyses, and their implication for the future
of real-time analysis at LHCb and elsewhere.
}

\begin{document}

\baselineskip=3ex

\pagestyle{empty}

\title{\LARGE{\thesistitle} \\[3cm]}

\author{\Large{\authorname} \\\\ \Large{\collegename} \\[3cm]}

\date{\DPhiltext \\[1cm] \DPhildate}

\maketitle

\cleardoublepage

\cleardoublepage

\begin{abstract}
\abstext
\end{abstract}

\cleardoublepage
\pagestyle{plain}
\pagenumbering{roman}
\raggedbottom

\vspace*{2cm}
\begin{textblock*}{8.5cm}(2.0cm,1.0cm)
\begin{flushleft}
\noindent For Paul Baily, who gave me the entry codes and never asked me to slow down.
\end{flushleft}
\end{textblock*}




\begin{textblock*}{11cm}(3.5cm,5.5cm)
\begin{flushright}
Ey Leute, ey Leute \linebreak
Wir bringen die Sachen, die Wirbel machen, wie 'ne Turbine \linebreak
Und nicht so 'ne Fa\"{u}le, so 'ne Fa\"{u}le \linebreak
Nee, Mann, alles chef, alles def, alles Routine \linebreak
Ey Leute, ey Leute \linebreak
Ja, wir basteln die ganzen Sachen, uns und euch zuliebe \linebreak
Und nicht so 'ne Fa\"{u}le, so 'ne Fa\"{u}le \linebreak
Nee, Mann, alles chef, alles def, alles Routine
\end{flushright}
\raggedleft
-- Absolute Beginner, Fa\"{u}le
\end{textblock*}

\begin{textblock*}{11cm}(3.5cm,12.5cm)
\begin{flushright}
Hic Rhodos, hic salta.
\end{flushright}
\raggedleft
-- Anonymous
\end{textblock*}

\chapter*{\vspace{-2.5cm}Acknowledgements}
\vspace{-1cm}
During the years in which the real-time analysis described in this thesis was
conceived, implemented, and commissioned I had the priviledge to work in
the High Level Trigger team of LHCb, and I suspect many years will 
pass before I am again fortunate enough to collaborate
with so many brilliant and generous people at once. 
None of the work described in this HDR would have been possible without them,
or without the reconstruction, alignment, calibration, online, and offline computing
teams of LHCb, who embraced the idea of real-time analysis and brought it to life. 
You know who you are, this work is ours,
and I simply hope to have done it justice in the writeup. 

That being said, I have used code written by my colleagues to produce many of the plots
in this document, and I wish to acknowledge those cases more specifically. The results presented in the
``haystack of needles'' chapter, which form the historical physics case for real-time analysis in LHCb,
were produced in collaboration with Conor Fitzpatrick, who wrote all the code. Similarly, Mike Sokoloff
wrote the code used to produce most of the plots in the ``making time less real''
chapter. The analysis of charm cross-sections was done in collaboration with a large number
of colleagues, and while I have highlighted some of my own technical contributions
most of the scripts and code used to produce the results were written by others,
in particular Christopher Burr, Dominik Mueller, Alex Pearce, and Patrick Spradlin.
A special thank you is due Alex and Dominik for maintaining a reproducible version
of the analysis framework such that I could rerun this code more than two years after the fact
and remake most of the analysis plots myself. 

I must also pay special respect to two people who really went above and beyond
in making real-time analysis possible in LHCb. Silvia Borghi led the development and deployment of the real-time detector alignment and calibration, 
and has spent much of her evenings since 2015 perfecting every detail
and training the next generation of LHCb reconstruction and calibration experts. 
Provela si ekipu kroz nevreme pravo, \v{s}ta da ti ka\v{z}em sem drugarice bravo!\footnote{This reference will make sense
to any ex-Yugoslav teenager who lived through the 90s.}
And Gerhard Raven... is simply the {\it nec plus ultra}
of real-time data processing in High Energy Physics. Gerhard
wrote much if not most of the code behind LHCb's High Level Trigger, and
laid the foundations on which all the work described in this HDR stands. Ave maestro.

Finally, I would like to acknowledge the LHC busbars, whose 2008 failure delayed
the start of datataking just long enough for me to find a use for my brain in the LHCb trigger. 
I've had a lot of lucky breaks and second chances in my life, but that
was an especially fortuitous roll of the dice --- cheers!
\vspace{0.5cm}

\noindent I acknowledge funding from the European Research Council (ERC) under the European Union's Horizon 2020 research 
and innovation programme, under grant agreement No 724777 \href{http://teamlhcb-lpnhe.in2p3.fr/rubriqueRecept.html}{``RECEPT''}.
All cited online content exists at the time of writing.
\cleardoublepage

\tableofcontents


\cleardoublepage
\pagestyle{headings}
\pagenumbering{arabic}
\flushbottom

\parskip=1.5ex

\chapter{Eastbound and down: an introduction to real-time analysis}
\pagestyle{myheadings}
\markboth{\bf Eastbound and down}{\bf Eastbound and down}

\begin{flushright}
\noindent {\it Eastbound and down, loaded up and truckin'\linebreak 
We gonna do what they said can't be done. \linebreak
We got a long way to go, and a short time to get there \linebreak
I'm eastbound just watch ol' Bandit run!\linebreak}
\linebreak
-- Jerry Reed, Eastbound and Down
\linebreak
\end{flushright}

When hearing someone claim to have performed a ``real-time analysis of the data'', three questions immediately spring to mind. First, what does
``real-time'' mean? Second, what does ``data'' mean? And third, what does ``analysis'' mean? All of these terms can,
and do, mean quite different things in different scientific domains. Even within the specific domain of High Energy Physics (HEP)
which will concern us here, real-time can mean anything from microseconds to days, data can mean anything from raw hits in a pixel detector
to reconstructed Higgs boson candidates, and analysis can mean just about any processing of this data from zero suppression operating on detector readout electronics to deep neural networks
running on distributed computing clusters. Moreover there is
hardly any pedagogical literature to light the way through this thicket of meanings: the entry level textbooks are too vague to be useful;
intermediate courses focus on statistical methods or, increasingly, trendy machine learning techniques\footnote{Revolutionize science with this one weird trick!};
and more specialist knowledge on the relationship between HEP detectors, data, and analysis is passed down from supervisor to supervisee without ever being written down anywhere. 
No wonder then that these semantic differences mean that a person speaking about real-time analysis can be, and often is, simultaneously confronted both with
colleagues claiming that real-time analysis is nothing more than a fancy name for what HEP has been doing for decades \textbf{and} 
colleagues claiming that real-time analysis is such a dangerous heresy that it must not be allowed to take root in HEP at all. 
In this HDR, I intend to explain why real-time analysis is a specific answer to a distinct and novel set of problems
faced by today's HEP experimentalists, and how real-time analysis can be used to increase the physics reach of our experiments without harming the fidelity of our results.
This document is intended to be readable by a typical graduate particle physics student,
so I will try to use words in their natural English meaning wherever possible, and define specialist terms or LHC jargon as I introduce them. 
Natural units are used and charge conjugation is implied throughout unless explicitly stated otherwise.  

Real-time data analysis is a concept which, at least within HEP, has arisen in the context of the Large Hadron Collider (LHC). It will therefore be useful
to begin with a brief overview of the LHC, its detectors, and the kinds of data analysis performed at them; a comprehensive overview can be found
in the LHC book~\cite{Voss:2009zz}. The LHC is a particle accelerator which
nominally\footnote{During the 2010-2012 period, the LHC ran at a reduced centre-of-mass energy of first 7, then 8~TeV, and a reduced collision frequency
of 15~MHz.} collides two beams of protons at a centre-of-mass energy of 14~TeV over the course of approximately 10-15 hour long ``fills'',
interspersed with 2-3 hour long pauses during which the machine is refilled with protons.
The LHC operates in blocks called ``Runs'': Run~1 took place between 2010-2012 and Run~2 started in 2015 and is scheduled to end in 2018. Two factoids can
help capture the scale of the problem: each beam of protons carries the energy of a fully loaded
TGV train travelling at $\approx 150$~kmh$^{-1}$, and the radiation pressure at the centre of the collision is similar to resting the sun on the head
of a pin.  Each proton beam is composed of individual proton bunches separated in time
by 25~ns, corresponding to a maximum of 40 million bunch crossings per second (MHz). For reasons of beam stability, not all 25~ns bunches can be filled with protons, and rate
at which non-empty bunches cross and collisions can occur is in practice limited to around 30~MHz.
Each bunch crossing results in one or more $pp$ collisions, which in turn produce particles and force carriers whose
properties we wish to study.\footnote{For brevity, both particles and force carriers will be referred to as particles in the rest of this document.} 
There are four major detectors of particles at the LHC: ALICE, ATLAS, CMS, and LHCb, whose goal is to record the trajectories
of the particles produced in these $pp$ collisions, identify these trajectories and use them to measure the particle four-momenta,
and consequently make inferences about the underlying physical properties of these particles.  
The work described in this HDR was done using LHCb, a general purpose detector which is particularly optimized for the study of ``heavy flavour'' particles
such as hadrons containing bottom or charm quarks, and which will be described more fully in the following chapter.

The essential difficulty faced by the LHC experiments is the volume of the data produced by the LHC. For example, a typical bunch crossing 
in the LHCb experiment, which produces 1 $pp$ collision on average, results in around 30~kB of data, or a nominal data rate of around 1~TB per second. For ATLAS and CMS these
data rates are between one and two orders of magnitude greater, both because these detectors are designed to work with a larger number of $pp$ collisions per bunch crossing,
and because these detectors are physically larger and record more information for each given $pp$ collision. 
As the LHC takes data for around $4\cdot 10^6$~seconds per year~\cite{Pojer:2012zz}, its detectors together process on
the order of 100~EB of data each year, on a par with today's largest commercial data processing applications. Two problems arise from these facts.
Firstly, economically transferring Terabytes of data per second from the detector electronics to an off-detector data processing centre
is only just becoming possible, while the detectors were designed two decades ago, when it was not.
And secondly, even if the data could be transferred off the detectors, it is not economically feasible to store and distribute
that much data to the thousands of physicists around the world waiting to analyse it. Furthermore,
a collider designed to find the Higgs boson and to search for putative new particles does not need
to analyze all bunch crossings in detail, because the overwhelming majority of these will only produce well known Standard Model processes
which are of interest only insofar as they may form a background to the more exotic processes which we do wish to study.
For these reasons, the LHC experiments use a process known as ``triggering''
in order to select roughly between $0.001\%$ and $0.01\%$ of the most interesting bunch crossings collected 
by the detectors for further analysis and \textbf{permanently discard} the rest. 

This physical constraint on how data can be processed naturally leads to a specific way of thinking about data and analysis.
First of all, the fact that not all data can be moved from the detector electronics to off-detector data processing centres means that
a significant part of a trigger's job must be performed by specialized electronics located close to the detector. This specialized electronics
typically has access to information from those parts of the detector which are either intrinsically low occupancy and can be fully read out, such as
the muon detectors, or which can be naturally aggregated into reduced-granularity regions of interest while keeping most of their power to
select interesting bunch crossings such as calorimeters.\footnote{Charged particle trajectory finding (tracking) systems are generally hard to aggregate into reduced-granularity regions of interest,
and absent this aggregation their data rate is often too high to allow a processing of the full data rate even in specialised electronics. This is particularly true of
those tracking systems located closest to the collision point, which have the highest occupancies and data rates. Even if enough high-speed data links could be placed
in the detector volume, and the entire system cooled adequately, the increase in the amount of material seen by the particles as they traverse the detector would lead
to an unacceptable degradation in the physics performance. Nevertheless there are counterexamples, notably
the tracking triggers used by CDF and H1 which processed significantly lower data rates, and the planned track trigger of the high-luminosity upgrade of the CMS experiment
which uses outer tracking layers where the readout issues are expected to be manageable a decade from now.} This first
part of a trigger system is usually called a ``hardware'' trigger. Because hardware triggers tell the detector electronics whether or not to read out a
bunch crossing, they must make their decision within the same fixed amount of time, known as ``latency'', for each bunch crossing. The latency 
is set by the depth of the buffer which stores the data in the readout electronics while waiting for a hardware trigger decision.

As they can only access information from a limited set of subdetectors, and only with reduced granularity and resolution, hardware triggers are
only able to reduce the rate of interesting bunch crossings by one to two orders of magnitude before they are saturated by backgrounds. These backgrounds,
at least in the case of the LHC experiments, are primarily caused by well-known Standard Model processes which, when described purely using the limited information
available to the hardware triggers, look very similar to the processes of interest which the trigger is trying to select. Fake or badly reconstructed particle trajectories\footnote{These fake particle
trajectories are sometimes called ``ghosts'' in the HEP literature.}
or energy deposits also contribute, but are generally subleading causes of background. It is therefore necessary to perform a more complete event reconstruction in order
to add information from the rest of the detector, and improve the granularity and resolution of this information, so that the interesting signals can be separated
from the Standard Model backgrounds. Because the hardware triggers have reduced the data rate by one to two orders of magnitude, this improved reconstruction is now
possible. The full information about those bunch crossings which the hardware trigger has flagged as interesting is sent for this more complete reconstruction,
which is generally performed in dedicated data centres located as close as possible to the detectors in order to minimize costs associated with transporting the data
over long distances.\footnote{It is possible that in the future, aggregating these data centres will save more money in maintenance, cooling, and power consumption than
it will cost in transporting large volumes of data far from the detectors where they are produced. This has been studied by CERN in the context of building a data centre in Prevessin~\cite{thebird}
which would serve the LHC experiments over the next decades, but it is not currently seen as a cost-effective solution.} These data centres may be equipped with a mixture of
off-the-shelf and bespoke data processing solutions, depending on the requirements of any given experiment, but regardless of the precise implementation they are usually called
``High Level Triggers'' (HLT). As they do not interact with the detector electronics, HLTs do not have to operate with a fixed latency --- instead, they are optimised so that
the average time taken to process a bunch crossing fits within the overall computing power available in the data centre and the number of bunch crossings arriving per second from
the hardware trigger. Local disk buffers are used to absorb temporary fluctuations in the load on the data centre, and bunch crossings which are particularly complicated
and time-consuming to reconstruct may be timed out in order to prevent a backlog from forming. If the rate at which these complicated bunch crossings occur is low enough,
they are typically also recorded to permanent storage for later inspection.

Although more complete, the detector reconstruction which can be performed in a traditional HLT is still somewhat less precise and complete than the reconstruction
which can be performed once a given bunch crossing has been recorded to permanent storage. This is partly a function of the finite processing power available within the data centre
implementing the HLT, and partially a function of the fact that the detectors are typically aligned and calibrated over weeks and sometimes months after the data has been
taken. While an HLT does not have to make decisions about each bunch crossing within a fixed latency, and the precise time constraints greatly vary depending on the detector,
the average processing times are still measured in milliseconds rather than days. This means that a fundamental assumption which underpins any traditional trigger system
is that the bunch crossings which the HLT identifies as interesting and records to permanent storage will be re-reconstructed once the ultimate detector calibrations are available.
This, in turn, means that once the HLT decides that a bunch crosing is interesting, all raw data from all the subdetectors must be saved to permanent storage in order to enable this re-reconstruction.
The hardware trigger and the HLT must also store information about how they reached their decision to keep a bunch crossing, to enable data-driven studies of their performance. 
By contrast to the trigger classification of bunch crossings, 
analysis is generally defined as the process of using the ultimate quality re-reconstructed data to make inferences about the particles produced
in each bunch crossing. It explicitly does not include the work done to classify the bunch crossing as interesting or otherwise by the trigger systems. The decisions made by the trigger
systems are rather seen as an empirical property of the data being analysed, and the efficiency of the trigger system to correctly identify interesting bunch crossings for later analysis
is treated by analysis in the same logical way as the hit efficiency of a layer of silicon within a given subdetector.

The traditional view of LHC data processing, then, is centered around the notion that the LHC produces far more data than either can be analyzed 
or which is interesting to analyze. Because of this, it is necessary to use real-time classifiers based on a fast and somewhat coarse detector 
reconstruction, called triggers, in order to spot the fraction of a permille of bunch crossings which are most likely
to contain interesting physical processes, and save them for later analysis. Because the trigger classification is fast and coarse, it cannot form the basis of
a final analysis of what actually happened in these bunch crossings: the detector must be recalibrated after the data has been permanently recorded, and this data
must then be re-reconstructed with the best available detector calibrations and performance. This in turn directly implies that while the real-time classification
algorithms can identify this or that bunch crossing as interesting, they must record the entire data for every selected bunch crossing for further analysis. As
most interesting bunch crossings, particularly in ATLAS or CMS, contain multiple independent proton-proton collisions, only one of which is the source of the physically
interesting process which led the real-time classifier to actually select this bunch crossing for permanent storage, this approach necessarily results in the saving of a great
deal of superfluous data related to the other uninteresting proton-proton collisions. 

By contrast to this traditional view of LHC data processing, real-time analysis affirms the feasibility of fully aligning and calibrating our detectors in real-time,
affirms the feasibility of fully reconstructing at least a large subset of all bunch crossings in real-time,
and asserts that most LHC bunch crossings produce interesting physical processes which are worth measuring. From these three axioms it follows that instead of triggering
a small fraction of bunch crossings to be fully preserved for later analysis, a large subset of the bunch crossings (ideally all) must be
analysed in real-time with the same precision as would be traditionally possible only once the data has already been recorded. And because this analysis is performed
in real-time with the best possible precision, it is then possible to select from each bunch crossing that small fraction of data which contains
the physical process of interest. This selection is desirable because by recording only this small fraction of each bunch crossing to permanent storage, we can dramatically increase the number
of bunch crossings for whom \textbf{some} information can be permanently recorded, and hence increase the physics output of our collaborations. 
Real-time analysis can therefore be thought of as a particular kind of trigger which
saves a little bit of information about many bunch crossings instead of a lot of information about a few bunch crossings, but as is hopefully clear the underlying
physics which motivates a real-time analysis system is fundamentally different from that which motivates a trigger system. 

We now also have the answers to the questions with which we began this chapter. Data means the signals recorded by a detector for each LHC bunch crossing.
Analysis means identifying the physically interesting subset of this data for each bunch crossing, and measuring its physical properties, which is to say
in most cases the four-momenta of the particles in question. And real-time means that this analysis of the data should be performed
rapidly enough that the available computing resources do not saturate; in other words, every bunch crossing should be analysed
on its merits, not discarded because the computing facilities were at capacity. It is important to note that real-time in this context does not generally
mean a commitment to analyse the data within any given time period: as we shall see in due course, by correctly structuring the problem it is possible
to stretch the concept of ``real-time'' quite a long way. The crucial feature of real-time analysis is not that it is fast, but that it is performed on data
which has not yet been recorded to permanent storage. By contrast term ``offline'' will be used throughout this document to mean any reconstruction or
analysis work which is performed on data which has already been recorded to permanent storage. 

The rest of this document will now describe the development of real-time analysis within LHCb. 
Similar developments occured around the same time within the ALICE~\cite{ALICEO2}, ATLAS~\cite{ATL-DAQ-PUB-2017-003}, and CMS~\cite{CMS:2012ooa} collaborations.
By comparison to LHCb the ATLAS and CMS approaches were restricted to a much smaller subset of analyses and did not benefit from the same kind of real-time detector alignment and calibration
as LHCb implemented and will be described later. On the other hand the ALICE schema very closely parallels that of LHCb in both motivation and execution:
a triggerless detector readout and a full real-time reconstruction, alignment, and calibration of the detector, driven by the fact that all collisions are to some extent interesting for analysis. 
This system, called O2, will not however come into being before the 2020s as it relies on a full upgrade of the ALICE detector.

Chapter~\ref{chpt:misenscene} introduces the LHCb detector and briefly describes
the way in which typical LHCb
analyses are carried out, with particular reference to how aspects of the analysis procedure will be affected by performing them in real-time.
Chapter~\ref{chpt:haystackofneedles} then uses LHCb as an example to demonstrate that most
LHC bunch crossings do contain physically interesting processes, and thus establishes the physics case for developing a real-time analysis infrastructure. 
Chapter~\ref{chpt:acunningplan} describes the requirements for
real-time analysis in LHCb, in particular the organisation of the real-time alignment and calibration of the detector. Chapter~\ref{chpt:unrealtime}
describes the optimization of the real-time reconstruction and of LHCb's two-stage HLT in order to fit the available computing power.
Chapter~\ref{chpt:uneprocedure} describes how the output of the real-time analysis must be formatted for permanent storage in order
to allow its use in later analysis. Chapter~\ref{chpt:charmxsec} describes the measurement of the cross-section to produce mesons containing charm
quarks, which was one of the first LHCb publications to use real-time analysis. Finally Chapter~\ref{chpt:lhcbrepent} discusses errors made
in this first real-time analysis, and their implications for future real-time analyses on LHCb and elsewhere.

I end this introduction by detailing which parts of the remaining document are my own original work, and which describe work
done by others. A trite, but accurate, answer is that everything documented in this HDR
is a non-linear combination of the author's own work and the contributions of numerous other collaborators.
Nevertheless it is fair to say that my role in this process was mainly organisational and supervisory,
which is reflected in the fact that most of the numbers and plots reported in this HDR were produced using scripts written by other people,
as noted in the Acknowledgements. I was deputy and subsequently project leader of the LHCb HLT at the time when the work documented in this HDR
was being performed, I proposed that we should attempt to fully reconstruct the detector in real-time and use the output of this reconstruction to
perform real-time analysis, and I personally recruited many new people to this project in order to make it happen. As a culmination of this process
I oversaw and coordinated both the implementation and commissioning of real-time analysis in LHCb. I also wish to acknowledge that much of the work
presented here is essentially a retelling of results which are already in the public domain, although almost all the text 
has been rewritten for the sake of a hopefully coherent presentation. The description of LHCb in Chapter~\ref{chpt:misenscene} relies on performance numbers
and plots reproduced from publically available sources\footnote{While I didn't make any of the plots, my work contributed to a few of them:  
between 2007 and 2009 I wrote the first implementation of the impact parameter resolution code
together with Michael Alexander, and I worked on a reoptimization of LHCb's primary vertex finding algorithms between 2013 and 2015
together with my PhD student Agnieszka Dziurda, which is also touched on in Chapter~\ref{chpt:unrealtime}}, however the description of LHCb's physics programme and associated analysis procedures is my own. 
Chapter~\ref{chpt:haystackofneedles} has substantial content overlap with the LHCb Upgrade Anatomy public note co-authored with Conor Fitzpatrick~\cite{Fitzpatrick:1670985}.
Chapter~\ref{chpt:acunningplan} has some content overlap with the internal LHCb note \textit{LHCb-INT-2013-031} which I edited and which documents the requirements for real-time analysis in LHCb. 
I would like to make it clear that while I coordinated the definition of these requirements, I did not myself work on either the organisation or implementation
of the real-time alignment and calibration of the detector. For this reason, the chapter does not go into a lot of detail about the actual performance of these.
Chapter~\ref{chpt:unrealtime} has some content overlap with the proceedings of a talk I gave at a workshop on machine learning in HEP~\cite{Gligorov:2015uja}.
Chapter~\ref{chpt:uneprocedure} documents work which was done under my supervision, and has content overlap with the commissioning paper for LHCb's real-time analysis~\cite{Aaij:2016rxn}
as well as two public notes on the topic to which I contributed~\cite{Anderlini:2199780,Lupton:2134057}. Chapter~\ref{chpt:charmxsec} has substantial content overlap with the LHCb
paper on charm cross-sections at 13~TeV~\cite{Aaij:2015bpa}, although it also contains some additional information about work done by me which were too detailed
for the journal paper. Chapter~\ref{chpt:lhcbrepent} draws on the published errata to this and several other LHCb papers which used real-time analysis,
although most of the commentary on the reasons for these errata and their implications for future real-time analyses is presented here for the first time. 

Having, I hope, taken care to not claim credit for things which I did not do, I do wish to end this introduction by taking credit for something
which I feel was quite a crucial part of why LHCb has a real-time analysis infrastructure in place today.
Although experimental high-energy physics is the domain of large collaborations, with hundreds if not thousands of researchers working together on every
published result, our work is still largely seen and judged in terms of individual achievement. We are, as a field, addicted to the question
of who had which idea, even though our everyday reality shows clearly that what really matters is which ideas found a solid enough team willing to coalesce around them and bring them to life.
I strongly believe that the vast majority of ideas neither emerge nor are developed in the minds of individuals. Ideas are produced through discussions,
and the precise individual in whose mind they happen to crystalize is generally as randomly selected among the interlocutors as the 
precise products of a given LHC collision. And ideas, particularly in collaborations of the scale which is now commonplace in high-energy physics, always require a vibrant community
to develop them from a virtual concept into a practical implementation which can yield publishable scientific results. We talk a lot about this community,
but in our concrete actions we almost always give the lie to our words by treating ideas as private property.
This is why I wish to explicitly state that my decisive contribution to the work presented here was not to have any particularly interesting ``physics idea'' but 
precisely to work over a number of years on building the HLT team of LHCb from private unanounced meetings in the office
of the so-called ``leader'' to a vibrant community exchanging and improving each other's ideas in the open. 
The work described is primarily, then, proof that such a team was built
and a selective record of that team's achievements. And insofar
as the phrase ``Habilitation a Diriger des Recherches'' means anything, I submit to you that this record demonstrates that I am
indeed ``habilit\'{e}''. 

\chapter{Mise-en-sc\`{e}ne: the LHCb detector and analysis methodology}
\label{chpt:misenscene}
\pagestyle{myheadings}
\markboth{\bf Mise-en-sc\`{e}ne}{\bf Mise-en-sc\`{e}ne}

\begin{flushright}
\noindent {\it Varburg Limuzina, komforna ma\v{s}ina,\linebreak
pravio je Pera, iz biv\v{s}eg DDR-a,\linebreak
i ne da ga je napravio, nego svaka \v{c}ast,\linebreak
stalno mu iz auspuha curi neka mast... \linebreak}
\linebreak
-- Atheist Rap, Varburg Limuzina 
\end{flushright}

\begin{figure}[ht]
\centering
\includegraphics[width=0.8\textwidth]{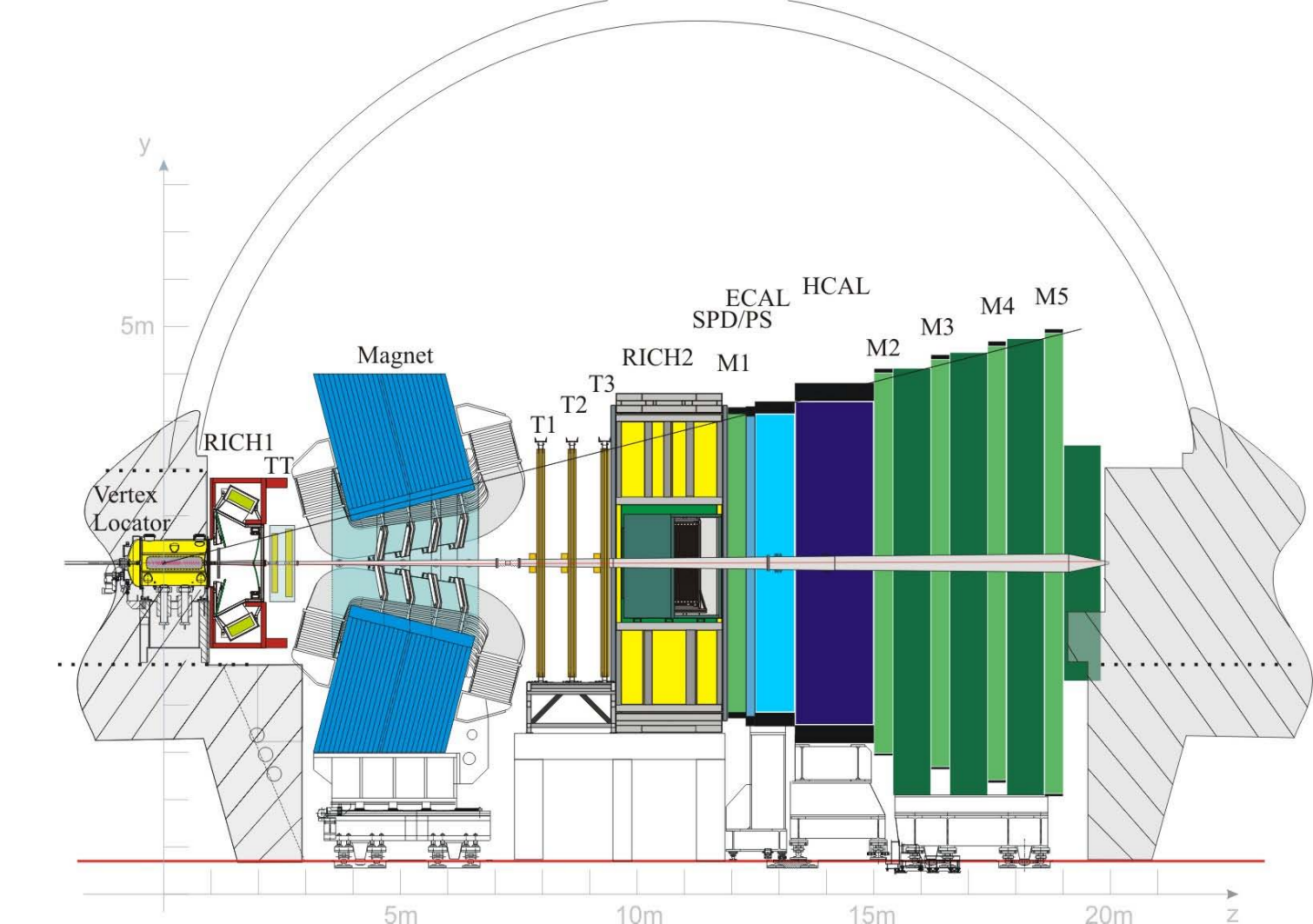}
\caption{\label{fig:lhcbdetector}The LHCb detector, reproduced from~\cite{Lindner:1087860}. The ``forward'' and ``backward'' directions
respectively refer to particles travelling from the VELO into, or out of, the LHCb acceptance.}
\end{figure}

The LHCb detector, shown in Figure~\ref{fig:lhcbdetector}, has been described in detail in the detector performance paper~\cite{Aaij:1978280}, as well as in the performance
papers of its subdetector components~\cite{LHCbVELOGroup:2014uea,Arink:2013twa,Adinolfi:2012qfa,Alves:2012ey,Aaij:2012me}. 
In addition, summaries of the detector and its components can be found in every PhD thesis
written by an LHCb member, including my own~\cite{Gligorov:2008zza}, and I do not wish to expand unnecessarily on this already voluminous body of paraphrased work.
Nevertheless, in order for this HDR to be a coherent standalone text, it is important that I give a brief summary of LHCb and its performance
here. Following this summary, 
I will describe the common LHCb analysis methodology, with particular reference to the way
it might be altered by being performed in real-time --- foreshadowing some of the discussion in later chapters.

\section*{The LHCb detector and its constituent parts}

LHCb is a forward spectrometer optimized for the study of particles between around 2~GeV and 200~GeV of energy. It has a 4~Tm dipole magnet which
bends charged particles in the horizontal plane, and whose polarity can be reversed to reduce charge asymmetries in the detector reconstruction.
LHCb has an outer acceptance of 300~mrad in the bending and 250~mrad in the non-bending plane. The inner
acceptance is around 5~mrad for the calorimeter and 10~mrad for the tracking system, corresponding to a pseudorapidity acceptance of $2<\eta<5$.
The system for finding charged particle trajectories (tracking) consists of a silicon-strip vertex detector (VELO) with $\textrm{\textbf{r}}-\phi$ geometry 
placed outside the magnetic field so that the tracks are straight lines, which surrounds the LHC interaction region and measures the location of the
$pp$ collisions (primary vertices), as well as the distance of closest approach between tracks and primary vertices, known as their impact parameter. Tracks
with small impact parameters are identified as direct products of a $pp$ collision, while tracks with large impact parameters are identified as decay products
of particles such as strange, charmed, or beauty hadrons whose lifetimes and boosts were sufficiently large to measurably displace their decay points from the primary vertex.
In addition, there are three stations of trackers (T1-T3, collectively the T-stations) 
after the dipole magnet which use silicon strips close to the beampipe and straw-tubes further away from it, and
a tracking station placed just before the magnet (TT) which uses silicon-strip sensors. The TT plays a crucial role in correctly matching track segments in the VELO to segments in the 
T-stations, and extending LHCb's acceptance for long-lived light particles such as $K^0_S$ mesons and $\Lambda$ baryons. 

\begin{figure}[htb]
\centering
\includegraphics[width=0.7\textwidth]{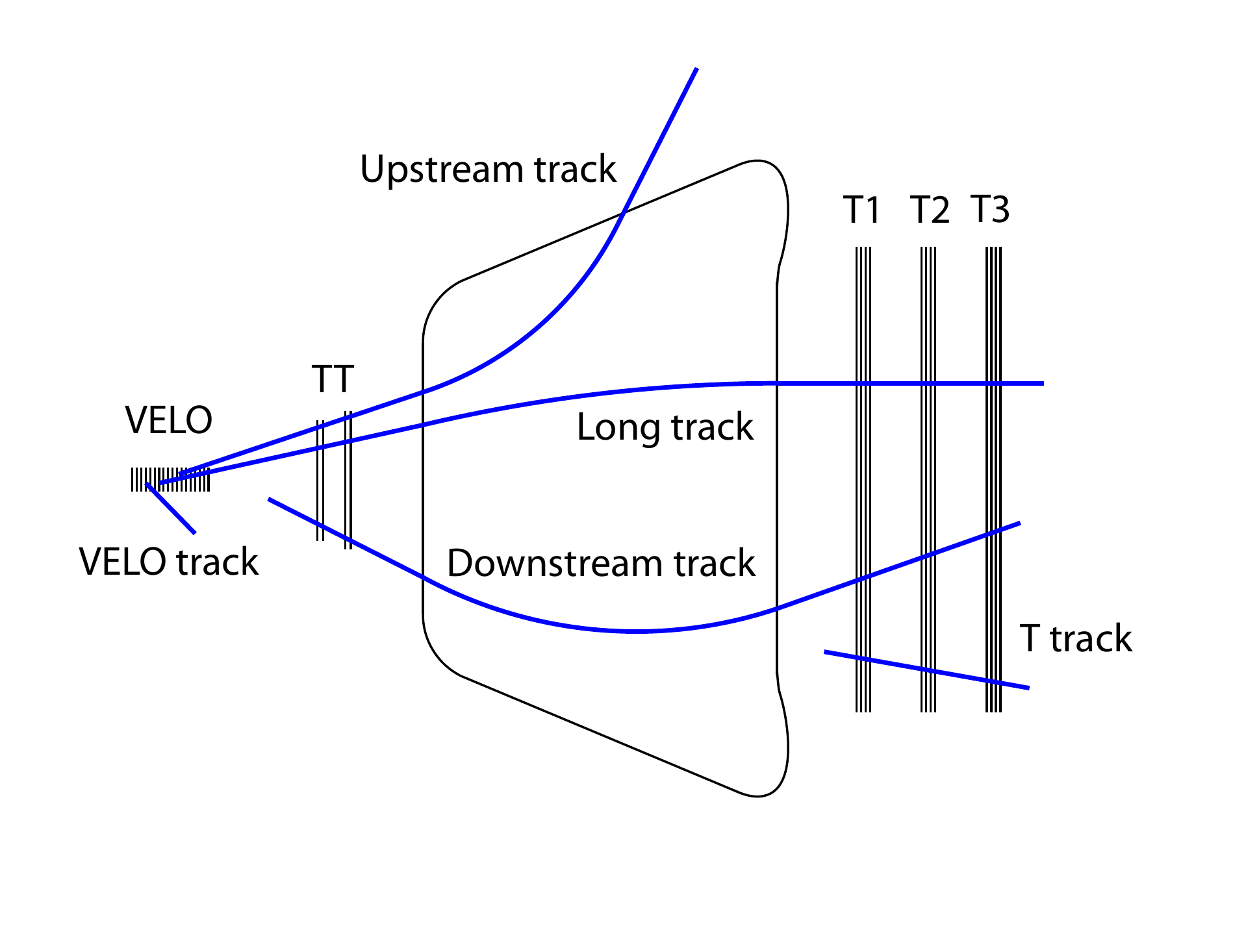}
\caption{\label{fig:tracktypes}Different kinds of tracks which can be reconstructed in LHCb, reproduced from~\cite{LHCbPerfPlots}.}
\end{figure}

LHCb uses a number of track finding algorithms, which are optimized for tracks originating from different points in the detector acceptance and provide a measure
of redundancy. A detailed summary of their performance can be found in a dedicated technical paper~\cite{Aaij:2014pwa}.
The different categories of tracks which can be reconstructed by these algorithms are shown in Figure~\ref{fig:tracktypes}. The most important
category is ``long'', which contains those tracks which originate in the vertex detector and traverse the full tracker acceptance including passing through
both the TT and T-stations.\footnote{LHCb track finding algorithms sometimes allow long tracks to be formed even without TT hits, because the TT acceptance is
smaller than that of the VELO or T-stations, particularly near the beampipe. However the percentage of fake tracks is much higher in this sample, partially
because of the higher occupancy near the beampipe and partially because without a TT hit it is much harder to match the correct VELO segment to the correct
T-station segment.} The second-most important category are ``downstream'' tracks, which consist of a T-station segment matched to hits in the TT. These tracks
more than double LHCb's acceptance for light long-lived particles, and the TT station hit is critical because it provides a measurement before the magnet
which allows a precise measurement of the downstream track momenta (though not as precise as for the long tracks). Upstream tracks and T-seeds play more peripheral roles
in the LHCb reconstruction. Upstream tracks consist of a VELO segment matched to hits in the TT, and in principle expand the detector acceptance, especially for low
momentum particles which would otherwise be swept out by the magnet. However because of the weak and non-uniform magnetic field around the TT\footnote{A consequence of the iron shielding
used to protect the photodetectors of LHCb's Ring Imaging Cherenkov detectors, which are described later.} the momentum resolution for these tracks is both
worse than $10\%$ and significantly non-Gaussian, making them difficult to use for precision measurements. T-tracks have an even worse momentum resolution, around 25\%, and
therefore cannot be used to further expand LHCb's acceptance for light long-lived particles. However, they play an important role in identifying
Electromagnetic Calorimeter (ECAL) clusters which originate from electrons and thus improving the purity of LHCb's photon reconstruction. This is particularly
relevant because of the many low momentum electrons produced in the region after the magnet which are not reconstructible as any other track type. The final
category are VELO tracks, for whom a momentum measurement is impossible but which are nevertheless very important for the primary vertex reconstruction, not least
because the VELO is the only LHCb subdetector able to find tracks in the backwards direction. VELO tracks also play an important role in classifiers designed
to select isolated particles (for example when searching for rare or forbidden decays), and the right-left or up-down imbalance in the number of reconstructed VELO tracks can
even provide a rudimentary measure of missing energy.\footnote{A rather cute point made to me by Mika Vesterinen, although he has so far sadly resisted the urge to turn it
into a single-author methodological paper. Get writing Mika, you've got research quality metrics to optimize!}

\begin{figure}[htb]
\centering
\includegraphics[width=0.5\textwidth]{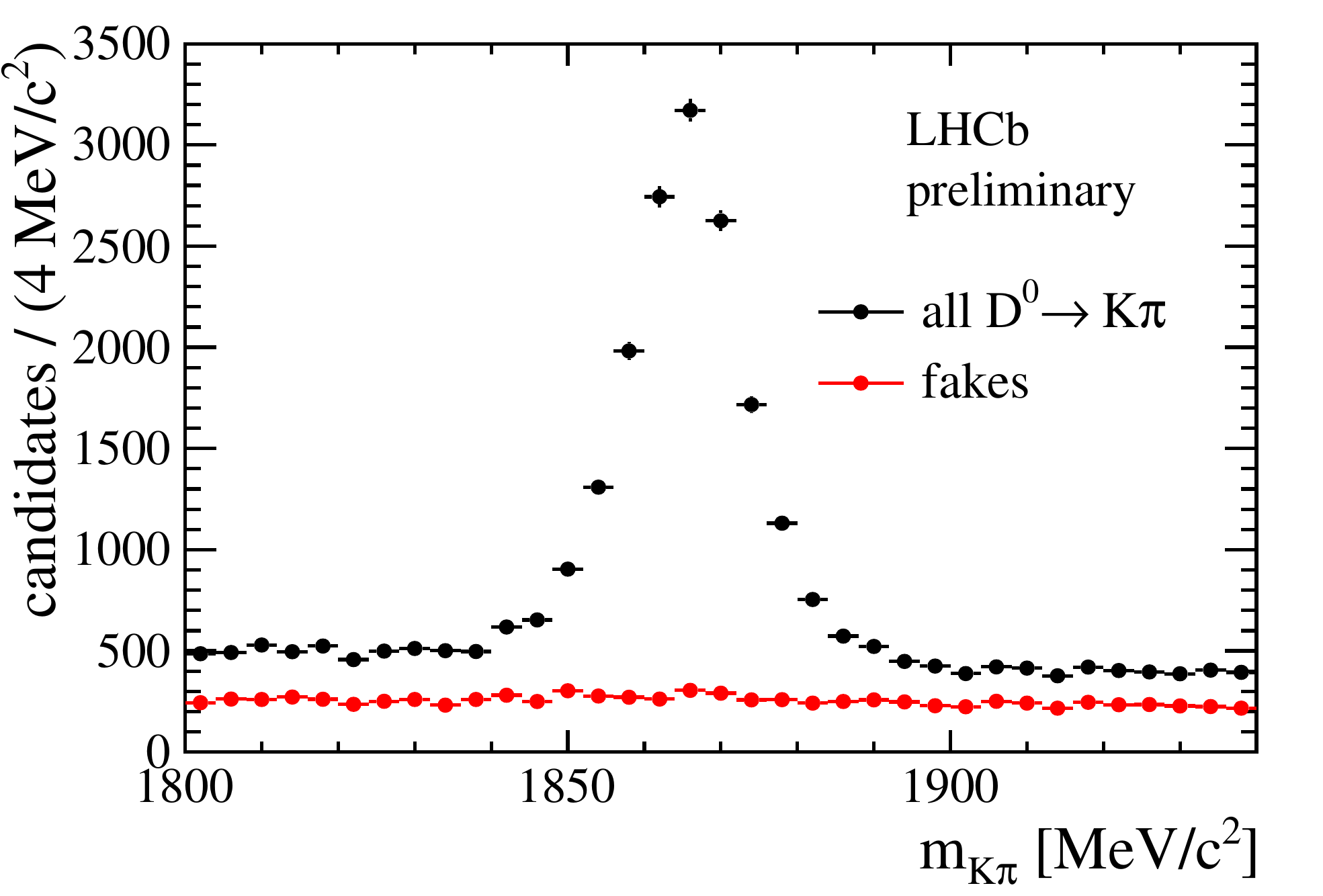}
\caption{\label{fig:ghostprob}Performance of the ghost probability neural network classifier on Run~2 data, reproduced from~\cite{LHCbPerfPlots}. Black are all $D^0$ candidates found in the
data sample, while red are $D^0$ candidates for which at least one child track has been identified as a fake.}
\end{figure}

\begin{figure}[htb]
\centering
\includegraphics[width=0.45\textwidth]{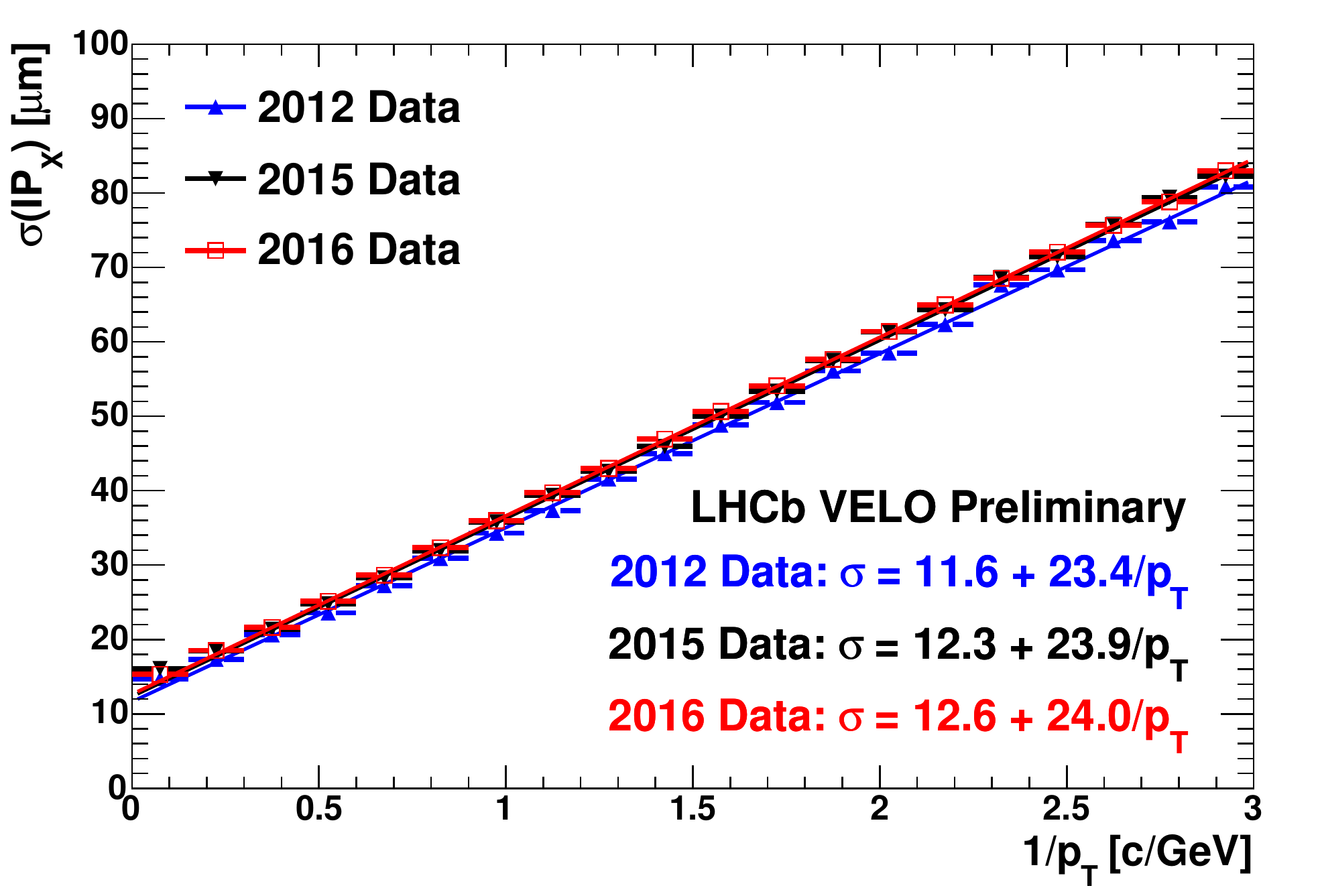}
\includegraphics[width=0.45\textwidth]{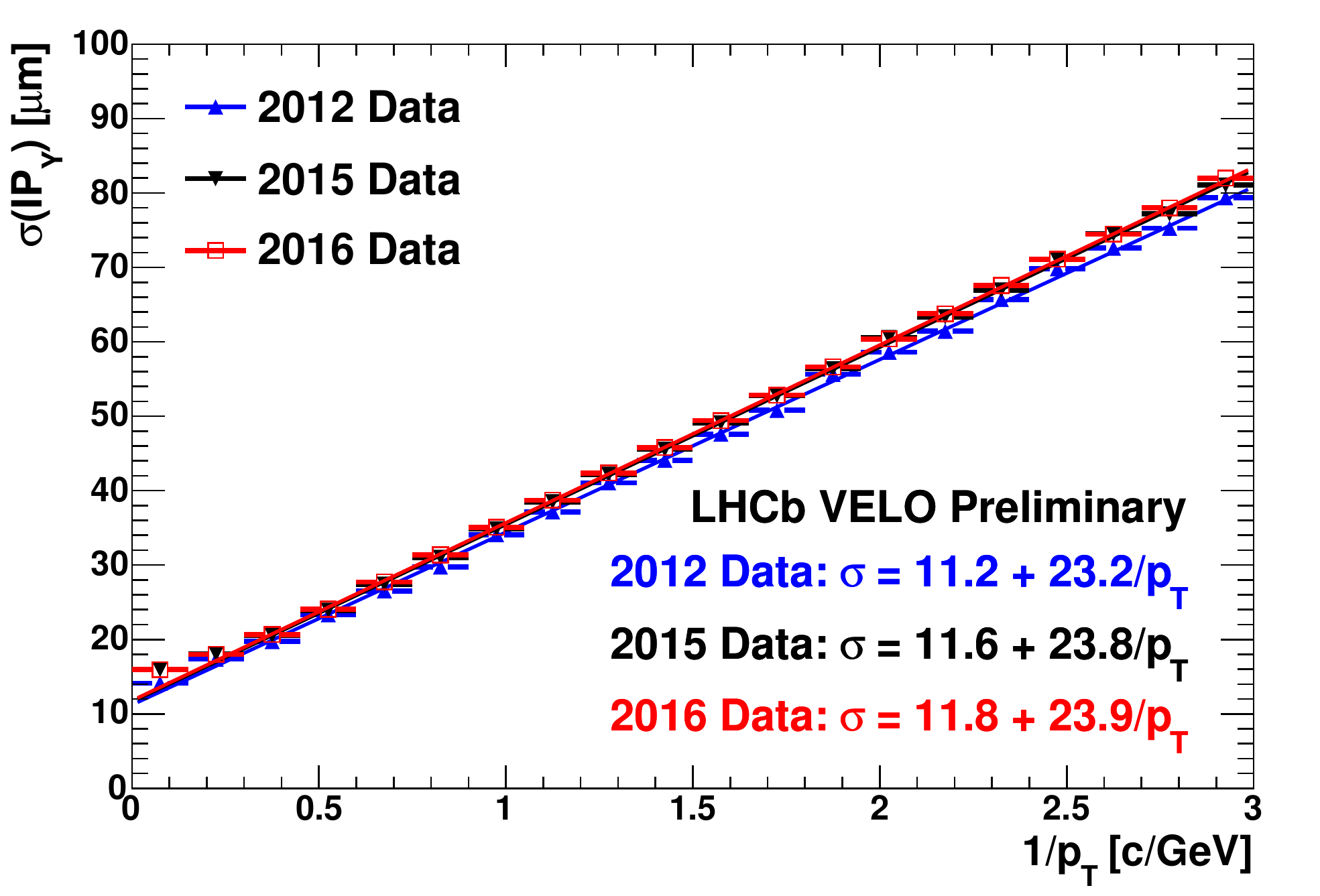}
\caption{\label{fig:ipres}The (left) \textbf{x} and (right) \textbf{y} impact parameter resolution of the LHCb detector for 2012, 2015, and 2016 data, reproduced from~\cite{LHCbPerfPlots}.}
\end{figure}

\begin{figure}[htb]
\centering
\includegraphics[width=0.45\textwidth]{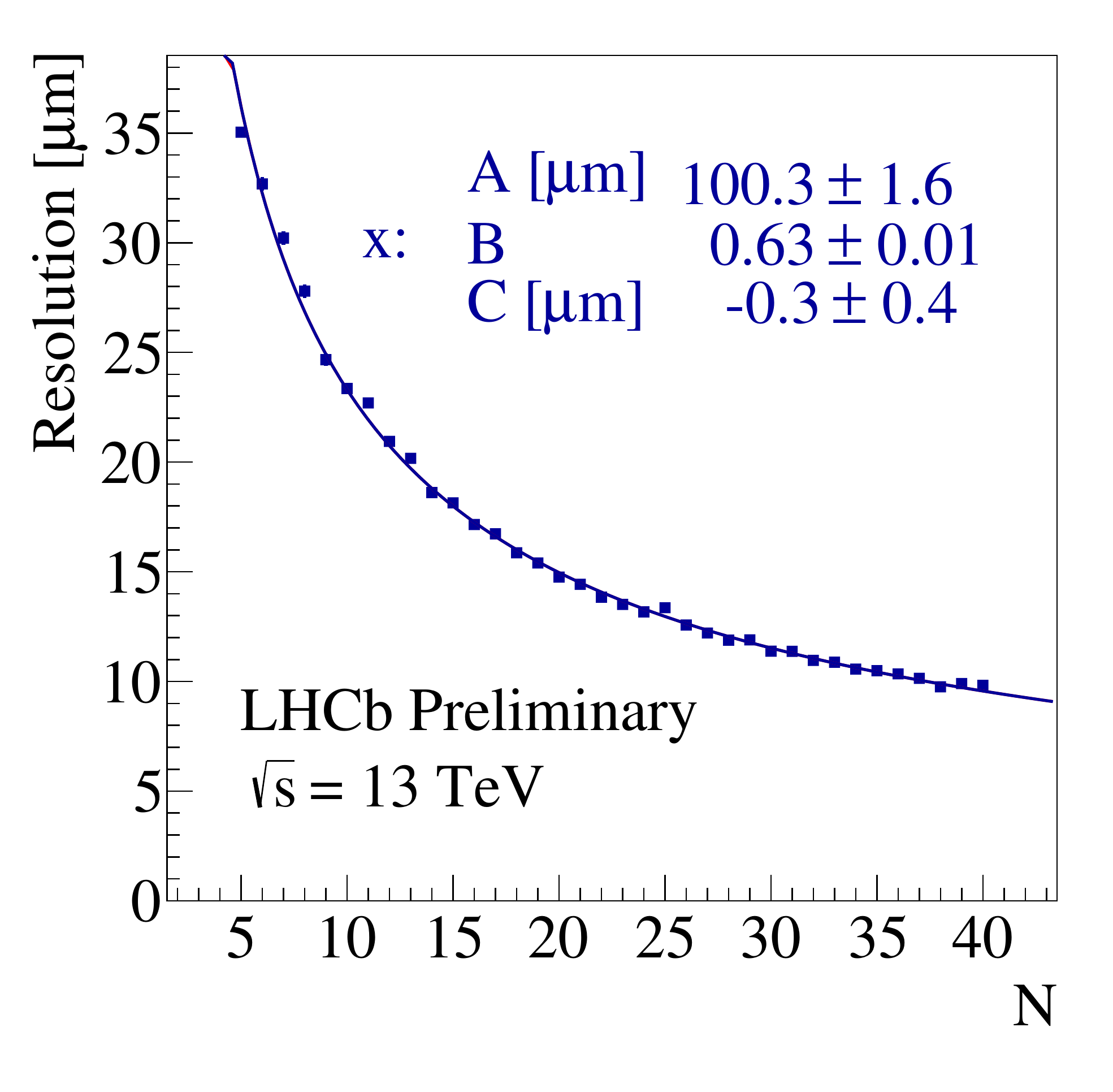}
\includegraphics[width=0.45\textwidth]{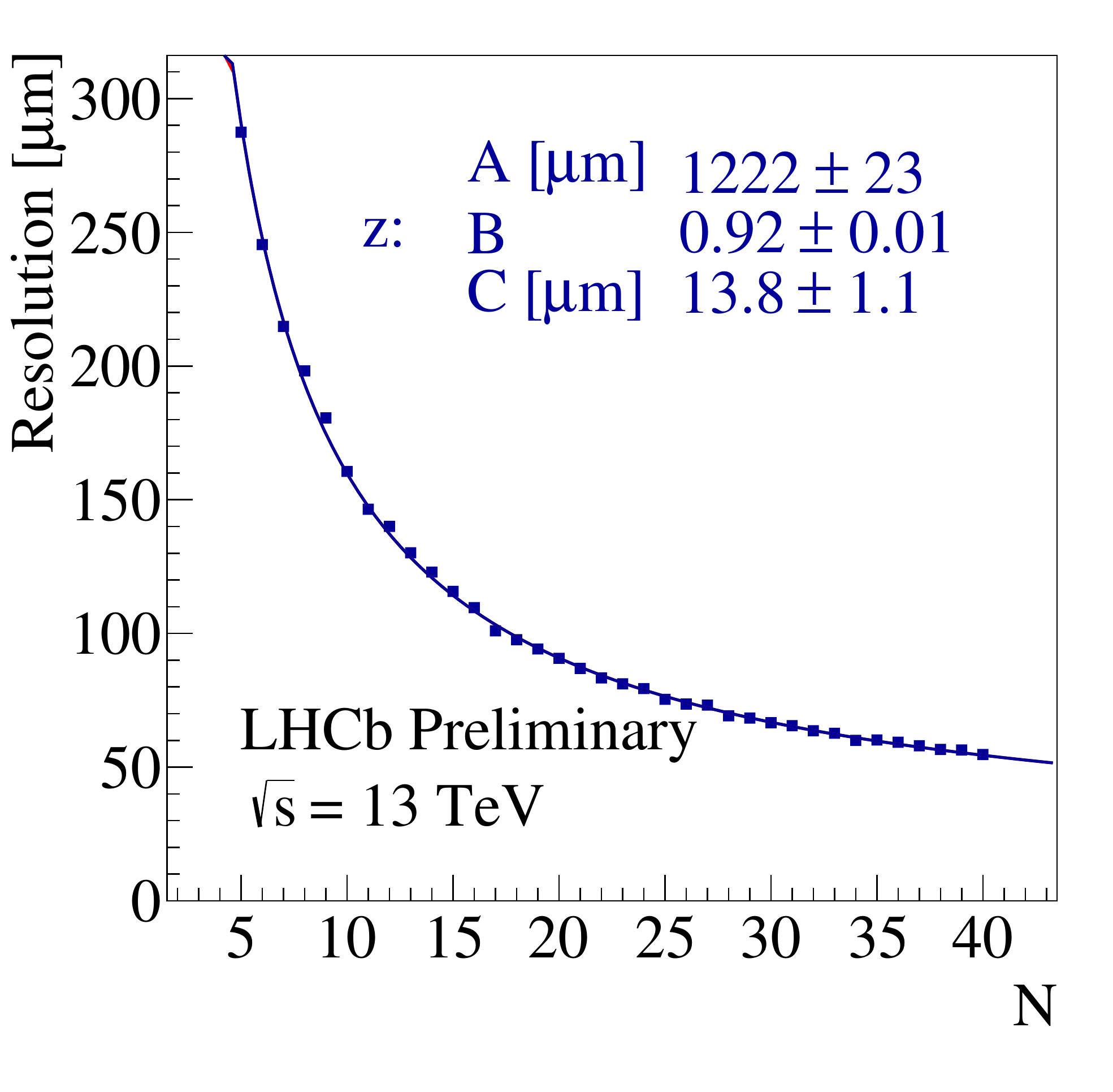}
\caption{\label{fig:pvres}The resolution on the (left) \textbf{x} and (right) \textbf{z} position of the primary vertex of the LHCb detector for 2015 data, reproduced from~\cite{LHCbPerfPlots}.}
\end{figure}

\begin{figure}[htb]
\begin{textblock*}{17cm}(3.5cm,3cm)
\raggedright
\includegraphics[width=0.55\textwidth]{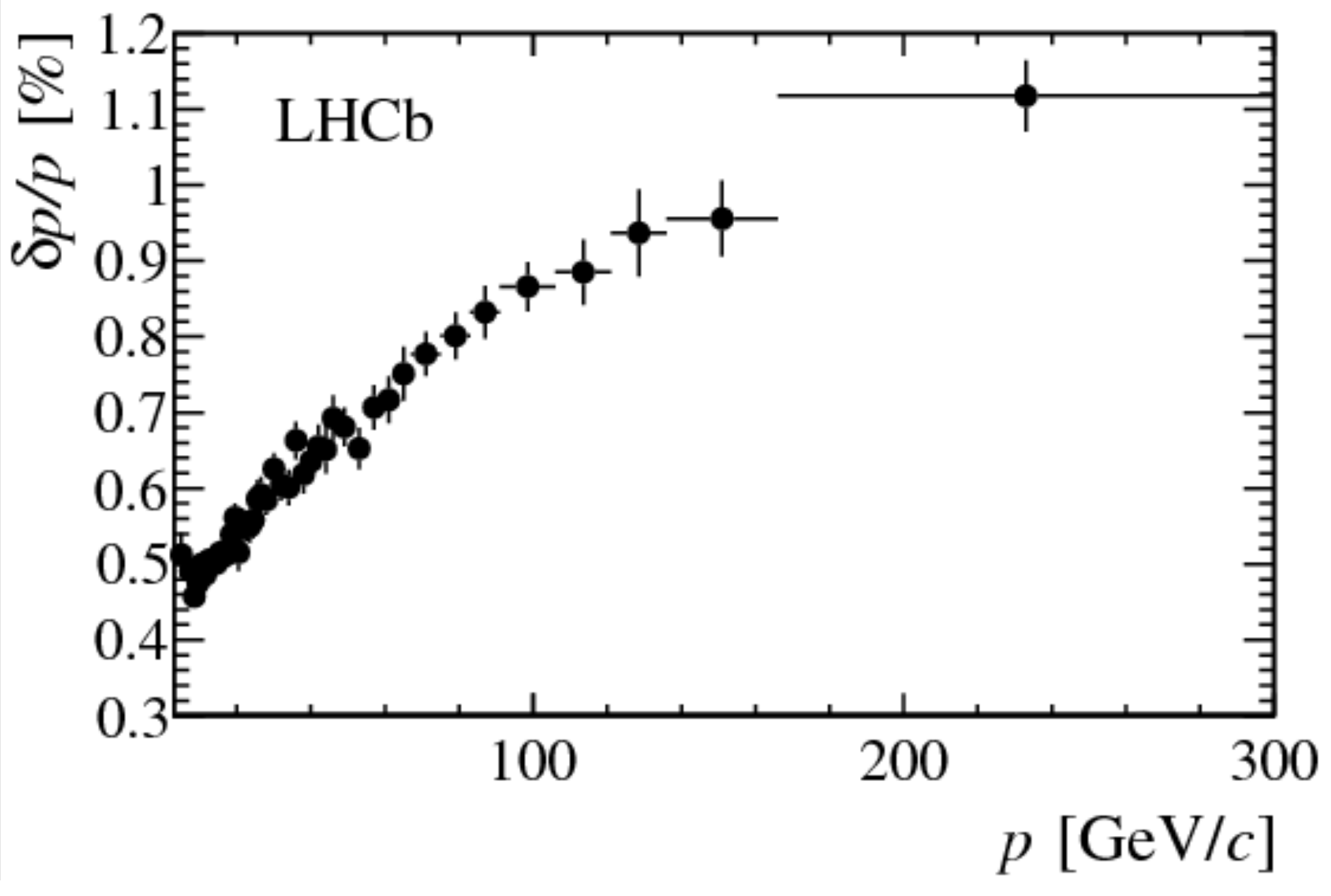}
\includegraphics[width=0.6\textwidth]{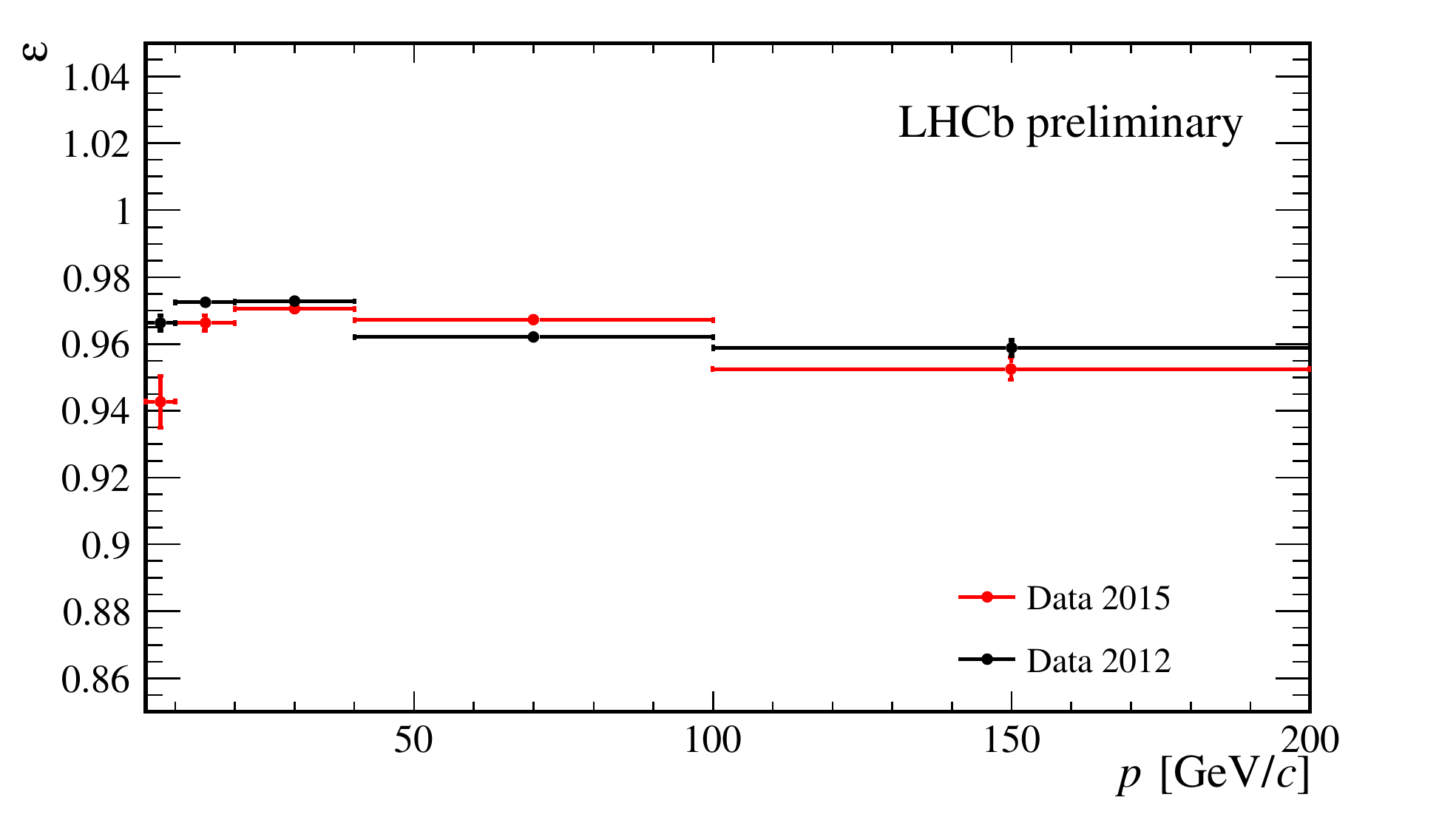}
\includegraphics[width=0.6\textwidth]{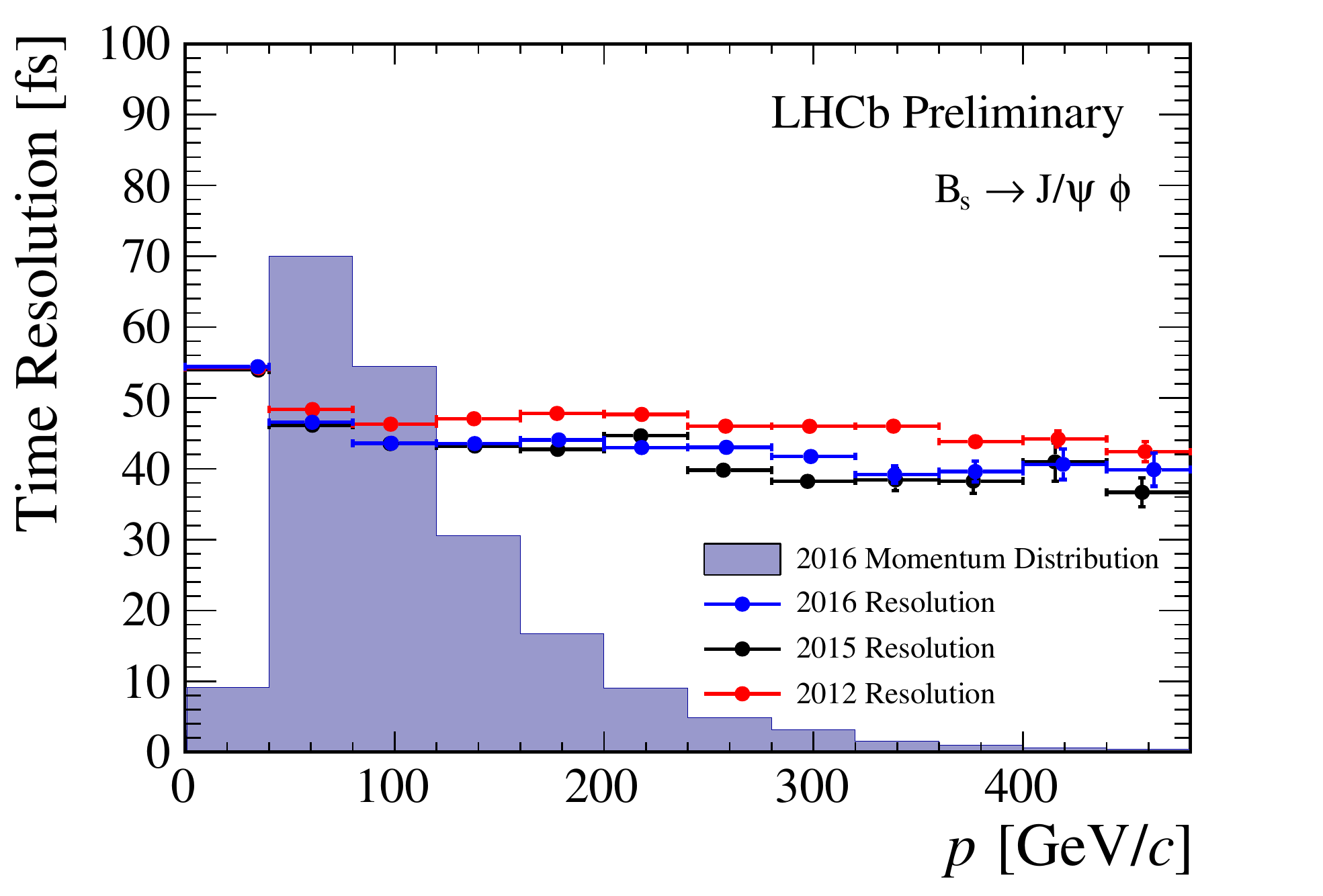}
\end{textblock*}
\caption{\label{fig:trackingbig}Top: the LHCb long track momentum resolution in Run~1, reproduced from~\cite{Aaij:1978280}.
Middle: the LHCb long track-finding efficiency for 2012 and 2015 data, reproduced from~\cite{LHCbPerfPlots}.
Bottom: the LHCb decay-time resolution for 2012, 2015, and 2016 data, reproduced from~\cite{LHCbPerfPlots}.
}
\end{figure}

LHCb's track finding algorithms use a Kalman filter based fit (Kalman fit) to improve the resolution on track parameters and to arbitrate between tracks which share detector hits. In addition,
the $\chi^2/\textrm{dof}$ of the Kalman fit is an important figure of merit and allows a substantial number of fake tracks to be rejected. Further fake track rejection
is provided by a neural network~\cite{DeCian:2255039} which uses information from all tracking subdetectors to assign a ``ghost probability'' to each track. Apart from the Kalman fit,
the ghost probability makes particular use of TT hits to ensure that the true VELO track segments are matched to the correct true T-station segments. The performance
of this fake track rejection algorithm is shown in Figure~\ref{fig:ghostprob} for a typical working point. 
LHCb's impact parameter resolution in shown in Figure~\ref{fig:ipres}, the primary vertex resolution in Figure~\ref{fig:pvres},
while the long track finding efficiency, long track momentum resolution and decay-time resolution
are shown in Figure~\ref{fig:trackingbig}. 
We can see that the momentum resolution is typically between 0.5 and 1 percent, the primary vertex resolution
is typically some tens of microns in the transverse direction and a couple of hundred microns along the beamline, and the impact parameter resolution has a constant
term of around 10 microns due to the primary vertex resolution, and a multiple scattering component of around 25 microns per unit of $1/p_{\textrm{T}}$, where $p_{\textrm{T}}$
is the track momentum transverse to the LHC beamline. These raw performances translate into mass resolutions of a few MeV for the decays of strange and charmed hadrons,
between roughly 10 and 25 MeV for the decays of beauty hadrons (depending on the mass difference between the parent and child particles), and a decay time resolution of
around 45~fs which is largely independent of particle species or momentum. They make it possible for LHCb to precisely measure the masses, lifetimes, and differential kinematic
distributions of both hadrons and electroweak bosons produced within its acceptance. 

Because the resolutions on quantities related to the displacement 
strongly depend on the transverse momenta of the particles, most analyses incorporate knowledge of the uncertainty dependence when using displacement
related quantites in classifiers or fits. This is most commonly done by measuring the change in the $\chi^2$ of a vertex fit when adding or removing particles from the vertex: for example,
the impact parameter $\chi^2$ of a track is the change in the primary vertex fit $\chi^2$ under these conditions. These quantities are preferred over the more historical ``significances'',
formed by dividing the impact parameter by its estimated uncertainty, because they properly take into account the diagonal terms in the track covariance matrix at the first measured point. 
Similar $\chi^2$ quantities are formed for flight distances of composite particles, distances of closest approach to secondary vertices, and so on.

\clearpage

\newpage

\begin{figure}[ht]
\centering
\includegraphics[width=0.45\textwidth]{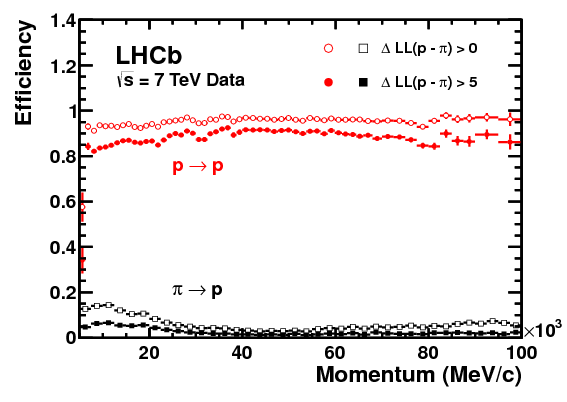}
\includegraphics[width=0.45\textwidth]{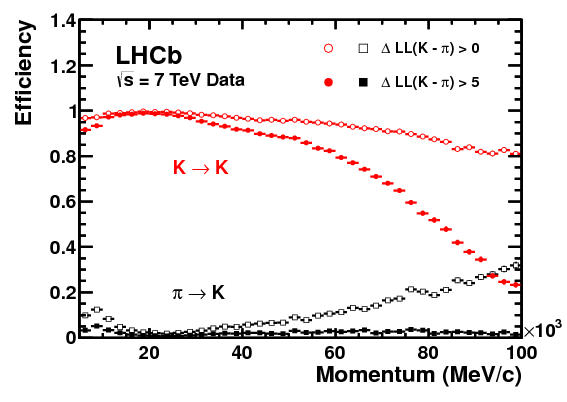}
\includegraphics[width=0.7\textwidth]{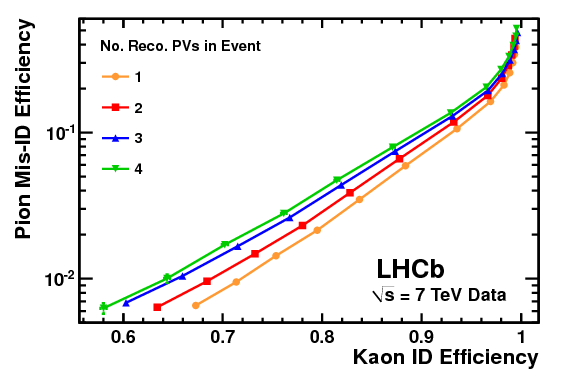}
\caption{\label{fig:richperf}Top: the (left) proton-pion and (right) kaon-pion identification efficiency and misidentification rate as a function of particle momentum, measured
with 2011 LHCb data. Bottom: the kaon-pion efficiency and misidentification rate as a function of the number of reconstructed $pp$ collisions (PVs) in a bunch crossing (event). 
Reproduced from~\cite{Adinolfi:2012qfa}.}
\end{figure}

In addition to its ability to find tracks, LHCb has a number of detector systems which can be used to identify the charged particle responsible for each track. Within
the LHCb acceptance, the stable charged particles are electrons, muons, pions, kaons, and protons.\footnote{There are exceptions, for example an LHCb analysis
which explicitly used the particle identification systems to search for long-lived stable charged particles~\cite{Aaij:2015ica}, but they are rare.} 
This identification can either be based on requirements within a specific subdetector, or on global likelihoods or neural networks which use information from
all subdetectors to assign a particle type to a track. These different identification methods serve different purposes: fast identification approaches which can be used
in the earliest stages of the LHCb trigger may be based on only a single subdetector, while the final identification algorithms used in analyses are almost
always of the global kind.  
The earliest, hardware, trigger identifies muons as stubs in the muon stations located behind the calorimeter (M2-M5), matched to a hit in the M1 station in front of the calorimeter
which is used to improve the momentum resolution by providing a measurement closer to the magnet. Electrons are identified as clusters in the ECAL which have
matching hits in the preshower system. Later on, hits or stubs in these subdetectors can be matched to the relevant tracks to identify them as muons or electrons independently of the rest
of LHCb. Hadron particle identification is primarily provided by two Ring-Imaging Cherenkov (RICH) detectors, one located in front of the magnet which identifies lower momentum particles, and one
located after the T-stations which identifies higher momentum particles. The second RICH has a reduced acceptance of 120~mrad in the bending and 100~mrad in the non-bending plane, an exercise in
cost-cutting justified by the fact that the highest
momentum tracks tend to be the most highly boosted and hence highest-$\eta$ ones. A dedicated algorithm associates tracks to rings reconstructed in the RICH system, and
assigns a likelihood to each track of being a given particle type. A representative sample of the RICH performance in Run~1 datataking is shown in Figure~\ref{fig:richperf}. Note in particular 
that the clearest separation between particle hypotheses is achieved between around 20 and 60~GeV of momentum, kaon-pion separation is not really possible above 100~GeV of momentum, 
and that the performance degrades with an increasing number of $pp$ collisions
and therefore an increasing detector occupancy. 

As well as identifying electrons, the ECAL is used to reconstruct photons. The ECAL energy resolution is $\sigma^{\textrm{ECAL}}_{\textrm{E}} = 10\%/\sqrt(E) \oplus 1\%$, where the energy is
given in GeV. In order to improve performance for high-occupancy events, the ECAL is divided into three granularity regions based on the proximity to the beamline, 
with cell widths of 40.4, 60.6, and 121.2~mm in the inner, middle, and outer regions respectively. Photons are principally identified as ECAL clusters which neither have associated tracks
pointing to them nor have associated hits in the preshower system. In addition, however, multivariate identification algorithms are used to separate ECAL clusters caused by single
photons from those which result from the merger of multiple nearby photons, for example from the decays of highly boosted $\pi^0$ mesons. In addition to being final state particles,
photons reconstructed in the ECAL are particularly important when used to recover the Bremsstrahlung radiation emitted by electrons through their interactions with the LHCb material.

By contrast to the ECAL, the LHCb Hadronic Calorimeter (HCAL) is primarily used to identify high energy hadronic clusters at the earliest hardware trigger stage. Its energy
resolution, around $\sigma^{\textrm{HCAL}}_{\textrm{E}} = 80\%/\sqrt(E)$ (no official LHCb number exists), is too poor to be of any use in identifying long lived neutral particles
which do not decay in the LHCb acceptance, although HCAL clusters are used as part of LHCb's particle flow jet building, briefly mentioned later. 

\section*{LHCb data processing before real-time analysis}

The LHCb Run~1 data processing consisted of three principal components: the trigger, which reduced the LHC bunch-crossing rate of 40~MHz to around 5~kHz for long term storage;
the reconstruction, which was executed offline once the best detector alignment and calibration had been made available; and the stripping, which was essentially an aggregation of
around 1000 individual analyses executed centrally on this reconstructed data. The objective of the trigger system was to fill the available long-term storage with the bunch crossings which were of interest
to the greatest number of analyses. The objective of the reconstruction was to provide the best possible measurement of the properties of photons and stable charged particles, and to reduce
as much as possible the number of fake particles. And the objective of the stripping was to use this reconstructed information in order to apply more stringent selections than could
be performed in the trigger, but also to distribute the so-selected bunch crossings into around a dozen streams, further reducing the number of bunch crossings which any one analyst
had to individually access or study.

\begin{figure}[t]
\centering
\includegraphics[width=0.35\textwidth]{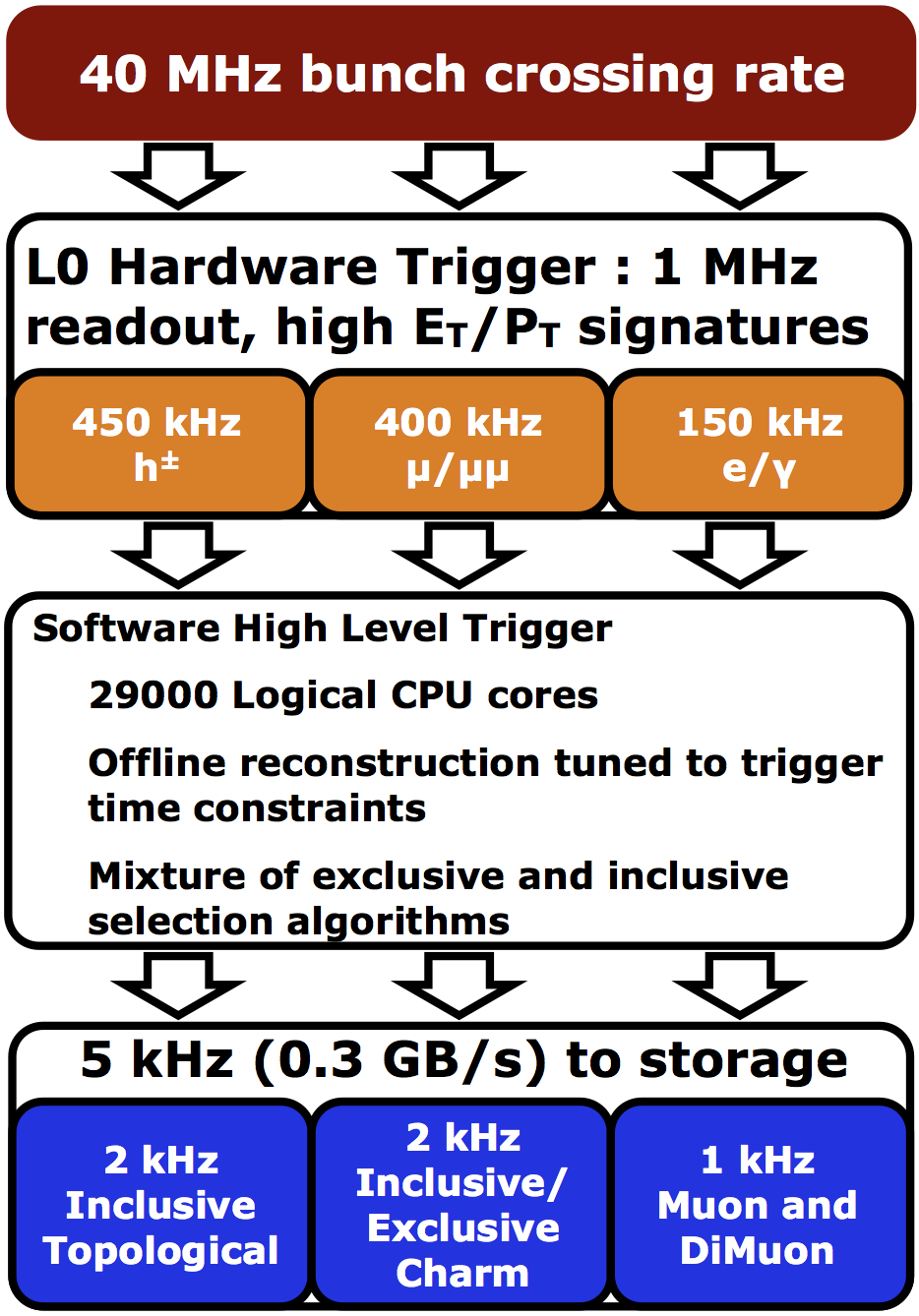}
\caption{\label{fig:lhcbtrig2012}A diagram of the 2012 LHCb trigger, see text for a description of the components. Inclusive and exclusive refer to triggers
which select a signal topology characteristic of a range of particle decays, or characteristic of a single particle decay, respectively.}
\end{figure}

A schematic of the LHCb Run~1 trigger system is shown in Figure~\ref{fig:lhcbtrig2012}, while a detailed summary of its performance can be found in~\cite{Aaij:2012me}. 
It consisted of a hardware trigger, whose job was to reduce the LHC bunch crossing rate to the maximal
rate at which all LHCb subdetectors can be read out (1~MHz), and an HLT which further reduced the rate for the 5~kHz which can be recorded to long-term storage. 
The hardware trigger was implemented in custom electronics and had access to information from the muon and calorimeter systems, and was therefore limited to selecting bunch
crossings which contain high-energy photons, electrons, or hadrons, or high transverse momentum muons or dimuon pairs. Because the hardware trigger had to reduce the rate by a factor 30, it
had particularly poor performance for analyses of charmed or strange hadrons, which are produced so abundantly that they would saturate the available bandwith if they were selected with full 
efficiency.\footnote{See also the discussion in Chapter~\ref{chpt:haystackofneedles}.} The HLT, implemented in a dedicated cluster of around 1500 servers with around 50000 logical
processes running in parallel, had access to information from all LHCb subdetectors and was equipped to run all of LHCb's
reconstruction algorithms, so long as they could be executed within the average processing time of around 50~ms available for each bunch crossing. During Run~1, as shown in the trigger diagram,
most of the HLT bandwidth was dedicated to three types of algorithms: a topological trigger~\cite{LHCb-PUB-2011-003,Gligorov:2012qt,Williams:1323557,Gligorov:1384380} 
which selected bunch crossings containing beauty hadrons,
muon and dimuon triggers, and a mixture of triggers for selecting bunch crossings which contained specific decays of charm hadrons. The HLT mainly used tracking and muon
identification information to make its decisions; electrons and photons were used for a small subset of specific triggers, as was hadronic particle identification information.
The information used by the HLT to make its decision, including the reconstructed objects and their parameters, was saved to long term storage alongside each bunch crossing, to allow
the HLT performance to be measured from the data itself later on.

Following the trigger, all selected bunch crossings were processed with a uniform ``offline'' reconstruction, with the objective of identifying photons and stable charged particles
in the LHCb acceptance and providing the most accurate estimate possible of their properties. Unlike the ATLAS or CMS reconstructions, the LHCb reconstruction did not build
more complex objects like $\tau$ leptons or jets; those were expected to be built later by individual analysts. There were two reconstructions: a prompt reconstruction executed
throughout any given year which used the best detector alignment and calibration available at that time, and a final end-of-year reconstruction which used the best possible
alignment and calibration parameters made available following a reappraisal of the detector's performance over the whole datataking year. Further discussion of LHCb's
alignment and calibration procedures is deferred to Chapter~\ref{chpt:acunningplan}, where the Run~1 strategy will be contrasted with the modifications which had to be made
to enable it to be used for real-time analysis.

Finally, the output of this reconstruction was used to strip the triggered bunch crossings into those which would be most interesting for currently ongoing LHCb analyses.
The basic problem which the stripping sought to address was that because the LHC
runs for a few million seconds per year, a 5~kHz trigger rate would mean that each analyst would have to sift through tens of billions of bunch crossings, quickly saturating
the computing resources available for analysis.
The idea behind the stripping was therefore twofold. First of all, it was assumed that the trigger would, and should, select a greater number of interesting bunch crossings than
LHCb authors were analysing at any given point in time --- partly because there are more possible analyses than people available to do them, partly because the trigger
can select bunch crossings with generically interesting topologies which might only become interesting after someone has a clever idea for how to analyse them. The stripping
could therefore reduce the data volume by applying somewhat loose versions of the optimal selection criteria for each analysis, safe in the knowledge that if the analyst made a mistake they could
go back and restrip the data some time later. And secondly,
the stripping was designed to provide a finer-grained streaming of the triggered bunch crossings, grouping bunch crossings of interest to related analyses into around
a dozen analysis streams of roughly comparable size, further reducing the volume of events which each analyst had to individually digest.
These stripping streams were then made available to the analysts, who could access the full raw detector information, full information from the detector reconstruction, as well as information on how the HLT
and stripping made their decisions to select a specific bunch crossing.\footnote{As Run~1 progressed, in order to save space, the stripping began reducing the amount of information
which the analysts could access, by progressively removing the raw detector and even reconstruction information which was not associated with particles selected in the stripping. This
idea of providing analysts with the minimal information needed for their analysis ($\mu$DST in LHCb jargon) was the forerunner of the TURBO stream described in Chapter~\ref{chpt:uneprocedure}
and laid key foundations for the eventual the design of real-time analysis in LHCb.}

\section*{LHCb analysis methodology}

LHCb's physics programme is briefly summarized in Table~\ref{table:lhcbphysprog}, roughly divided into the ``roadmap'' analyses which were considered~\cite{Adeva:1224241} as the most critical
objectives at the start of LHCb datataking, and analyses which emerged as interesting since Run~1 datataking began. 
The reason this divide is highlighted is because LHCb's analysis strategy and the ``event model'' governing
how analysis information is structured and persisted to storage were optimized for these roadmap measurements. This did not always make them optimal for all other analyses which
emerged throughout Run~1 datataking, in large part thanks to the performance of the inclusive HLT selections and the ease with which the HLT could be adapted to new analysis ideas.
When designing the real-time analysis infrastructure described in this document, however, the full breadth of this physics programme had to be considered and optimized for. 

\begin{table}\centering
  \begin{tabular}{llr}
    Subject & Analyses & Historical? \\
    \hline
    \multirow{7}{*}{\Pqb-hadrons}                  & Searches for rare decays & \multirow{6}{*}{YES} \\
                                                   & Time-integrated $CP$ violation & \\
                                                   & Time-dependent $CP$ violation & \\
                                                   & Dalitz measurements & \\         
                                                   & Angular measurements & \\
                                                   & Radiative decays & \\    
                                                   & Searches for forbidden decays & NO \\                                                                                
    \hline
    \multirow{5}{*}{\Pqc-hadrons}                  & Searches for rare decays & \multirow{5}{*}{NO} \\
                                                   & Searches for forbidden decays & \\                                          
                                                   & Time-integrated $CP$ violation & \\
                                                   & Time-dependent $CP$ violation & \\
                                                   & Dalitz measurements & \\
    \hline
    \multirow{2}{*}{\Pqs-hadrons}                  & Searches for rare decays & \multirow{2}{*}{NO} \\
                                                   & Searches for forbidden decays & \\                                          

    \hline
    \multirow{6}{*}{Spectroscopy}                  & Hadron masses & \multirow{5}{*}{NO} \\
                                                   & Hadron quantum numbers & \\
                                                   & Penta and tetraquark searches & \\
                                                   & Hadron differential cross-sections & \\ 
                                                   & Exclusive production of hadrons & \\
                                                   & Hadron widths or lifetimes & YES \\
    \hline
    \multirow{3}{*}{Electroweak and top}           & EW boson differential cross-sections & \multirow{4}{*}{NO} \\
                                                   & EW boson forward-backward asymmetries & \\
                                                   & Single and double-top differential cross-sections & \\
    \hline
    \multirow{1}{*}{Exotica}                       & Direct searches for new particles& \multirow{1}{*}{NO} \\
    \hline
    \multirow{2}{*}{Ion and fixed target physics}  & Hadron differential cross-sections & \multirow{2}{*}{NO} \\
                                                   & EW boson differential cross-sections & \\
    \hline
  \end{tabular} 
  \vspace{0.5cm}
  \caption[LHCb physics programme]{\label{table:lhcbphysprog}A brief and condensed summary of today's LHCb physics programme, for the full list of published papers divided by
physics working group see~\cite{LHCbPapers}. The phrase ``Dalitz measurements'' is used as shorthand for measurements of the resonant structure of multi-body decay processes,
and these measurements are often combined with searches for matter-antimatter asymmetry ($CP$-violation). 
The ``ion and fixed target physics'' part of the programme refers to measurements made in heavy ion collisions (for example lead-lead),
or in the collision of protons or heavy ions with a gas injected into the area of the vertex detector. All other measurements are made in $pp$ collisions. 
Historical refer to analyses which were included in the roadmap of
key LHCb measurements before Run~1 datataking began~\cite{Adeva:1224241}. This does not mean that analyses not listed in this column were not planned before LHCb datataking began, simply that they were
not seen as being quite so critical to the success or failure of the experiment as the historical measurements.}
\end{table}

All LHCb analyses are ultimately either measurements of the fundamental properties (masses, lifetimes, or quantum numbers) of specific particles, or 
else differential measurements of the rate at which a certain specific process occurs in LHC collisions. All analyses require the ability to select a high-purity
signal sample, which relies on having a good underlying detector performance (as detailed above) and a reasonable agreement between data and simulation, so that
simulated signal events can be used as a proxy when training selections. 
Because most LHCb analyses select signal candidates by fully reconstructing one of their decay modes,
they use the mass of the signal candidate to discriminate between the signal and residual backgrounds passing the selection.
The mass of the signal candidate is also typically used to define control regions known to be populated entirely by backgrounds (often called sidebands),
and events in these regions are then used as background for training the selection.
Except for a few of the very first LHCb papers, almost all LHCb analyses use multivariate selection techniques,
and techniques such as cross-validation\footnote{The most common use case within LHCb is that we wish to train a classifier for a rare signal process, and cannot generate sufficient simulated background
samples for the training. In this case, the backgrounds are normally taken from a preselected control region in the data, which is known not to contain signal. In order not to bias the eventual search
by training on the same data as we will be evaluating, the dataset is split into several subsets -- typically between two and ten. A separate classifier is then trained for each subset by excluding that subset
from the training sample, thus ensuring that the trained classifier cannot be influenced by the data it will be used on.} are used to ensure that training the classifier on data background samples
does not introduce biases. 
Once the signal is selected, different measurements are particularly sensitive to different aspects
of the detector performance:
\begin{itemize}
\item Measurements of particle masses or lifetimes are performed using likelihood fits to distributions of the masses or decay-times measured for each signal candidate.
They require a precise and accurate knowledge of the resolution, and any intrinsic bias, with which the detector measures these per-candidate quantities. 
\item Measurements of particle quantum numbers are typically performed by studying angular correlations in a specific decay process of these particles.
They require a precise and accurate knowledge of the way the trigger, reconstruction, and selection efficiencies varies with the kinematic and geometric properties of the decay products,
for example the transverse momenta and pseudorapidity of tracks, or the point at which the track originated in the detector. 
\item Knowledge of these properties is also particularly important for 
measurements of the absolute rate of a certain process, for example the rate at which a given hadron is produced inside the LHCb acceptance,
or the rate at which a given hadron decays into a specific final state. 
\item Measurements of $CP$-violation involve a comparison of the rates for processes involving particles and antiparticles. While many detector effects cancel in such
a comparison, they are especially sensitive to knowledge of trigger, reconstruction, and selection asymmetries between positive and negative stable particles.
\item Searches for rare and forbidden processes require a precise understanding of the way the trigger, reconstruction, and selection efficiencies vary with the mass
of the selected candidates, to ensure that backgrounds cannot be sculpted into signal-like shapes by kinematic correlations between the selection quantities.
\end{itemize}
Despite these differences, all these cases ultimately come down to knowing the efficiency for LHCb to reconstruct and identify stable charged particles and photons,
whether using the trigger or offfline reconstruction, split by particle charge, kinematics, and position in the detector. It is also necessary to know the precision and accuracy
with this LHCb's reconstruction determines track properties --- essentially the slope and covariance matrix at the first measured point --- which also allows us to determine
the efficiency for LHCb to combine stable charged particles and photons into signal candidates. The single most important aspect of LHCb's analysis methodology
is that in all these cases, any absolute efficiencies should be measured in a data-driven way, using appropriate control samples, and that simulation should only trusted for relative efficiencies.
As we shall see in Chapter~\ref{chpt:acunningplan}, this requirement drove much of the design of real-time analysis in LHCb. 

In addition to this conceptual overview, it is also worth touching on certain mechanical aspects of LHCb's analysis methods.
As the earlier overview of the LHCb detector systems hinted, the easiest processes to study at LHCb are those which produce a number of charged hadrons
or leptons in the final state. A good rule of thumb when designing an LHCb analysis is that for every $K^0_S$ or $\Lambda$ in the process being studied, the efficiency will be a factor
3 lower than for an equal multiplicity fully charged process. The situation is even more challenging for low-momentum photons or $\pi^0$ particles, where the efficiency is roughly a tenth
of that achievable for an equal multiplicity fully charged process. On the other hand, efficiencies for high-energy photons, such as those produced in radiative decays of B-hadrons e.g. $B^0_s\to \phi\gamma$
are not all that different from charged particle efficiencies, particularly once photons which convert in the detector material are included in the analysis. 
While high-momentum hadronic particle identification is difficult, LHCb has the ability to reconstruct and cleanly identify high momentum muons and electrons, 
allowing measurements of electroweak boson properties.
Jet reconstruction in LHCb is based on a particle-flow paradigm, in which momentum information from the tracking
system is used in preference to calorimetry wherever available.
Although the limited acceptance makes jet physics quite challenging, LHCb's precise vertex detector gives it an unparalleled~\cite{LHCb:2016yxg}  
ability to separate light-quark jets from charmed and beauty jets, which 
makes precision measurements of top-quark production and properties possible. 

Because it is optimized for processes involving charged particles, and because of the ``roadmap'' analyses listed earlier,
LHCb's analysis paradigm is based around combining stable charged particles into particle (typically hadron) candidates. Most LHCb analyses therefore 
begin by taking bunch crossings selected by the trigger system and forming combinations of reconstructed tracks (and/or photons), before applying selection criteria to them. Depending on the
precise signal topology there may be several combinatoric and selection stages which are interleaved into a more complex analysis sequence, but 
it is important in any case to understand that LHCb's software assumes that a reconstructed track will be identified
with a particular stable charged particle. That is to say that each track must be assigned a specific mass hypothesis before being combined with other tracks. This approach assumes that particle identification
information is available before any combinatorics, and that the majority of analyses will use particle identification information at the earliest possible stage to reduce the number of track
combinations. This reduction is not only important for controlling backgrounds, but is critical in reducing both the time taken to build particle candidates and the number of such candidates which
have to be saved for further analysis --- and hence the disk space used. 
As we shall see in due course, this reliance on particle identification criteria at the earliest stages of signal selection,
combined with the requirement to measure absolute efficiencies in a data-driven manner,
had important consequences for the design of LHCb's real-time analysis infrastructure.

\chapter{A haystack of needles: the necessity of real-time analysis in LHCb}
\label{chpt:haystackofneedles}
\pagestyle{myheadings}
\markboth{\bf A haystack of needles}{\bf A haystack of needles}

\begin{flushright}
\noindent {\it Kada primitivac ovlada tehnologijom, a nema \v{s}iru sliku\linebreak 
o posledicama te upotrebe tehnologije, to je turbo folk.\linebreak}
\linebreak
-- Rambo Amadeus
\end{flushright}

In order to establish the necessity of real-time analysis in LHCb, we must show that the LHC produces a significantly greater number of interesting
bunch crossings than can be fully recorded to long-term storage. Two simulated LHCb minimum bias datasets are used for this purpose: one corresponds to the 2011 datataking conditions (Run~1) while the second 
corresponds to the nominal data taking conditions expected in the upgrade of the LHCb experiment which is expected to come online in 2021.  
The relevant conditions are: the average number of both inelastic and elastic proton-proton collisions per event, referred to as $\nu$, the instantaneous luminosity, \luminosity, and the collision energy, $\sqrt{s}$. Table~\ref{table:genconf} describes the conditions and naming conventions of these samples. 
Two kinds of samples are used: generator level samples which can tell us how much of a given kind of signal\footnote{The word signal is used to refer to any physical process of interest,
including processes used to calibrate the detector.} is produced in the LHCb detector acceptance, and a fully simulated sample of events
which is passed through the detector digitization and reconstruction, and which
tell us how much of this signal is realistically reconstructible with the LHCb detector. 
In the case of the Run~1 dataset, the reconstructed sample is big enough (around 200k bunch crossings) that it can be used for both purposes. 
For the upgrade dataset the fully simulated sample is only around 100k
bunch crossings, and is therefore complemented with a sample of 10M generated bunch crossings.

\begin{table}\centering
  \begin{tabular}{lcc}
    \hline Property & Run~1, 2011  & Upgrade, nominal luminosity \\
    \midrule $\sqrt{s}$ [TeV] & 7 & 14 \\
    \nu & 2 & 7.6 \\
    \luminosity $10^{33}$cm$^{-2}$s$^{-1}$ & 0.4 & 2 \\ 
    Spillover & N & Y \\ \midrule
    Number of events & 200k & 10M (100k)\\
    \hline
  \end{tabular}
  \caption[Generator configuration for minimum bias datasets used in this report]{\label{table:genconf}Generator configuration for minimum bias datasets used in this report; 
    the second number for upgrade events indicates the sample of fully simulated data. Reproduced from~\cite{Fitzpatrick:1670985}.}
\end{table}

What, then, constitutes an interesting bunch crossing? In some sense, all bunch crossings are interesting by definition, because we could always want to study
the properties of any (or all) the lightest hadrons produced in LHC collisions as a function of their momentum or rapidity. On the other hand, most studies of this type
are ultimately systematics limited around the percent level by knowledge of the luminosity and detector efficiencies, so they could be performed using relatively small
samples of randomly triggered bunch crossings; they do not \textbf{need} real-time analysis as such. On the other end of the interest scale, ATLAS and CMS have used
real-time analysis to search for new particles (typically interpreted in terms of dark matter candidates) which primarly decay to pairs of 
light-quark jets~\cite{CMS:2015neg,Khachatryan:2016ecr,Isildak:2017brw,ATLAS:2016xiv}, where the irreducible QCD background greatly exceeds the trigger
output rate which would be allowed in a traditional analysis strategy. Similar searches, although based around decays of dark photons to dileptons, were both
proposed and carried out using real-time analysis at LHCb~\cite{Ilten:2016tkc,Aaij:2017rft}, but while these searches greatly benefit from real-time analysis, they were not the underlying
reason for its development at LHCb. 

Rather than enabling a specific kind of search or measurement, real-time analysis in LHCb was explicitly motivated from the very beginning by a belief that this was the only way to
enable the whole of LHCb's then-existing physics programme to reach its full potential, while \textbf{also} enabling this physics programme to be expanded as often as 
new analysis ideas emerged.\footnote{I have always passionately believed that LHCb's strength is the breadth and diversity of its physics programme, and I saw and continue to see 
real-time analysis as a way of making sure that any signals which could be selected and reconstructed in LHCb would also be analysable in LHCb. 
For what it is worth, debates over trigger bandwidth and whose analysis is the most important, which raged early in LHCb's life, have largely been absent
since 2015, although we'll have to wait for the much greater instantaneous luminosities of the upgraded detector before declaring victory in this cultural battle.} 
Therefore an interesting bunch crossing is one which enables any part of the collaboration's physics programme,
summarized earlier in Table~\ref{table:lhcbphysprog}, to be studied.
If we are to establish the necessity of real-time analysis, the key constraints will come from the highest rate, or production cross-section, signals. In other words, 
the interesting bunch crossing is one which produces a reconstructible beauty or charm hadron, or a light long-lived particle
such as a $\PKshort$ or a hyperon, as listed in Table~\ref{table:partclass}.
It would not make sense to try and enumerate all interesting decay modes
of these particles here. For one thing, the evolution of LHCb's physics programme during Runs~1~and~2 of the LHC has already shown that many analyses of decay modes which were once considered 
impossible or uninteresting can become very interesting once they are performed. Secondly and perhaps even more importantly, the most abundant Cabibbo-favoured decays of \Pqb- and \Pqc-hadrons,
while not themselves used to measure interesting physical parameters, are critical for understanding the detector efficiencies and asymmetries and controlling the systematic uncertainties in measurements
of rarer decay modes. For these reasons, we will simply take the mixture of decays in our minimum bias simulated sample as topologically
representative of all interesting particle decays. So how many interesting bunch crossings are there? 

\begin{table}\centering
  \begin{tabular}{cccccc}
    \hline \multicolumn{2}{c}{\Pqb-hadrons} & \multicolumn{2}{c}{\Pqc-hadrons} & \multicolumn{2}{c}{Light long-lived} \\
    \hline Particle & $|$PDG ID$|$ & Particle & $|$PDG ID$|$ & Particle & $|$PDG ID$|$ \\
    \PBzero & 511 & \PDplus & 411 & \PKshort & 310 \\
    \PBplus & 521 & \PDzero & 421 & \PLambda & 3122 \\
    \PBs & 531 & \PDsplus & 431 & \PSigmaplus & 3112 \\
    \PBc & 541 & \PLambdac & 4122 & \PXizero & 3322 \\
    \PLambdab & 5122 & \PXiczero & 4132 & \PXiminus & 3312 \\
    \PgXbp & 5312 & \PXicplus & 4232 & \POmegaminus & 3334 \\
    \PgXb & 5322 & \PgOc & 4332 & & \\
    \PgOb & 5332 & & & & \\
    \hline
  \end{tabular}
  \caption[Particle classification used in this report]{\label{table:partclass}Classification of particles into those containing bottom quarks, those containing charm quarks, and light long-lived
particles. Reproduced from~\cite{Fitzpatrick:1670985}.}
\end{table}

In order to be interesting, a bunch crossing must produce a particle listed in Table~\ref{table:partclass} and satisfying the following generator-level criteria designed
to ensure that its decay products lie within the LHCb detector acceptance.
\begin{itemize}
  \item Photons are not required to be within the acceptance unless they come from \Peta, \Ppizero decays.
  \item All charged decay products must be within 10-400~mrad, corresponding to the acceptance of the LHCb tracking system.
  \item All neutral decay products must be within 5-400~mrad, corresponding to the acceptance of the LHCb calorimeter system.
  \item \PKshort, \PLambda decay products must lie within the acceptance.
\end{itemize}
In addition, candidates are categorized according to their true decay-time, transverse momentum, and whether or not they have a reconstructible vertex within the LHCb vertex
detector. A decay product is defined as being within the \VELO acceptance if it has positive momentum in the forward direction and traverses at least three \VELO stations. 
Candidates can be classified as either partially in the \VELO acceptance, if at least two of their decay products satisfy this requirement, or fully within the \VELO acceptance,
if all of their decay products satisfy the requirement. 
The reason for this distinction is that traditional inclusive triggers for the kinds of particles listed in Table~\ref{table:partclass} 
are generally based on finding a decay vertex displaced from the primary $pp$ collision, and two is the minimum number of tracks needed to form a vertex.

Table~\ref{table:mc11genyields} presents the per-bunch-crossing yields in Run~1 conditions for \Pqb, \Pqc, and light, long-lived hadrons respectively.
Table~\ref{table:upggenyields} presents the same information for events generated assuming upgrade conditions.
Both tables also show the percentage of these candidates that leave two tracks in the \VELO, and the percentage that meet both this \VELO requirement and that have all decay products
fully contained within the LHCb acceptance.
These tables give some indication of the relative complexity of proton-proton collisions pre- and post-upgrade. In particular, notice that even in the Run~1 datataking collisions,
over a third of all bunch crossings produce a light long-lived hadron which is partially reconstructible in the vertex detector, while in the upgrade conditions over a fifth
of bunch crossings produce a charm hadron which is partially reconstructible in the vertex detector! 
\begin{table}\centering
  \begin{tabular}{lccc}
     Category & In 4\pi & $\epsilon(\text{\VELO})$ &  $\epsilon(\text{\VELO})\times\epsilon(\text{\LHCb})$ \\ \midrule
    \Pqb-hadrons & $0.0258\pm0.0004$ &      $30.5\pm0.6\%$ & $11.1\pm0.4\%$\\
    \Pqc-hadrons & $0.297\pm0.001$ & $21.9\pm0.2\%$ & $14.2\pm0.1\%$\\
    light, long-lived hadrons & $8.04\pm0.01$  &  $6.67\pm0.02\%$ & $6.35\pm0.02\%$\\
  \end{tabular}
  \caption[Candidates per event in MC11a]{\label{table:mc11genyields}Candidates per event and efficiencies for generator-level events in LHCb. 
   $\epsilon(\text{\VELO})$ is the efficiency for candidates having at least two tracks traversing at least three modules in the current \VELO. $\epsilon(\text{\LHCb})$ is the efficiency for candidates having all daughter tracks contained in the \LHCb acceptance. Reproduced from~\cite{Fitzpatrick:1670985}.}
\end{table}

 \begin{table}\centering
  \begin{tabular}{lccc}
    Category & In 4\pi & $\epsilon(\text{\VELO})$ &  $\epsilon(\text{\VELO})\times\epsilon(\text{\LHCb})$ \\ \midrule
    \Pqb-hadrons & $0.1572\pm0.0004$ & $34.9\pm0.1\%$ & $11.9\pm0.1\%$\\
    \Pqc-hadrons & $1.422\pm0.001$ &  $24.73\pm0.04\%$ & $15.12\pm0.03\%$\\
    light, long-lived hadrons & $33.291\pm0.006$ &   $7.022\pm0.004\%$ & $6.257\pm0.004\%$\\
  \end{tabular}
  \caption[Generator-level candidates candidates per event after the upgrade]{\label{table:upggenyields}Candidates per event and efficiencies of generator-level events after the upgrade.   $\epsilon(\text{\VELO})$ is the efficiency for candidates having at least two tracks traversing at least three modules in the upgrade \VELO. $\epsilon(\text{\LHCb})$ is the efficiency for candidates having all daughter tracks contained in the \LHCb acceptance. Reproduced from~\cite{Fitzpatrick:1670985}.}
\end{table}

\begin{figure}\centering
  \includegraphics[width=0.75\textwidth]{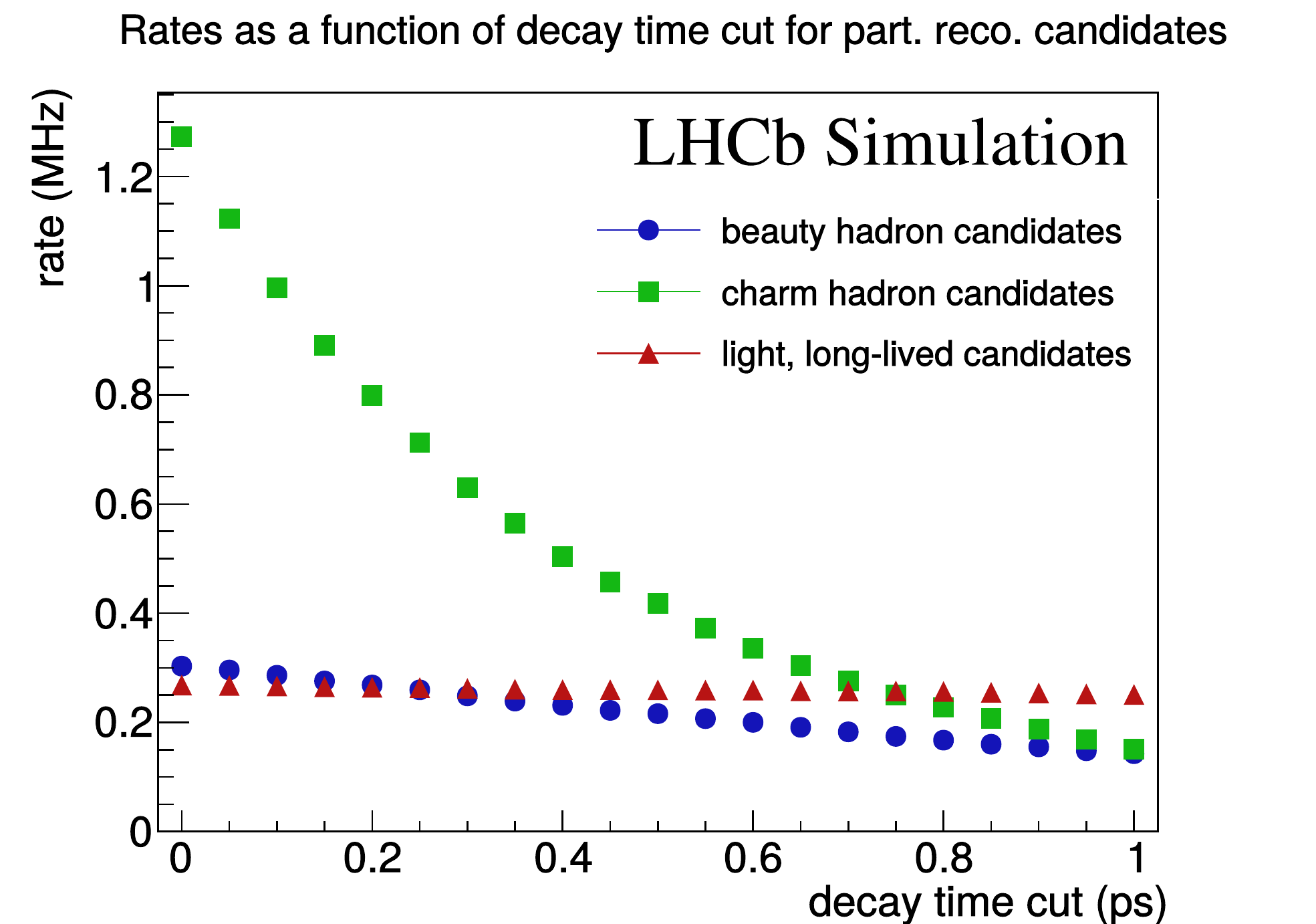}
  \includegraphics[width=0.75\textwidth]{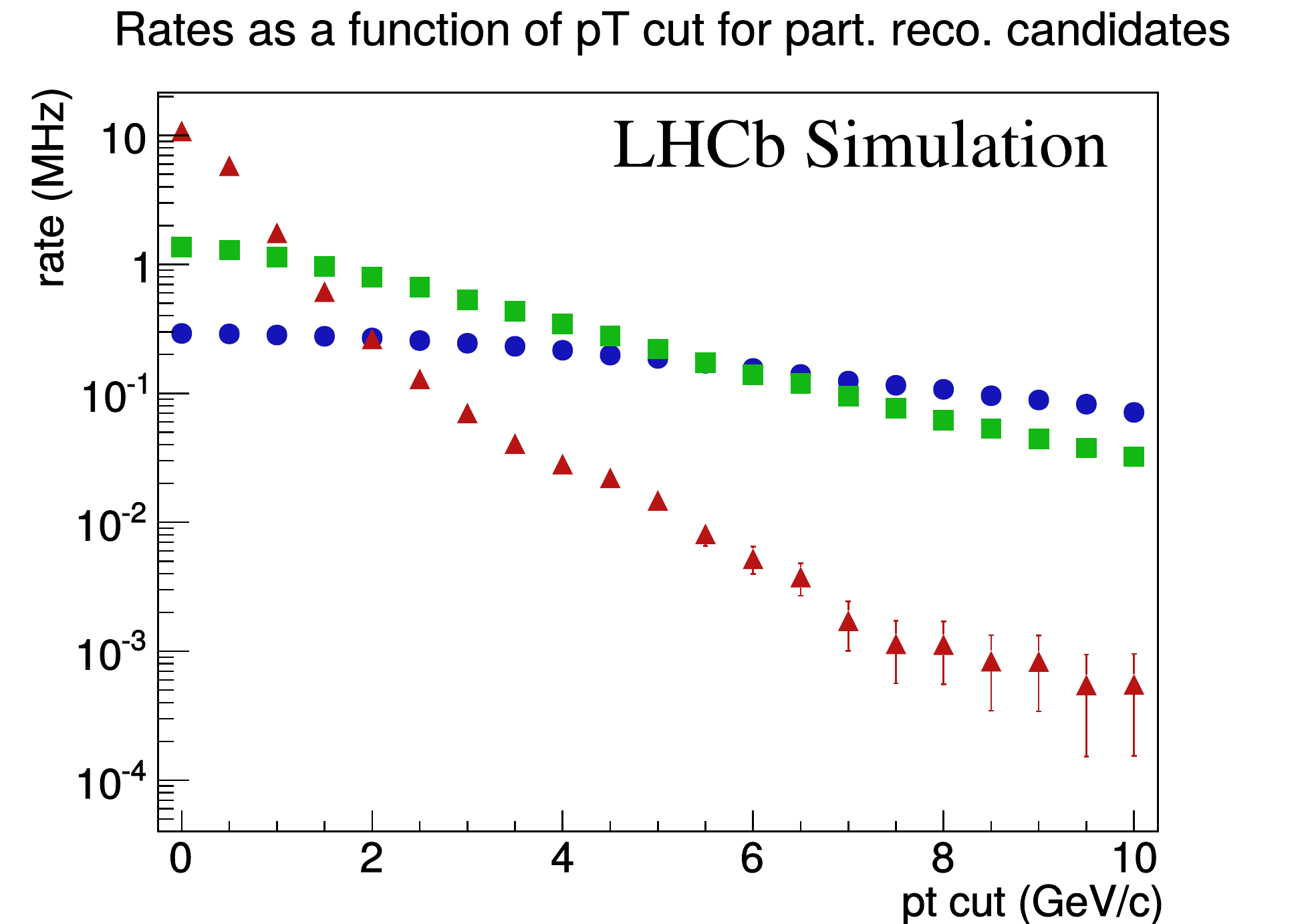}
  \caption[HLT signal output rates as a function of decay time or \pT cuts]{\label{fig:ptlt} HLT partially reconstructed (but fully reconstructible) signal rates as a function of (top) decay time for candidates with $\pT>2~\GeVoverc$ and (bottom) transverse momentum cuts for candidates with $\tau>0.2$~\ps. The rate is for two-track combinations that form a vertex only for candidates that can be fully reconstructed offline, ie: All additional tracks are also within the LHCb acceptance. Reproduced from~\cite{Fitzpatrick:1670985}.}
\end{figure}

In order to reinforce this point, Tables~\ref{table:mc11recoyields}~and~\ref{table:upgrecoyields} gives the yields of fully reconstructible signals
which might be kinematically interesting for analysis, while Tables~\ref{table:mc11fullrecoyields}~and~\ref{table:upgfullrecoyields} give the same information
broken down by the type of decaying parent hadron. Of course any specific analysis will have a particular set of kinematic selection criteria,
tuned to its own optimal efficiency-purity working point, and often tuned in order to optimally reject a specifically dangerous kind of background.
However experience with LHCb analyses as a whole shows that candidates with a \pT above $2~\GeVoverc$ and a decay-time above $0.2~\ps$ have a purity
which is sufficient for a wide range of LHCb analyses, and so these two criteria are used to illustrate one specific set of interesting signals. 
The last row in these tables presents the output rate of an ideal trigger, which selects signal with 100$\%$ efficiency and purity in these scenarios, while
Figure~\ref{fig:ptlt} shows how the rates of fully reconstructible signal candidates varies as a function of \pT and decay time cuts in the upgrade scenario.
Recall that under the traditional triggering paradigm, the permitted output rate of the Run~1 LHCb detector is in the 7~kHz range, while for the upgrade detector this number is expected to be
somewhere between 20 and 50~kHz. 
\begin{table}\centering
  \begin{tabular}{lccc}
    	& \Pqb-hadrons & \Pqc-hadrons & light, long-lived hadrons\\
    \midrule
    Reconstructed yield & $(4.0\pm0.1)\cdot 10^{-3}$ & $0.0196\pm0.0003$ &$0.0792\pm0.0006$\\
    $\epsilon(\pT>2\GeVoverc)$ & $83\pm 1\%$ & $47.2\pm0.8\%$ &  $2.0\pm0.1\%$ \\
    $\epsilon(\tau>0.2~\ps)$ & $89\pm1\%$ & $64.2\pm0.7\%$ & $99.53\pm0.05\%$ \\
    $\epsilon(\pT)\times\epsilon(\tau)$ & $73\pm2\%$ & $30.2\pm0.7\%$&  $1.9\pm0.1\%$ \\
    $\epsilon(\pT)\times\epsilon(\tau)\times\epsilon(\text{\LHCb})$& $29\pm1\%$ & $22.3\pm0.6\%$ & $1.9\pm0.1\%$ \\
    \midrule
    Output rate & 17.3~\kHz & 66.9~\kHz & 22.8~\kHz \\
  \end{tabular}
  \caption[Per-event yields of candidates in offline-reconstructed Run~1 events]{\label{table:mc11recoyields}Per-event yields determined from 0.21M of Run~1 minimum-bias events after partial offline reconstruction. The first row indicates the number of candidates which had at least two tracks from which a vertex could be produced. The last row shows the output rate of a trigger selecting such events with perfect efficiency, assuming
an input rate of $15~\MHz$ from the LHC, as during 2012 running. A breakdown of each category is available in Table~\ref{table:mc11fullrecoyields}. Reproduced from~\cite{Fitzpatrick:1670985}.}
\end{table}

 \begin{table}\centering
  \begin{tabular}{lccc}
    	& \Pqb-hadrons & \Pqc-hadrons & light, long-lived hadrons\\
    \midrule
    Reconstructed yield &  $0.0317\pm0.0006$ &  $0.118\pm0.001$ &  $0.406\pm0.002$ \\
    $\epsilon(\pT>2\GeVoverc)$ & $85.6\pm0.6\%$ & $51.8\pm0.5\%$ & $2.34\pm0.08\%$ \\
    $\epsilon(\tau>0.2~\ps)$ & $88.1\pm0.6\%$ &  $63.1\pm0.5\%$ & $99.46\pm0.03\%$  \\
    $\epsilon(\pT)\times\epsilon(\tau)$ & $75.9\pm0.8\%$ & $32.6\pm0.4\%$& $ 2.30\pm0.08\%$ \\
    $\epsilon(\pT)\times\epsilon(\tau)\times\epsilon(\text{\LHCb})$& $27.9\pm0.3\%$ &  $22.6\pm0.3\%$ & $2.17\pm0.07\%$ \\
    \midrule
    Output rate & 270~\kHz & 800~\kHz & 264~\kHz \\
  \end{tabular}
  \caption[Per-event yields of candidates offline-reconstructed in post-upgrade MC]{\label{table:upgrecoyields}Per-event yields determined from 100k of upgrade minimum-bias events after partial offline reconstruction. The first row indicates the number of candidates which had at least two tracks from which a vertex could be produced. The last row shows the output rate of a trigger selecting such events with perfect efficiency, assuming
an input rate of $30~\MHz$ from the LHC, as expected during upgrade running. A breakdown of each category is available in Table~\ref{table:upgfullrecoyields}. Reproduced from~\cite{Fitzpatrick:1670985}.}
\end{table}

As a final example of this point, we can consider the rate of one particularly important signal: \HepProcess{\PDzero\to\PKplus\PKminus} decays with \pT above 2~\GeVoverc~and a decay-time
above 0.2~ps. This decay is one of the most important signals for the upgraded LHCb detector because it is sensitive to the $A_\Gamma$, the parameter
which quantifies \CP violation in the interference of decay and mixing for $\PDzero$ mesons.
The observation of a non-zero value of $A_\Gamma$ is one of the most plausible ways to see time-dependent \CP violation in charm hadrons, in particular
because there is no foreseen systematics limitation which cannot be controlled with sufficiently large control samples~\cite{Aaij:2017idz}.
From Table~\ref{table:upgfullrecoyields} we can see that 8$\%$ of all upgrade
events will contain a \PDzero meson whose products are in the detector acceptance.
The branching fraction of this decay is $0.4\%$ and 
the efficiency of the listed \pT and \tau cuts is 31.4$\%$, resulting in a total output rate
of approximately $3.1~\kHz$ for an LHC interaction rate of $30~\MHz$.
This rate can be significantly decreased by requiring that the \PDzero came from a \HepProcess{\PDstarplus\to\PDzero\Ppiplus} decay chain, so that the slow pion tags its flavour,
a requirement which would anyhow be applied by the offline analysis. From current experience this can be used to decrease the event yield by a factor of around 5, still leaving us with around $0.6~\kHz$ of signal, or about $3\%$ of the baseline trigger rate! A similar calculation yields a signal rate of
around $1.5~\kHz$ for the \HepProcess{\PDsplus\to\Ppiplus\Ppiminus\Ppiplus} decay mode which is also very important as it is highly sensitive to direct \CP violation: we have now used around $10\%$ of the trigger bandwidth simply to select \textbf{the signals} for two analyses of \CP violation in Cabibbo-suppressed decays of charmed hadrons. 

In summary, the large cross-section to produce reconstructible and selectable beauty, charm, and light flavour hadrons,
combined with the large branching fractions for these hadrons to decay into physically interesting final states, makes it impossible to efficiently store
all raw detector information for all interesting bunch crossings. Moreover, these interesting
final states are often topologically identical to other even more abundant signals, differing only in swapping one kind of hadron for another --- a typical example being the above mentioned Cabibbo-suppressed
signal \HepProcess{\PDsplus\to\Ppiplus\Ppiminus\Ppiplus} and the corresponding Cabibbo-favoured control-mode $D^{\pm}_{s}\to \pi^\pm K^+ K^-$. 
Even with the present LHCb detector, which takes data at an average of $\sim 1$ $pp$ collisions per bunch crossing, the rate of interesting charm signals saturates the available
output rate of a traditional trigger. And this problem will only become worse in the upgraded LHCb detector, partly because of an increased
instantaneous luminosity leading to an average $\sim 5$ $pp$ collisions per bunch crossing, and partly because the removal of the hardware
trigger will greatly increase the efficiency and hence signal rate! 
It was therefore necessary to enable the real-time analysis of LHCb data. Once this necessity became clear, it was decided 
to enable real-time analysis in time for Run~2 of the LHC both to extend LHCb's immediate physics reach and to gain experience before 
having to tackle the much more challenging environment of the LHCb upgrade.
Because the fundamental motivation for real-time analysis comes from the large cross-section to produce
interesting signals, it is not a curiosity of LHCb or indeed the LHC, or a technique restricted to a few exotic searches in specific regions of parameter space. All the arguments
presented in this chapter will be even more true of the high-luminosity or high-energy upgrades of the LHC, but they could
also be applied to expand the precision physics reach of the next generation of fixed-target experiments searching
for long-lived particles, such as SHIP or NA62. 

\begin{sidewaystable}\centering
  \begin{tabular}{lccccc}
    Cand. & Reco. yield & $\epsilon(\pT>2~\GeVoverc)$ & $\epsilon(\tau>0.2~\ps)$ & $\epsilon(\pT)\times\epsilon(\tau)$ & $\epsilon(\pT)\times\epsilon(\tau)\times\epsilon(\text{\LHCb})$ \\
    \midrule \PBzero & $(1.77\pm0.092)\e{-3}$ & $85.4\pm1.8\%$ & $86.5\pm1.77\%$ & $72.8\pm2.31\%$ & $28.2\pm1.8\%$\\
    \PBplus &$(1.50\pm0.085)\e{-3}$&$81.8\pm2.3\%$ & $91.7\pm1.56\%$ & $74.8\pm2.45\%$ & $29.8\pm2.1\%$\\
    \PBs & $(3.91\pm0.43)\e{-4}$ & $73.2\pm4.9\%$ & $87.8\pm3.61\%$ & $65.9\pm5.24\%$ & $27.2\pm3.5\%$\\
    \PBc & 0 & - & - & - & - \\
    \PLambdab & $(3.53\pm0.41)\e{-4}$& $85.1\pm4.1\%$ & $89.2\pm3.61\%$ & $  77\pm4.89\%$ & $29.3\pm4.6\%$\\
    \PgXbp & 0 & - & - & - & - \\
    \PgXb & 0 & - & - & - & - \\
    \PgOb & 0 & - & - & - & - \\
    \midrule Total & $(4.01\pm0.14)\e{-3}$ & $82.9\pm 1.3\%$ & $88.8\pm1.09\%$ &  $73.2\pm1.53\%$ & $28.8\pm1.2\%$\\
    \midrule \PDzero &  $0.014\pm0.00026$ & $46.9\pm0.92\%$ & $61.8\pm0.897\%$ &       $28.7\pm0.834\%$ & $21.7\pm0.68\%$\\
    \PDplus & $0.00308\pm0.00012$ & $47.8\pm2.0\%$ & $83.6\pm1.46\%$ & $40.6\pm1.93\%$ & $30.1\pm1.6\%$\\
    \PDsplus & $(1.80\pm0.093)\e{-5}$  & $49.3\pm2.6\%$ & $69.5\pm2.37\%$ &  $33.2\pm2.42\%$& $21.8\pm1.8\%$\\
    \PLambdac & $(1.19\pm0.075)\e{-5}$ & $47.6\pm3.2\%$ & $37.2\pm3.06\%$ & $18.8\pm2.47\%$ & $13.1\pm1.8\%$\\
    \PXiczero & $(3.34\pm1.3)\e{-5}$ & $14.3\pm13\%$ & $   0\pm   0\%$ & $   0\pm   0\%$ & $0\pm0\%$\\
    \PXicplus & $(8.1\pm2.0)\e{-5}$ & $41.2\pm12\%$ & $58.8\pm11.9\%$ & $17.6\pm9.25\%$ & $9.5\pm5.5\%$\\
    \PgOc & 0 & - & - & - & - \\
    \midrule Total & $0.020\pm0.00031$ & $47.2\pm0.767\%$ & $64.2\pm0.737\%$ & $30.2\pm0.706\%$& $22.3\pm0.6\%$\\
    \midrule \PKshort & $0.068\pm0.00057$ & $1.69\pm0.11\%$ & $99.5\pm0.0608\%$ & $1.66\pm0.107\%$ &$1.63\pm0.11\%$\\
    \PLambda & $0.010\pm0.000223$ & $3.63\pm0.40\%$ &$99.8\pm0.0918\%$&$3.63\pm0.401\%$ &$3.58\pm0.40\%$\\
    \PSigmaplus & 0 & - & - & - & - \\
    \PXizero  & $(1.43\pm0.83)\e{-5}$ & $   0\pm   0$ & $ 100\pm   0$ & - & - \\
    \PXiminus & $(3.05\pm0.38)\e{-4}$ & $3.1\pm2.2\%$ & $ 100\pm   0\%$ & $3.12\pm2.17\%$  &$3.12\pm3.80\%$\\
    \POmegaminus & $(2.4\pm1.1)\e{-5}$ & $  20\pm18\%$ & $ 100\pm   0\%$ & $  20\pm17.9\%$  &$20\pm21\%$\\
    \midrule Total & $0.079\pm0.00061$ & $1.96\pm0.11\%$ & $99.5\pm0.0539\%$ & $1.93\pm0.107\%$ &$1.90\pm0.11\%$\\
    \midrule
  \end{tabular}
  \caption[Per-event yields of candidates in offline-reconstructed Run~1 conditions]{\label{table:mc11fullrecoyields}Per-event yields determined from 0.21M of Run~1 minimum-bias events after partial offline reconstruction. The first column indicates the number of candidates which had at least two tracks from which a vertex could be produced. The second column indicates the efficiency of a $\pT>2\GeVoverc$ cut applied to the fully reconstructed candidate. Reproduced from~\cite{Fitzpatrick:1670985}.}
\end{sidewaystable}

\begin{sidewaystable}\centering
  \begin{tabular}{lccccc}
    Cand. & Reco. yield & $\epsilon(\pT>2~\GeVoverc)$ & $\epsilon(\tau>0.2~\ps)$ & $\epsilon(\pT)\times\epsilon(\tau)$ & $\epsilon(\pT)\times\epsilon(\tau)\times\epsilon(\text{\LHCb})$ \\
    \midrule \PBzero & $0.014\pm0.00038$ & $  86.0\pm0.95\%$ & $  87\pm0.917\%$ & $75.4\pm1.17\%$ & $27.7\pm0.45\%$\\
    \PBplus & $0.013\pm0.00036$ & $86.2\pm0.99\%$ &   $88.3\pm0.925\%$&       $76.2\pm1.22\%$& $28.1\pm0.47\%$\\
    \PBs & $(3.42\pm0.19)\e{-3}$ & $85.8\pm1.9\%$&       $  91\pm1.57\%$&       $  78\pm2.27\%$ & $28.3\pm0.86\%$\\
    \PBc & $(4.1\pm2.1)\e{-5}$ & $  75\pm22\%$&       $  50\pm  25\%$& $  25\pm21.7\%$& $8.3\pm7.4\%$\\
    \PLambdab & $(2.36\pm0.16)\e{-3}$ & $80.3\pm2.6\%$&       $90.8\pm1.91\%$&       $74.7\pm2.87\%$& $27.2\pm1.1\%$\\
    \PgXbp & 0 & - & - &- & -\\
    \PgXb & 0 & - & - & - & -\\
    \PgOb & 0 & - & - & - & -\\
    \midrule Total &  $0.032\pm0.00058$ & $85.6\pm0.63\%$ & $88.1\pm0.58\%$ &       $75.9\pm0.77\%$& $27.9\pm0.29\%$\\
    \midrule \PDzero & $0.080\pm0.00091$ & $51.3\pm0.57\%$ & $60.7\pm0.554\%$ & $31.4\pm0.526\%$ & $22.4\pm0.38\%$\\
    \PDplus & $0.0182\pm0.00043$ & $51.2\pm1.2\%$ & $82.2\pm0.908\%$ & $40.9\pm1.17\%$ & $27.6\pm0.79\%$\\
    \PDsplus & $0.0118\pm0.00035$ & $54.4\pm1.5\%$ & $67.8\pm1.38\%$ & $37.3\pm1.43\%$ & $23.5\pm0.90\%$\\
    \PLambdac & $(7.50\pm0.28)\e{-3}$ & $54.6\pm1.8\%$ & $35.4\pm1.77\%$ & $18.2\pm1.43\%$ & $12.1\pm0.95\%$\\
    \PXiczero &  $(2.16\pm0.47)\e{-4}$ & $38.1\pm11\%$ & $4.76\pm4.65\%$ & $4.76\pm4.65\%$ & $3.2\pm3.2\%$\\
    \PXicplus & $(2.26\pm0.48)\e{-4}$ & $40.9\pm11\%$ & $59.1\pm10.5\%$ & $36.4\pm10.3\%$ & $23.6\pm6.7\%$\\
    \PgOc & $(1.0\pm1.0)\e{-5}$ & $ 100\pm   0\%$ & $   0\pm   0\%$ & $   0\pm   0\%$ & $0\pm0\%$\\
    \midrule Total & $0.118\pm0.0011$ & $51.8\pm0.47\%$ & $63.1\pm0.451\%$ & $32.6\pm0.438\%$ & $22.6\pm0.30\%$\\
    \midrule \PKshort & $0.35\pm0.0019$ & $1.99\pm0.076\%$ & $99.5\pm0.039\%$ & $1.95\pm0.0751\%$ & $1.84\pm0.071\%$\\
    \PLambda & $0.055\pm0.00075$ & $4.47\pm0.28\%$ &       $99.8\pm0.0647\%$&       $4.43\pm0.281\%$ & $4.18\pm0.27\%$\\
    \PSigmaplus & $(2.1\pm1.5)\e{-5}$ & $  50\pm35\%$$ 100\pm   0\%$ &$  50\pm35.4\%$ &  $47\pm33\%$\\ 
    \PXizero & $(1.0\pm1.0)\e{-5}$  & $   0\pm   0\%$ & $ 100\pm   0\%$ & $   0\pm   0\%$ & $0\pm0\%$\\
    \PXiminus & $(1.52\pm0.13)\e{-3}$ & $4.7\pm1.7\%$ $ 100\pm   0\%$ & $4.73\pm1.74\%$ & $4.3\pm1.6\%$\\
    \POmegaminus & $(1.65\pm0.41)\e{-5}$ & $6.3\pm6.1\%$ &  $ 100\pm   0\%$  & $6.25\pm6.05\%$ & $4.8\pm4.7\%$\\
    \midrule Total &  $0.406\pm0.0020$ & $2.34\pm0.076\%$ &  $99.5\pm0.0347\%$  & $ 2.30\pm0.076\%$ &  $2.17\pm0.071$\\
    \midrule
  \end{tabular}
  \caption[Per-event yields of candidates offline-reconstructed post-upgrade]{\label{table:upgfullrecoyields}Per-event yields determined from 100k of upgrade minimum-bias events after partial offline reconstruction. The first column indicates the number of candidates which had at least two tracks from which a vertex could be produced. The second column indicates the efficiency of a $\pT>2\GeVoverc$ cut applied to the fully reconstructed candidate. Reproduced from~\cite{Fitzpatrick:1670985}.}
\end{sidewaystable}

\chapter{A cunning plan: the requirements for real-time analysis}
\label{chpt:acunningplan}
\pagestyle{myheadings}
\markboth{\bf A cunning plan}{\bf A cunning plan}

\begin{flushright}
\noindent {\it Am I jumping the gun, Baldrick, or are the words ``I have a cunning plan'' \linebreak
               marching with ill-deserved confidence in the direction of this conversation?\linebreak}
\linebreak
-- Blackadder
\end{flushright}

Having established the physics interest in analysing a far greater fraction
of LHCb bunch crossings than possible with a traditional triggering strategy, we now turn to the question of the
requirements for a working real-time analysis of LHCb data. To answer this question, we must first restate
the objectives of real-time analysis in the LHCb context: to efficiently reconstruct and select signal candidates of interest,
as well as all control samples needed for a data-driven measurement of the efficiency of this reconstruction and selection,
and to record only these signal candidates to permanent storage, discarding all other raw and reconstructed detector information. From this
objective follow the general requirements to:
\begin{itemize}
\item align and calibrate LHCb and all its subdetectors in real-time;
\item propagate the resulting alignment and calibration constants in real-time;
\item execute the full detector reconstruction in real-time;
\item execute selections for all analyses of interest in real-time;
\item execute selections for all relevant control samples in real-time;
\item and monitor the data quality and quickly correct any problems.
\end{itemize}
I will now discuss each of these requirements, and then we will see how they were addressed in a coherent manner, and in particular
how the HLT software was split into two distinct processes in order to allow the HLT computing cluster to be used
as a temporary buffer while the alignment and calibration was performed. Details of the performance of the real-time
alignment and calibration in Run~2, in particular its stability over time, are beyond the scope of this chaper, and can be found elsewhere~\cite{Dujany:2015lxd,Borghi:2017hfp}.

\section*{Automated alignment and calibration}
Different parts of the LHCb physics programme are sensitive to different aspects of the detector alignment and calibration,
and at the same time different subdetectors have intrinsically different sensitivities to how often they have to be
aligned and calibrated. During the LHC shutdown between Run~1 and Run~2, it was therefore important to study how the
detector alignment and calibration evolved over Run~1, in order to establish guidelines on the required frequency of
updates for Run~2. The picture which emerged was as follows:
 
\begin{itemize}
\item The VELO alignment constants were stable to within a few microns during Run~1, but nevertheless when looked at over a long
period of time certain trends in the alignment constants became apparent, caused by the fact that the precision of the VELO opening and closing
mechanism is around 10~$\mu$m, while the alignment precision is around 2~$\mu$m. From this it was concluded that the VELO alignment
would be executed at the start of each fill, which naturally coincided with the moment when the VELO is mechanically moved closer
to the beams. The expectation was that this fill-by-fill alignment would trigger an update of the constants once every few fills.
This alignment requires using around 100,000 randomly triggered bunch crossings, and hence presents no special
difficulty for the datataking.
\item The TT and T-station alignment is, on the other hand, more sensitive to magnet polarity switches, temperature changes,
and detector interventions, with both trends and sharp isolated changes in the alignment parameters seen over the course
of Run~1. It would therefore also run once per fill, requiring a sample of 200,000 $D^0\to K\pi$ decays. A separate one-off alignment
would be performed at the start of each year of datataking using magnet off data, which is particularly helpful for establishing
the vertical alignment constants of the tracking system. In addition the drift-time of the straw-tube tracker would need to be calibrated
on a fill-by-fill basis.
\item The muon system alignment does not need to performed more than a couple of times per year, if the chambers move, and is mainly important for the
performance of the hardware trigger rather than analysis, since hits in the muon system do not contribute significantly to the
measurement of track parameters. It requires a sample of requiring a sample of around 250,000 $J/\psi\to \mu\mu$ decays.
\item The alignment of the RICH mirrors is fairly stable over time, but can nevertheless benefit from updates after
magnet polarity changes. It requires a special sample of around 3,000,000 bunch crossings selected in such a way 
to uniformly populate the RICH acceptances with tracks. This alignment would also be automated and performed once per fill,
although updated much less frequently.
\item The RICH image and refractive index calibrations are particularly sensitive to natural variations of temperature and pressure
in the LHCb cavern, and must be performed and updated on an hourly basis. They can be performed using a sample of randomly triggered bunch
crossings. 
\item The calorimeter requires calibration constants to follow the radiation-induced ageing, and to maintain a stable relationship
between the raw signal amplitude and the particle energy. During Run~1 this was a particularly
significant problem for the hardware ECAL triggers, whose rates would sometimes vary by up to $30\%$ between calibration updates.
A coarse automatic calibration, good to about $5\%$, was therefore implemented for Run~2 based purely on the evolution of the observed
calorimeter occupancy in randomly triggered bunch crossings. Unlike any other LHCb calibration, this coarse calorimeter calibration
did not produce constants but rather directly adjusted the detector High Voltage to maintain the relationship between particle
energy and the recorded signal. An absolute cell-by-cell calibration, based on reconstructed $\pi^0$ candidates and divided by calorimeter region, 
would be needed several times per year.
\end{itemize}
A summary of these requirements can be found in Table~\ref{table:aligncalibsummary}. 
A crucial common requirement for all these alignment and calibration tasks was that the time to execute them was significantly shorter
than the frequency with which they had to be executed. This was achieved through a combination of code optimization and by implementing
a way of distributing these jobs across the HLT computing cluster, in which each server performed a small part of the alignment or calibration
job using a subset of bunch crossings, and these results were then aggregated into the overall result. In addition, it was necessary to devise
a lightweight mechanism for propagating the updated alignment and calibration constants to the HLT processes. These constants are normally
loaded from a versioned database\footnote{There are actually two databases used by LHCb: the detector database holds information about
the detector geometry, which detector channels are masked or disabled, and other parameters which are expected to change very rarely. The
conditions database holds the alignment and calibration constants, as well as other detector related parameters which are expected to be updated
more frequently.}, but since this database is large, and since at least some of the calibration parameters would now be updated on an hourly basis, or more than a thousand times
per year, distributing a different versioned database for each of these hour-long blocks would represent a significant computational overhead on the computing cluster. 
The solution was for the HLT to use a single baseline calibration, read from the database, and a small number of versioned updates of only those
parameters which had changed, read from lightweight \textit{xml} files distributed across the computing cluster.

\begin{table}[htb]\centering
  \begin{tabular}{lll}
      Alignment/Calibration task & Sample & Size of sample \\
      \hline
      VELO  & random triggers & $\mathcal{O}$(100k) events \\
      Tracker & $D^0\to K^-\pi^+$, high momentum tracks & $\mathcal{O}$(200k) events\\
      Tracker vertical alignment  & magnet off tracks & 5-10M events\\
      Muon system & $J/\psi\to\mu^+\mu^-$ & $\mathcal{O}$(250k) events \\
      RICH mirrors & equal occupancy triggers & $\mathcal{O}$(3M) events\\   
      RICH image &  random triggers & \\
      RICH refractive index & random triggers & \\
      CALO coarse & random triggers & $\mathcal{O}$(100k) events \\
      CALO fine & $\pi^0\to\gamma\gamma$ & \\
      \hline
  \end{tabular} 
  \vspace{0.5cm}
  \caption[Summary of alignment and calibration tasks]{\label{table:aligncalibsummary}A summary of the alignment and calibration tasks and the required types and numbers of events for each one.}
\end{table}

\vspace{-5mm}
\section*{Reconstruction}
In its purest form, real-time analysis requires the ability to execute the full detector reconstruction for all bunch crossings passing
the hardware trigger. This, however, was obviously not going to be possible in the Run~2 LHCb, because the HLT had around 50~ms available
to process each bunch crossing while the offline reconstruction needed around half a second. What was possible, however, was to execute
a subset of the full reconstruction (in practice a subset of the track finding), use this to preselect a fraction of the bunch
crossings, and then execute the rest of the reconstruction on these. It was however important to show that the reconstruction could be performed
in this two-stage approach without  reducing its performance.
The optimization of the partial and full reconstruction stages to fit into the HLT computing budget is discussed further in Chapter~\ref{chpt:unrealtime}.
\vspace{-5mm}
\section*{Signal and control sample selections}
Because the HLT used the same underlying codebase as the stripping and offline analysis software, once the full offline reconstruction
is available in the HLT the only mechanical requirement for signal
selections was to move all signal selections for real-time analyses from the stripping to the HLT. In some cases this meant
having to reoptimize the selections in order to reduce the amount of time spent making particle combinations, but in most cases
the stringent computational requirements which had already been imposed in the stripping made this relatively easy.
The more subtle requirement imposed by real-time analysis is that not only the signal, but all relevant control modes, must 
be selected in real-time. This means both the control modes used for studying specific backgrounds, and the control modes
used for controlling detector efficiencies. 

In the case of backgrounds, most of the analyses in question already implemented
such selections within the stripping, so it was once again a case of transferring them over to the HLT. Although this seems like a potentially
dangerous requirement, in practice it has helped focus analysts on thinking about their backgrounds right from the earliest stages of analysis,
and has helped improve analysis documentation, a point discussed further in Chapter~\ref{chpt:conclusion}. 

LHCb measures detector efficiencies in a data-driven way using tag-and-probe selections. The ``tag'' identifies a bunch crossing
as containing a particle of interest without reconstructing that particle in any way, which is then ``probed'' using the reconstruction. 
The difference in the number of signal bunch crossings in the ``tag'' and ``probe'' samples gives the efficiency to reconstruct the specific particle
in a given way: to find the track, to correctly identify the particle, etc. Real-time analysis meant that 
all the different tag-and-probe selections summarized in Table~\ref{table:tagandprobe} had to be transferred to the HLT and executed in real-time.\footnote{There
is a physics hierarchy to how easily different efficiencies can be measured for different particle species. Tracking efficiencies can most easily be measured
for muons, because $J/\psi$ mesons are abundantly produced and because muons can be cleanly identified by the muon system without any requirements in the rest of
the detector. Furthermore, muons undergo the fewest destructive material interactions in the detector, a component of the reconstruction efficiency which is particularly
hard to measure since it requires the signal to be cleanly identified without any information whatsoever about the probe. 
Pion and kaon tracking efficiencies are harder, protons harder still, and electrons hardest of all, since unlike with all other particle species
electron tracking inefficiencies in LHCb are mainly caused by bremsstrahlung sending the particle out of the detector acceptance. Nevertheless, LHCb is in the process
of developing tag-and-probe methods for measuring both hadron and electron tracking efficiencies, at least up to the uncertainty on the frequency of destructive hadronic
interactions in LHCb's vertex detector. For particle identification
efficiencies, on the other hand, muons, pions, and kaons are all similarly straightforward to measure. Electrons are harder because of the variation of calorimeter response over time, 
but in this case $J/\psi\to e^+e^-$ decays can nevertheless be used without too much difficulty. Protons are on the other hand hard, as there is no single calibration
sample which gives a good coverage of the interesting proton kinematics and which can be cleanly selected without applying particle identification requirements to the proton.}
However, it was also important to maintain the ability to perform tag-and-probe measurements for the part of the physics programme which would still
be done in a traditional manner, so a dedicated calibration data stream was created, for which both the real-time analysis information and the
more traditional reconstruction information was written to permanent storage. This also allowed the calibration stream to be used for validating the
real-time analysis in early 2015 datataking by comparing its output directly to the output of the traditional offline analysis, as described
further in Chapter~\ref{chpt:uneprocedure}. 

\begin{table}[htb]\centering
  \begin{tabular}{lll}
      Species & Soft & Hard \\
      \hline
      $e^\pm$ & \multicolumn{1}{c}{---} & $J/\psi\to e^+e^-$ \\
      $\mu^\pm$ & $D^+_s\to \mu^+\mu^-\pi^+$ & $J/\psi\to \mu^+\mu^-$  \\
      $\pi^\pm$ & $K^0_S\to\pi^+\pi^-$ & $D^{*+}\to D^0\pi^+$, %
                                      $D^0\to K^-\pi+$ \\
      $K^\pm$ & $D^+_s\to K^+K^-\pi+$ & $D^{*+}\to D^0\pi^+$, %
                                        $D^0\to K^-\pi+$ \\
      $p^\pm$ & $\Lambda\to p\pi^-$ & $\Lambda\to p\pi^-$, %
                                      $\Lambda^+_c\to p K^- \pi^+$ \\ 
      \hline
  \end{tabular} 
  \vspace{0.5cm}
  \caption[Tag-and-probe samples]{\label{table:tagandprobe}A summary of the tag-and-probe samples used for measuring detector efficiencies.}
\end{table}

Another subtlety related to real-time selection concerns global information, such as particle isolation or information about other
particles in the event which may be correlated to the signal. A good physics example is the study of short-lived resonances containing
a charm quark, which typically decay into a combination of a \Pqc-hadron and some number of accompanying pions, kaons, protons, or photons.
Because neither the number nor the mass of possible resonances are a priori known, the \Pqc-hadron could in principle be combined with
almost every track coming from the same $pp$ collision, thus enormously increasing the number of candidates which have to be persisted and negating
most of the space saved by using real-time analysis.
Aside from the space taken, defining many different real-time
selections, one for each combination of signal particle and a specific companion track, can quickly become operationally unwieldy. 
Finally, the selection used and the additional information stored about the bunch crossing had to be able to evolve over time. While real-time
analysis requires the analyst to fully define what they want to do before taking data, it obviously also has to allow the analyst to learn
and improve the analysis as they go along. It also must be possible to easily analyse data taken with an older, less sophisticated, real-time analysis
together with the data taken with the latest and greatest version.
Further discussion of a scalable solution to these problems can be found in Chapter~\ref{chpt:uneprocedure}.
\vspace{-5mm}
\section*{Data quality and monitoring}
During Run~1, LHCb made a distinction between detector monitoring, performed by shifters in the control room during datataking,
and data quality verification which was performed only once the data had already been fully reconstructed. The main job of detector
monitoring was to spot major problems with the detector hardware or trigger configuration in real-time, using a set of monitoring
histograms aggregated across the HLT computing cluster and automatically provided to the shift crew by a dedicated algorithm.
This monitoring could spot if a part of the detector had malfunctioned and switched off, if the datataking conditions were significantly different
than expected (e.g. the LHC was providing the wrong number of $pp$ collisions per bunch crossing), or if the wrong trigger thresholds had
been loaded. The data quality verification was not performed in real-time, and had the goal of catching up any issues which might
have been missed by the real-time monitoring; following up on data which had been flagged as problematic in real-time and deciding
if it could be salvaged; and validating the alignment, calibration, and full detector reconstruction itself. It was performed by
a dedicated shift crew using a much larger list of histograms than that available online, and in the case of doubts the data quality
shifters could flag the data as requiring further attention by alignment, calibration, or reconstruction experts.

This structure was kept in place for Run~2, however it was necessary to augment it to account for the fact that part of the analysis
would now be performed in real-time. Each of the automated calibration and alignment tasks was equipped with a dedicated monitoring
algorithm which tracked the evolution of the constants produced by the task in question. Each of these constants had an associated threshold, 
which represented the maximum allowed variation from one iteration to the next, and which was set based on detailed studies of how these constants had varied during Run~1.
If a constant changed by more than this threshold from one variation to the next, it would be flagged as requiring an update --- in the early
parts of Run~2 these updates would trigger a manual verification by an expert, but the system was automated very quickly, within weeks in the case
of some algorithms and by 2016 for most of them. Nevertheless,
even in fully automatic mode, the monitoring programs have a second set of ``sanity'' thresholds, so that if a constant changed by too much
from one iteration to the next, an expert was called to verify the output of the alignment or calibration task. In addition to this automated monitoring,
it was necessary to perform at least some of the data quality checks before executing the real-time analysis, since the real-time analysis
would throw away all information about the bunch crossing except for the signal candidate, making later data quality checks impossible.
A shortage of people made it impossible to implement any sophisticated data quality solution in time for the start of Run~2, and these data
quality checks were performed manually by the HLT team (very much including myself) by executing the HLT on a subset of buffered data.
This also meant that information normally been thrown away by the real-time analysis had to be archived and kept
``just in case'' while further validation took place, and it wasn't until the start of 2017 that the additional data archived in this way for 2015 datataking was permanently deleted.
A more detailed discussion of this validation of the real-time data can be found in Chapter~\ref{chpt:uneprocedure}.
\vspace{-5mm}
\section*{Software validation and maintenance}
Analogously to the monitoring and validation of the data, real-time analysis requires a much more stringent and comprehensive
monitoring, testing, and validation of the HLT software. This is partially because the HLT now
throws away most of the information even about the saved bunch crossings, so it is much more difficult or impossible to correct any
mistakes by re-analyzing the data later. And it is partly because real-time analysis implies both a great deal of extra complexity in the code,
and an enormous expansion in the number of people working on this code. During Run~1, LHCb's HLT software was maintained by a team
of around a dozen people, of which maybe five or six were responsible for the vast majority of the actual coding and testing. On the other hand
LHCb is a collaboration of a thousand people which publishes 50-60 analyses per year. Most of those people have had no exposure not only to the HLT
software, but have never coded software for a critical system, either in terms of the speed of the code or its reliability. It would have been
impossible to implement real-time analysis by making all these analysts submit their optimized selections to a small handful of HLT experts for implementation;
even if we had wanted to go down this road, the personpower simply did not exist. 

For these reasons, the shift to real-time analysis imposed specific requirements on the way code was written and maintained in LHCb. Firstly,
all central code, including real-time analysis code and selections, would have to undergo review before being put into production.\footnote{To be clear,
this did not happen during Run~1; code was largely written, tested, and committed by a single expert, put into production, and then debugged as
we went along. To be fair, however, it was only with the switch to git as a versioning system, which was not at all driven by LHCb or our real-time analysis,
that the code review approach became practically possible.} Secondly, best practices for writing and deploying real-time analysis code would have to be documented
in a way which could be followed by the average doctoral student, as independently as possible of their skill as a programmer. And finally, 
all code would have to be instrumented with automated tests, and this test suite would have to be updated as new bugs and failure modes were discovered.
These requirements were all met, in large part helped by the CERN-wide switch to git versioning and the associated central gitlab and JIRA infrastructure,
but also because LHCb's students and postdocs enthusiastically embraced these concepts and became their most strident advocates. I will not describe
details of this further, but will touch on some of its long term implications in the conclusion.
\section*{Addressing the requirements by splitting the HLT}
A central theme running through the above requirements is that the data processed by the HLT must be buffered while most of
the detector calibration is performed and validated. In particular, while the tracker alignment needs to be performed at most once
per fill, and updated much less frequently, the RICH detector calibrations must be updated on an hourly basis
in order to achieve the best physics performance. In addition, being able to buffer the data helps the data quality and monitoring,
as it allows a grace period during which any unexpected output of the automated alignment and calibration routines can be checked by experts
and corrected, if necessary, before it impacts on the datataking. 

\begin{figure}[t]
\centering
\includegraphics[width=0.35\textwidth]{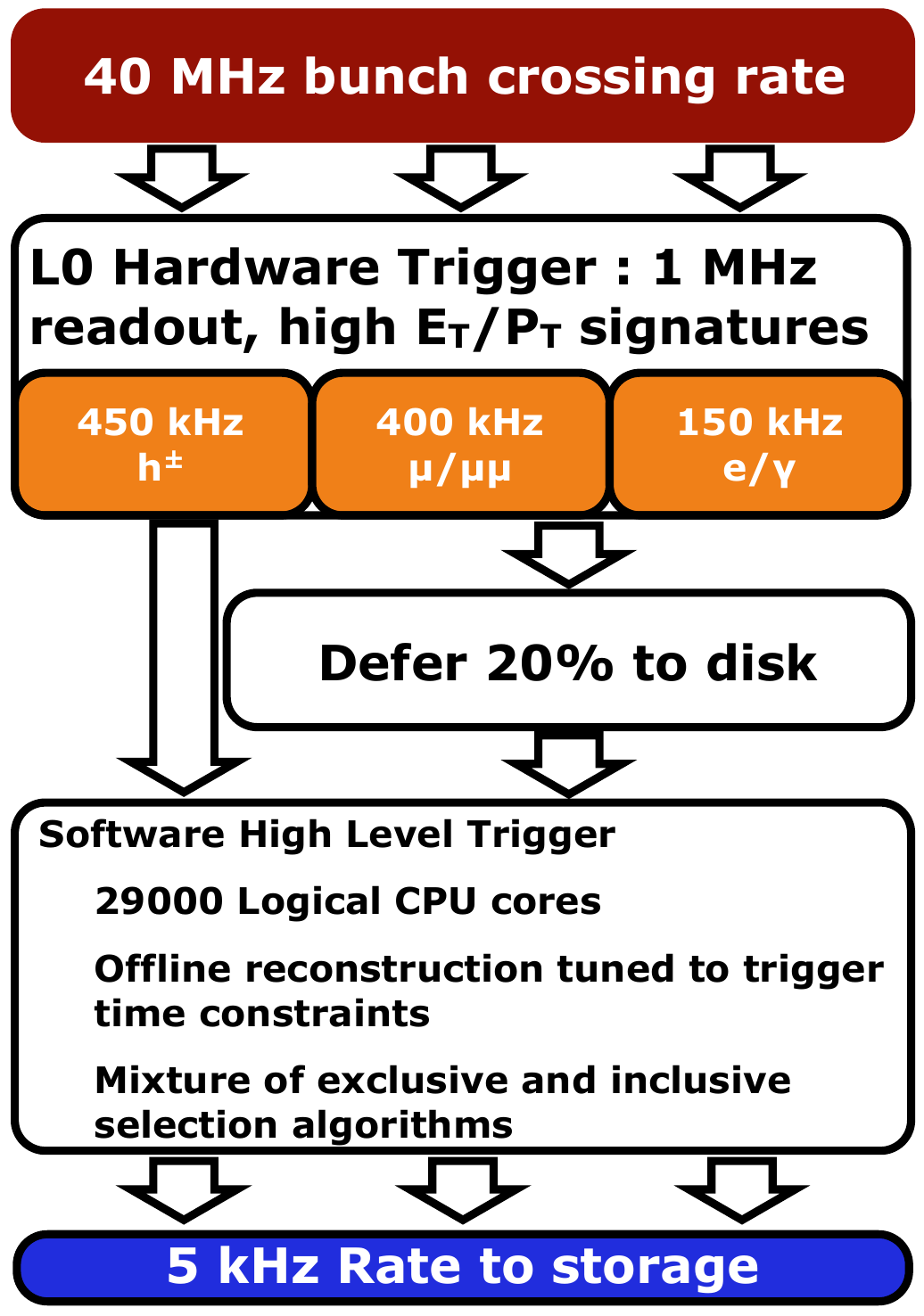}
\hspace{1.5cm}
\includegraphics[width=0.35\textwidth]{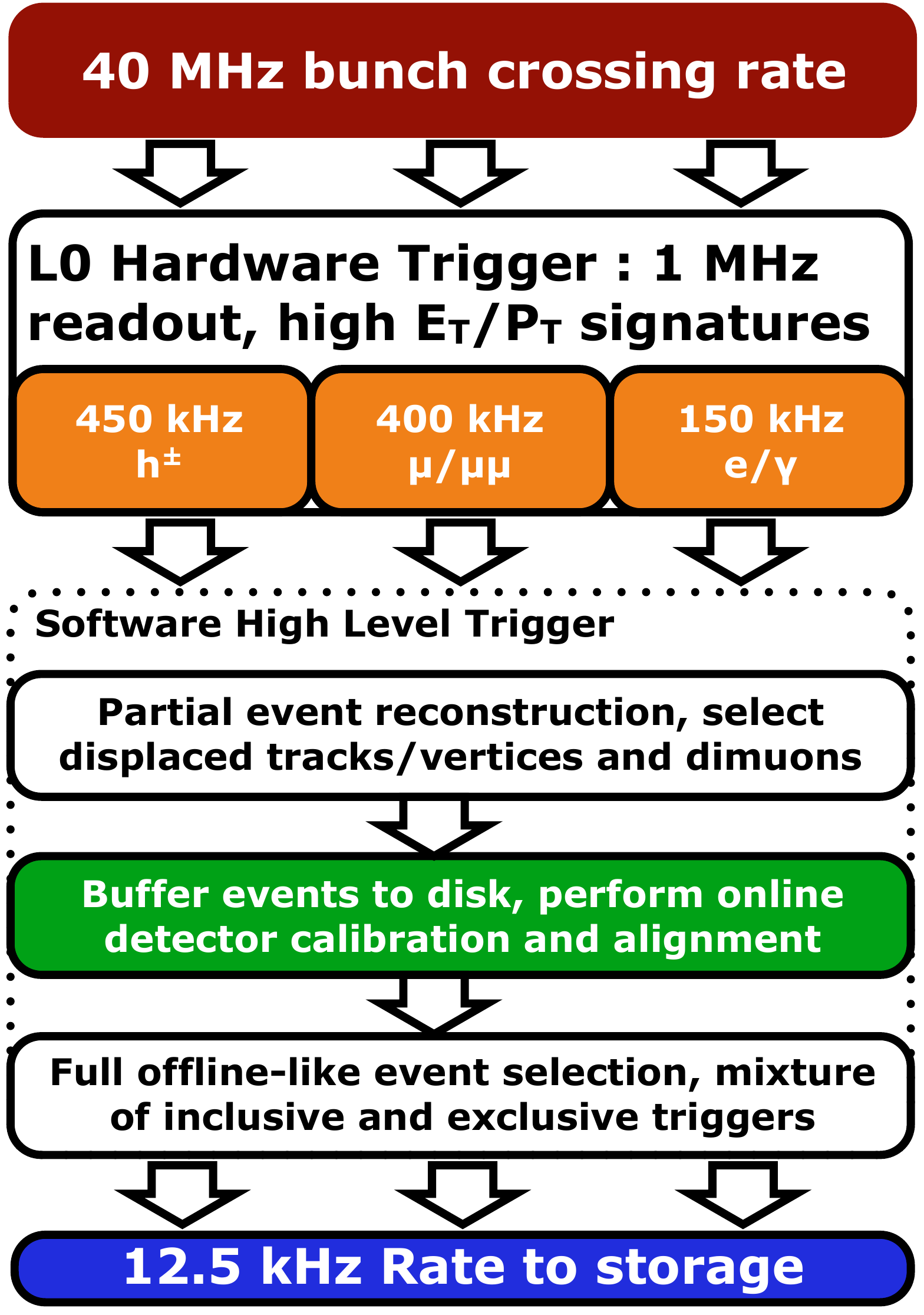}
\caption{\label{fig:lhcbtrig2015}Left: a diagram of the deferred Run~1 LHCb trigger, showing the buffering of the bunch crossings
coming from the hardware trigger. Right: a diagram of the 2015 LHCb trigger, showing the split nature of the HLT and the buffering after the first
HLT stage.}
\end{figure}

The HLT computing cluster provides a natural home for such a buffer, as was realized already during Run~1 datataking. At that time,
the goal of this disk buffer was to allow the HLT to take longer to process each bunch crossing: those which the HLT was too slow to process
while the LHC was running could be buffered to disks installed in each server node (with a capacity of 5-10~PB for the whole farm) and then processed in the period between LHC fills.
This configuration is shown on the left plot in Figure~\ref{fig:lhcbtrig2015}, and as can be seen around $20\%$ of bunch crossings selected
by the hardware trigger were buffered in this way.
This configuration of the disk buffer could not respond to the needs of real-time alignment and calibration, however, because this would
have required all bunch crossings selected by the hardware trigger to be buffered. At a hardware trigger output rate of 1~MHz, and around
50~kB per bunch crossing, this would have meant a data rate of around 180~TB per hour of LHC datataking, or around 2.7~PB of data for a typical
15 hour LHC fill. If everything was going well, this would not be a problem, because the data would only be buffered for a brief period of time
while the automated alignment and calibration were performed. But in the event of a problem in the automated alignment and calibration,
we would only have enough buffer space for around 55 hours of datataking before the data would have to be processed with an imperfect set of
detector constants. 

This seemed like an unacceptably risky approach, particularly given LHCb's experience with alignment and calibration in Run~1, which sometimes took 
weeks to get right after datataking started, and which required a good deal of manual intervention to achieve the best possible performance. Of course
all these algorithms would have to be much faster, more automatic, and less reliant on human intervention if any of this was to work, but
we nevertheless wanted to find a way to give ourselves a grace period of a couple of weeks, rather than days, in which to fix any problems. 
In addition, the need to verify the data quality before running the real-time analysis, discussed earlier, also spoke in favour of a deeper buffer.   
The solution was to split the HLT into two separate processes, as shown on the right plot in Figure~\ref{fig:lhcbtrig2015}.\footnote{The split between HLT1 and HLT2 already 
existed logically in Run~1, but the physical separation into separate applications was a major reworking of the codebase which consumed a great deal of effort
during 2013 and 2014, and which was critical to enable real-time analysis and a reliable distribution
of updated detector calibration constants during datataking.}
The first process (HLT1) would reduce the data
rate by a factor 7-10 using only tracking information, which was expected to be relatively insensitive to calibration updates beyond the very initial datataking period.
The selected bunch crossings would then be buffered to the disks in the server nodes, which could now hold hundreds instead of tens of hours of datataking before overflowing.
HLT1 would also implement specific selections for the different kinds of alignment and calibration samples listed earlier in this chapter, all of which required only tracking information.
Finally, the second process (HLT2) would implement the full detector reconstruction and real-time analysis, using the best possible detector alignment and calibration.
This setup had the additional benefit that LHCb could use not only the periods between individual LHC fills to run HLT2, but could actually buffer the data
long enough to also profit from the longer LHC technical stops, significantly increasing the available processing time and making it possible for HLT2 to execute
the full detector reconstruction needed for real-time analysis. We will now see how this buffer was optimized and used to make time a good deal less real than it would
otherwise have been.  

\chapter{\`{A} la recherche du temps r\'{e}el: optimizing the cascade buffers}
\label{chpt:unrealtime}
\pagestyle{myheadings}
\markboth{\bf \`{A} la recherche du temps r\'{e}el}{\bf \`{A} la recherche du temps r\'{e}el}

\begin{flushright}
\noindent {\it We are an empire now, and when we act, we create our own reality. And while \linebreak
you're studying that reality... we'll act again, creating other new realities. \linebreak}
\linebreak
-- Karl Rove
\end{flushright}

In order for real-time analysis to be implemented in LHCb, it is necessary to split the HLT into two different processes, the first
of which executes a partial detector reconstruction and allows a reduction of the bunch crossing rate with minimal loss of physics,
and the second of which executes the full detector reconstruction and real-time analysis. In between these two steps the data are buffered
while the real-time alignment and calibration of the detector is performed. This structure is an example of a ``cascade buffer'', which is a typical
structure of particular use in real-time data processing. Here I will give a pedagogical explanation of how a cascade buffer works
before describing the optimization of this buffer in the specific case of LHCb's Run~2 HLT and real time analysis. 

\section*{The concept of a cascade buffer}

The typical real-time data processing cascade used in high-energy physics interleaves reconstruction stages, which obtain information about a given
bunch crossing, with selection stages which decide whether to keep or discard a specific bunch crossing based on the information obtained in the preceeding
reconstruction stages. Every processing cascade relies on a data buffer, which can hold unprocessed bunch crossings while the selection stage is deciding
whether or not to keep them for further processing. We have already seen an example of such a cascade structure with the hardware and HLT triggers, which
is mandated by the inability to read out the full detector information for each bunch crossing. However the HLT can itself be subdivided in a series
of processes, and the optimal number of cascade steps depends on the balance between available processing power, buffer space, and the fraction of time
when the LHC is colliding and producing data. 

The reason for this cascade is that the computing budget available for real-time analysis is too small to
allow the full reconstruction to be performed upfront. At the same time, the reconstruction which we \textbf{can} afford to perform
upfront does not allow for an efficient data reduction to the final target, but it \textbf{does allow} for an efficient partial data reduction.
The cascade leverages this limited ability of the fast analysis to perform a partial data reduction and consequently make more time
available for more complex analysis steps. To give a simple example, we might have an overall budget of 10~ms in which to reduce the data
by a factor 100, and two analysis steps: a fast analysis which takes 5~ms and provides information which can efficiently reduce the data by a factor 10, 
and a slow analysis which takes 50~ms and provides information which can reduce the data by a factor 100. Running the fast analysis reduces the overall
time budget by 5~ms, but also reduces the data volume by a factor 10. It thus leaves an effective budget of 50~ms for processing the data which
survives the first analysis step, enough to run the slow analysis and perform the final data reduction.

While the described cascade allows us to stretch the concept of real-time to a certain extent, it remains constrained by the logic
that the real-time analysis should run while the collider is colliding particles. However, a typical collider will only be running
around $20-30\%$ of the time. This fairly universal ceiling is driven by the maintenance and commissioning needs of the machine,
and opens the possibility to stretch real-time even further by temporarily buffering the data on hard drives installed in the HLT computing cluster while the collider is running and
completing the processing and data reduction during the collider downtime, whether between fills or even during longer maintenance stops and shutdowns. 

In order to understand how such a data buffer can help to make time less real, let us revisit our earlier example.
Since the data buffer's performance depends on the volume of data, we will have to add these parameters to the model: a 50~GBs$^{-1}$
input data rate, and an overall data buffer of 5~PB.
Our first instinct might be to attempt to run the slow analysis from the beginning,
which would mean that $80\%$ of the input data rate, or 40~GBs$^{-1}$ would need to be buffered. The 5~PB buffer would consequently allow
for around 35~hours of buffering before it filled up, while it could be emptied at a rate of 10~GB$^{-1}$. It might seem
that this approach works, since the collider is only running at $30\%$ of the time, but this neglects the fact that the collider runtime and
downtime are not uniformly distributed throughout the year: most of the LHC downtime occurs during a several-month long winter shutdown
and two-week-long ``technical stops'' distributed throughout the year. The runtime is consequently also concentrated, with a peak structure
of repeated 15~hour long collision periods with breaks of 2--3~hours. In this context our naive approach is clearly suboptimal.

\begin{figure}
\centering
\includegraphics[width=0.8\linewidth]{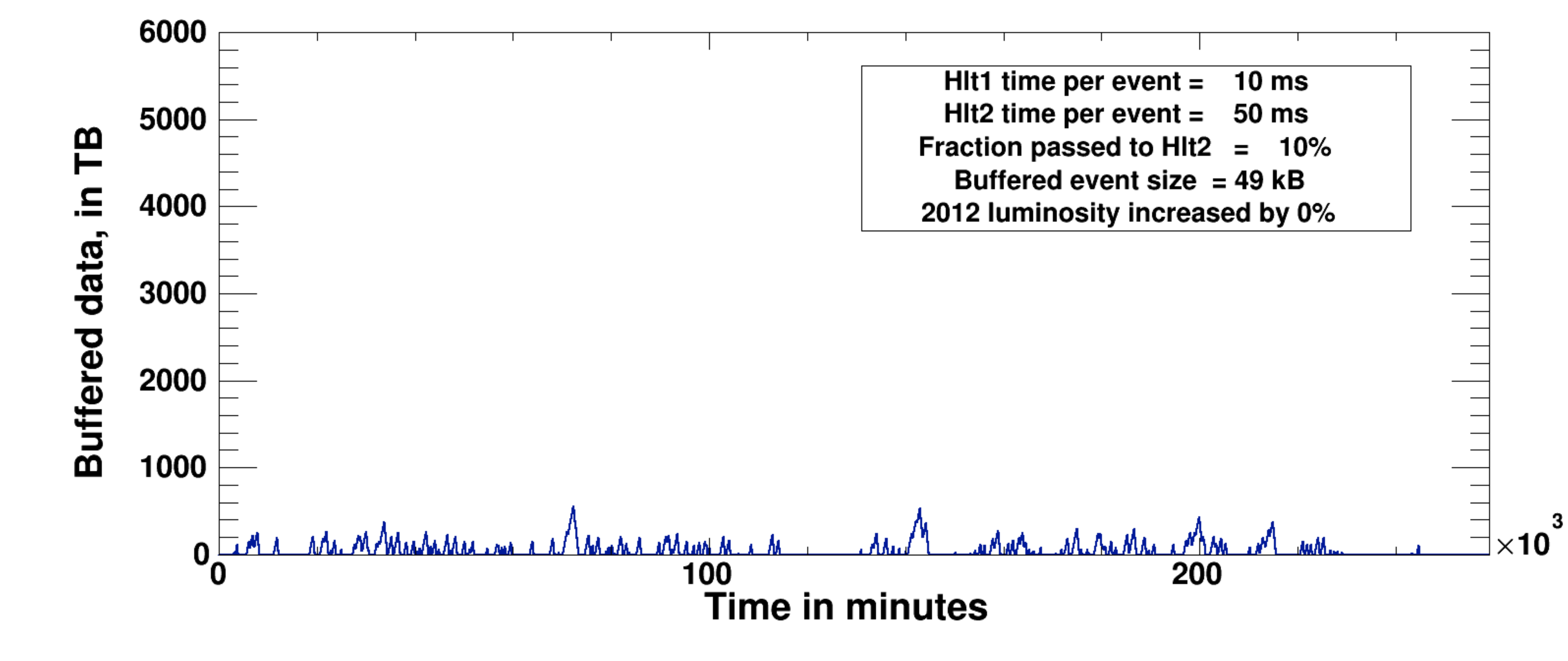}
\includegraphics[width=0.8\linewidth]{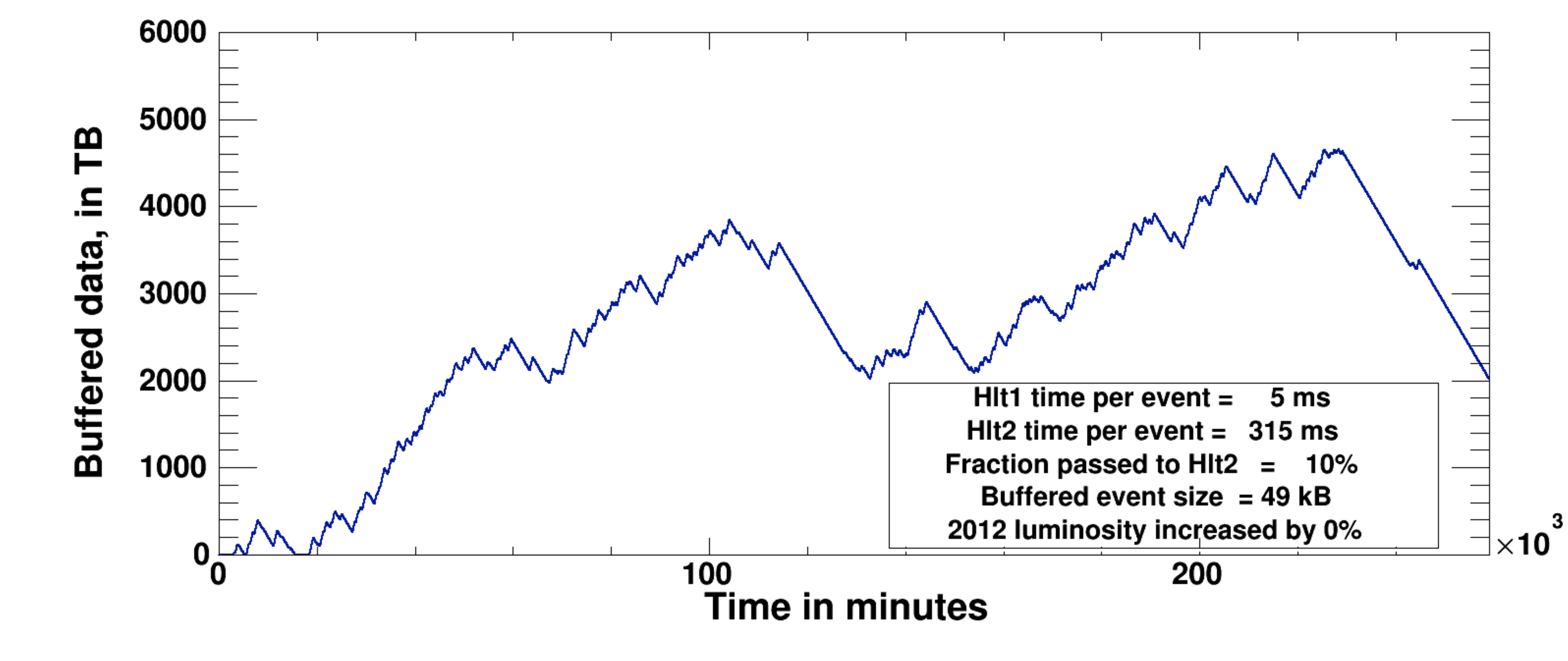}
\includegraphics[width=0.8\linewidth]{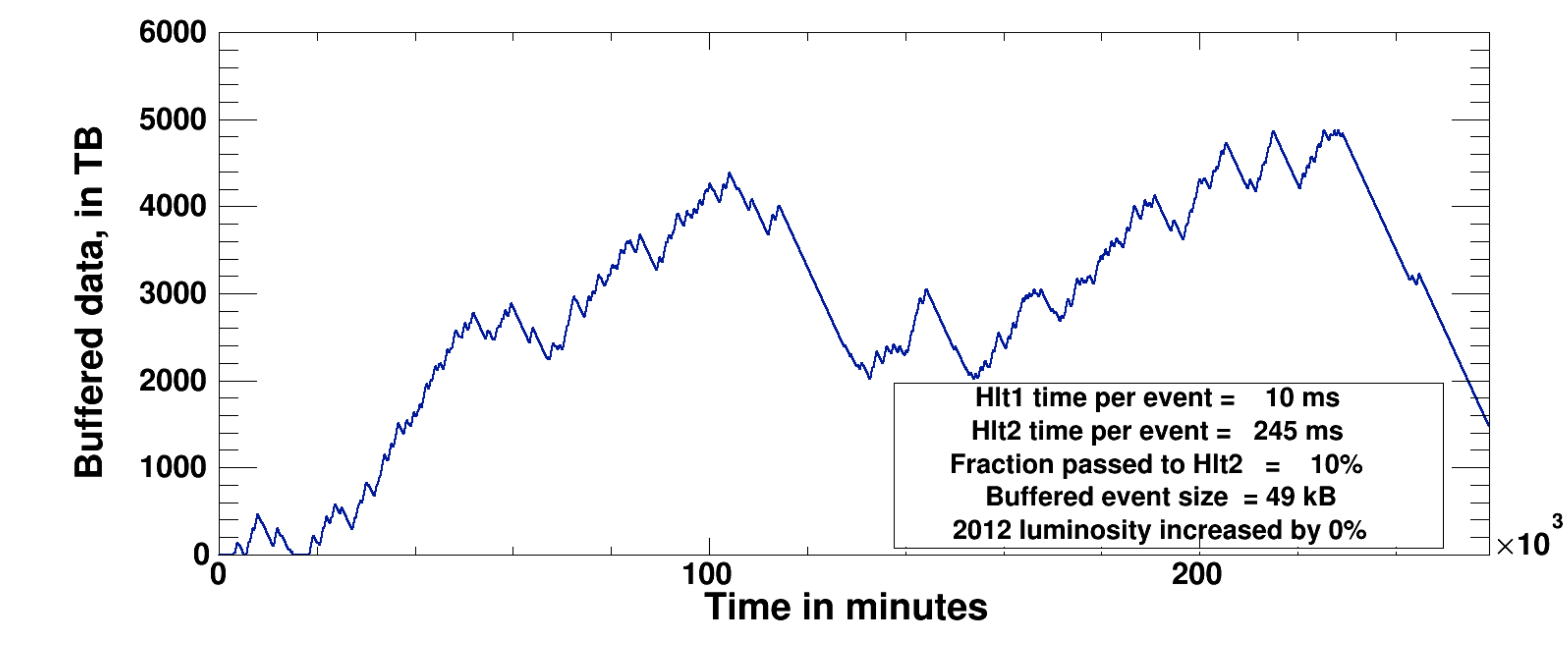}
\caption{Three possible ways to use the disk buffer: give more time to the fast analysis (top), give more time to the slow analysis (middle),
or share additional time between the two (bottom).}
\label{fig:diskbuffer}
\end{figure}

The problem with the naive approach is that it requires us to buffer too much data, so let us now try to combine our earlier analysis cascade
with the buffer by buffering only that data which survives the first analysis stage. We can then use the buffer to allow either the fast analysis,
slow analysis, or both, to take longer. In order to simulate the possible gains
we use the observed LHC fill structure in 2012, and show the buffer usage in three possible scenarios in Figure~\ref{fig:diskbuffer}. 
When giving all the additional time to the fast analysis, the buffer is hardly used, while the amount of additional time which can be given 
to the slow analysis varies between a factor of 5 and 6, depending on whether the fast analysis is also given more time. The precise optimum
will of course be problem dependent, but it should be clear that this buffering approach allows for non-linear gains in time for the final 
analysis stages. 

This specific pedagogical example was chosen to closely reflect the key cascade buffer which was implemented to enable real-time analysis in LHCb, but 
in reality the components of a cascade buffer are of course conceptual rather than physical entities. The buffer can
just as easily be in memory or on a cloud computing site halfway across the world, while the slow and fast analyses can just as easily be 
a reconstruction algorithm for high-momentum and low-momentum tracks. What is important is the structure of fragmenting work by preprocessing the
data with fast algorithms to reduce the volume of data which needs to be processed by slower ones, which is central to just about any real-time
data processing. We now turn to the optimization of LHCb's cascade buffers for Run~2, and how the offline reconstruction was made executable within
the HLT's resource constraints.

\section*{Optimizing the Run~2 cascade buffer and splitting the reconstruction}

The requirements presented in Chapter~\ref{chpt:acunningplan} can be translated into three key objectives for the design of HLT1, HLT2,
and the buffer connecting them. First of all, HLT2 must execute the full offline reconstruction, and in particular, not only tracking
but also all particle identification information must be made available from the start of the real-time analysis. Secondly, this full
reconstruction must be factorized into two parts: a fast reconstruction executed in HLT1, and a slower reconstruction which is executed
in HLT2 and which does not redo any of the work already done in HLT1 but instead builds on it. That is to say, the full HLT2 reconstruction
must not be some more sophisticated version of what was done in HLT1, but must supplement it in such a way that all objects found in HLT1
remain available for analysis in HLT2. The physics performance of this factorized reconstruction must also be equivalent to the physics
performance of a single reconstruction optimized without taking into account resource constraints, such as was executed offline on the Run~1 data.
Thirdly, HLT1 and HLT2 must fit into the resource constraints of the HLT computing cluster, optimized taking into account the available disk
buffer and the typical LHC fill structure.

Before discussing how these objectives were achieved, it is worth discussing in more detail the requirement that the HLT1 and HLT2 reconstructions
factorize. This requirement can be restated as follows: once a piece of reconstruction has been used to analyse the data, its output can be calibrated
and understood, and its output may be used to seed further reconstructions, but it should never be superseeded. In one sense, this is of course
just another version of the general constraint facing real time analysis: once the data has been analysed, all information which is not critical
to understanding the performance of this analysis or to inferring the physical observable of interest must be thrown away. But by implementing this approach
already in the HLT, where it would in principle be possible to run some different reconstruction in HLT2, we are in fact elevating this principle
of ``do not re-reconstruct'' from something done out of necessity to a virtue. I am explicitly underlining this point because it runs contrary to 
almost all existing HEP philosophy and practice, which takes it as a given that we will keep re-reconstructing our data as we understand it better.
Why, then, did we decide to do things differently in LHCb?

\begin{figure}
\centering
\includegraphics[width=0.48\linewidth]{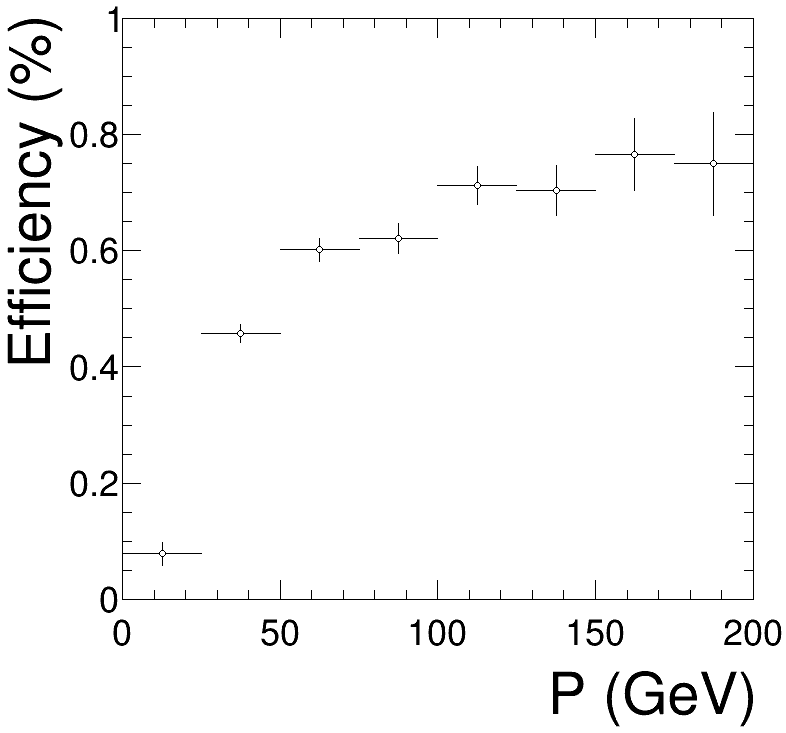}
\includegraphics[width=0.48\linewidth]{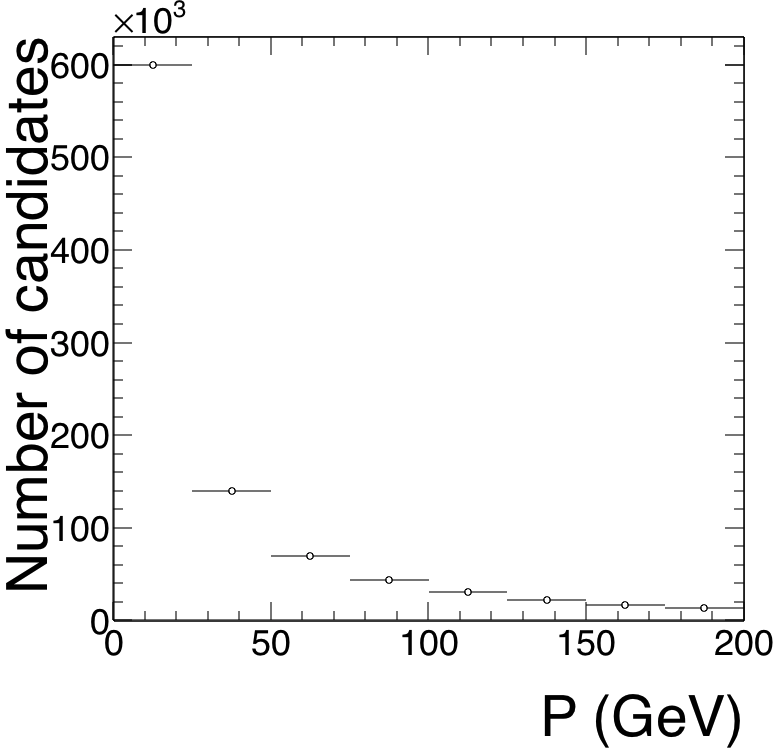}
\caption{Left: typical \Pqc-hadron efficiency curve for the Run~1 HLT1 track-based trigger. Right: typical kinematic
distribution of signal \Pq-hadrons, generated using the RapidSim package~{\cite{Cowan:2016tnm}}.}
\label{fig:trigeffcurve}
\end{figure}

A short answer could be that factorizing the reconstruction did not lead to any significant drop in performance,
but this glosses over the underlying reason for why the factorized reconstruction works in LHCb. 
LHCb is uncommon among HEP experiments in being neither a triggerless experiment, like most of those located
at $e^+e^-$ machines, nor an experiment whose analyses work on the trigger's efficiency plateau, like most general purpose detectors at hadron colliders.
The second point is illustrated in Figure~\ref{fig:trigeffcurve}, which shows a typical efficiency curve of the Run~1 HLT1 track-based trigger as a function of the
\Pqc-hadron transverse momentum, as well as the typical kinematic distribution of \Pqb-hadrons in the LHCb acceptance. As we can see, LHCb cannot
work on the so-called efficiency plateau where the trigger is close to 100$\%$ efficient and where this high efficiency would be constant
no matter what reconstruction was applied to the data, as most general purpose detector analyses do; it would simply thrown away far too much signal. 
Instead, LHCb analyses have to understand the efficiency across the full turn-on curve, and because of this they are extremely sensitive to differences between the reconstruction
applied in the trigger and any subsequent reconstruction. This leads, as LHCb analysts learned over and over again during Run~1, to one particularly important
effect. Any improvement in the reconstruction is stochastic in nature: if you improve the track finding efficiency by $1\%$, you will
typically find $4\%$ of tracks which you hadn't found before, and lose $3\%$ of the tracks which you had already found.\footnote{At this point you may be wondering
why, if this is true, LHCb does not simply run these multiple reconstructions in parallel and combine their output in some way. This is an interesting point
which parallels typical working practices in much of data science and machine learning, where it is known that combining the output of several different
analyses or classifiers, known as ensembling in the jargon of that field, systematically leads to better performance than any single analysis can achieve
on its own. It may be understandable in terms of genuine underlying physical effects: for example, a real particle trajectory may be more stable
against misalignments in the detector than a random collection of hits, because a real trajectory can tolerate one or two hits becoming outliers
before the fit $\chi^2/\textrm{ndof}$ becomes unacceptably large, while random collections of hits already exist at the edge of the $\chi^2/\textrm{ndof}$ distribution
and any change in the alignment may be enough to push them over the threshold. In practice, however, this approach would require
vastly greater computing resources than even traditional, nevermind real-time, analysis and for this reason nobody in LHCb has worked on it to the best of my knowledge.}  
However since the trigger already used tracks to make a decision, any tracks which were so used can only be lost by further reconstruction. This
places a natural ceiling on what can be gained by re-reconstructing, and in particular in the case of exclusive triggers which reconstruct the full signal
candidate before making their decision, no improvement is possible by definition. This holds between the Run~1 trigger and the then-offline reconstruction,
but the same argument would equally well apply to the HLT1 and HLT2 reconstruction, or indeed any other data processing cascade, and is at the heart
of LHCb's decision to almost entirely do away with re-reconstructing the data.\footnote{Almost, because for the minority of Electroweak, top, or Higgs analyses which
are able to work on an efficiency plateau and where the very-high momentum tracks make it particularly challenging to obtain the best possible detector
alignment in real-time, we will keep doing analysis in the traditional way.}

\begin{figure}
\centering
\includegraphics[width=0.95\linewidth]{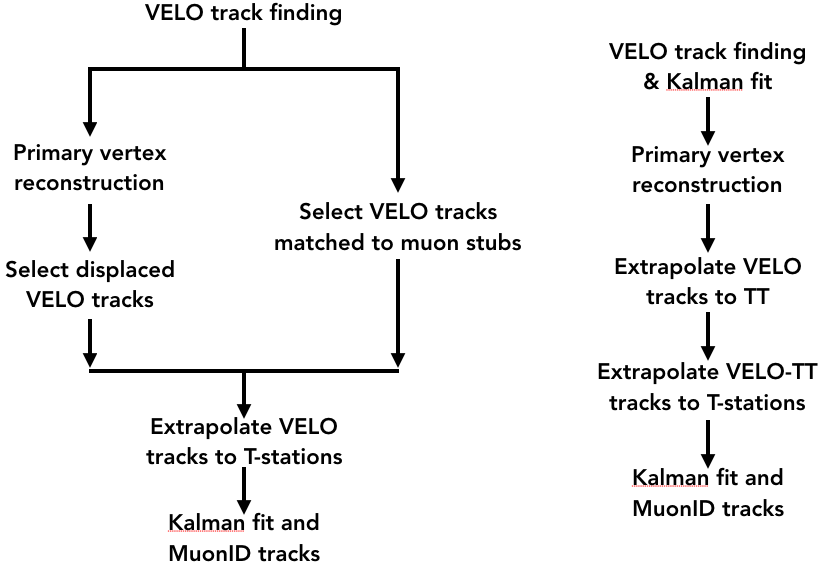}
\caption{Left: the main HLT1 reconstruction chain used in Run~1. Right: the main HLT1 reconstruction chain used in Run~2. Note that other more specialized
reconstructions were used for events selected by specific L0 triggers, which have been omitted for simplicity.}
\label{fig:hlt1recochains}
\end{figure}

The actual factorization of the reconstruction was performed based on experience gained during Run~1, when the split between HLT1 and HLT2
already existed within the trigger although both ran within a single software process. We knew that HLT1 could achieve the required rate reduction
using only tracking information, and that this strategy would remain feasible in Run~2; the task was to, as much as possible, remove or reduce simplifications
in the HLT1 reconstruction, so that it could be executed as the first stage of a factorized reconstruction without leading to a loss in performance.
The HLT1 reconstruction chain used during Run~1 is shown in Figure~\ref{fig:hlt1recochains}, and it departed from the full offline reconstruction in three major respects.
First of all, the primary vertices were made with VELO tracks which have not been Kalman fitted, rather than with the full set of tracks found and Kalman fitted by
the reconstruction. This meant that both the efficiency to find a primary vertex, the efficiency to misidentify a secondary decay vertex as a primary vertex,
and the primary vertex position resolution were worse in HLT1 than in the full reconstruction. Secondly, only a subset of VELO tracks which passed impact
parameter cuts, or had matching hits in the muon system, were propagated through to the T-stations in order to estimate their momentum. And finally, the search
windows for this track-finding were tightened compared to the search windows used in the full reconstruction. This made the reconstruction significantly faster, but
also meant that it could only find tracks above the transverse momentum thresholds listed in the figure.
It was clear from the beginning that HLT1 would never be able to find tracks over the whole momentum range covered by the full reconstruction, 
and therefore that the primary vertex reconstruction could also not be performed using all the tracks in the same way as had been done offline during
Run~1. The strategy adopted was to understand if the primary vertex reconstruction could be changed to use only VELO track segments without worsening its performance,
and if it was possible to factorize the track finding into a high-momentum search followed by a low-momentum search without losing performance. 
The specific threshold for this high-momentum search would be set in such a way that all VELO tracks could be processed, without requiring a
large impact parameter or matching muon hits.\footnote{This decision led to other incidental benefits, for example the fact that we were able to select
hadronic signatures without introducing any trigger-level bias on their decay-time, a first for a triggered hadron collider experiment.} 

The reoptimization of the primary vertex finding was largely performed by my student Agnieszka Dziurda and is
documented in her PhD thesis~\cite{Dziurda:2015xfb} as well as in an internal LHCb note which we coauthored with Mariusz Witek.
Briefly, however, we reoptimized the primary vertex finding to use
Kalman fitted VELO tracks; as a VELO track has no momentum estimate, the Kalman fit was told to assume the same with the same assumed average momentum (400~MeV)
in all cases. Because the VELO tracks are straight lines, this Kalman fit could be performed in a fraction of the time of the full Kalman fit including
the T-stations, and therefore could fit into the HLT1 resource requirements. Not only did this not degrade the primary vertex finding and resolution but it actually improved it.
The reason for this improvement is that tracks which come from the decays of long-lived particles like \Pqb- and \Pqc-hadrons tend to have higher momenta than those produced in
the $pp$ collision itself, and consequently suffer less multiple scattering and have smaller estimated uncertainties on their slopes and impact parameters.
This meant that on the occasion that a \Pqb- or \Pqc-hadron decayed close enough to the primary vertex for its decay products to be included in the primary
vertex by the algorithm\footnote{We carried out a dedicated reoptimization of the thresholds used to include a track within a given primary vertex, but because
the lifetime of particles follows an exponential distribution there is no threshold which can entirely avoid this effect.}, those decay products would
have smaller uncertainties and hence higher weights in the primary vertex finder, thus biasing the primary vertex position towards the long-lived particle
and eventually measurably biasing the estimated decay-time of that particle. Using only VELO tracks and assigning the same momentum to each one significantly
reduced and in many cases eliminated this effect. 

The reoptimization of the track-finding is documented elsewhere~\cite{Storaci:2015rma}, and relied on a more sophisticated use of the TT. In Run~1, TT hits were added
to tracks at the end of the full reconstruction sequence, and mainly used by the ghost probability algorithm to help identify fake tracks, or in helping
to find decays in flight from kinks in their trajectory. For Run~2, as shown in Figure~\ref{fig:hlt1recochains} the TT was placed at the heart of the HLT track finding, with all VELO tracks first
extrapolated to the TT in order to obtain a rough momentum estimate, before being further extrapolated to the T-stations. The momentum estimate from the VELO-TT
tracking helped to define much more precise T-station search windows than the hypothetical momentum cutoffs used in Run~1, and meant that the efficiency
of the HLT1 tracking as a function of momentum turned on very sharply. This, in turn, made it possible for the HLT2 reconstruction to begin from the high-momentum tracks
already found in HLT1, and add lower-momentum tracks using a separate set of search windows executed only on those detector hits not already used in HLT1. This
achieved the stated objective of factorizing the HLT1 and HLT2 reconstructions. 

\begin{figure}
\centering
\includegraphics[width=0.9\linewidth]{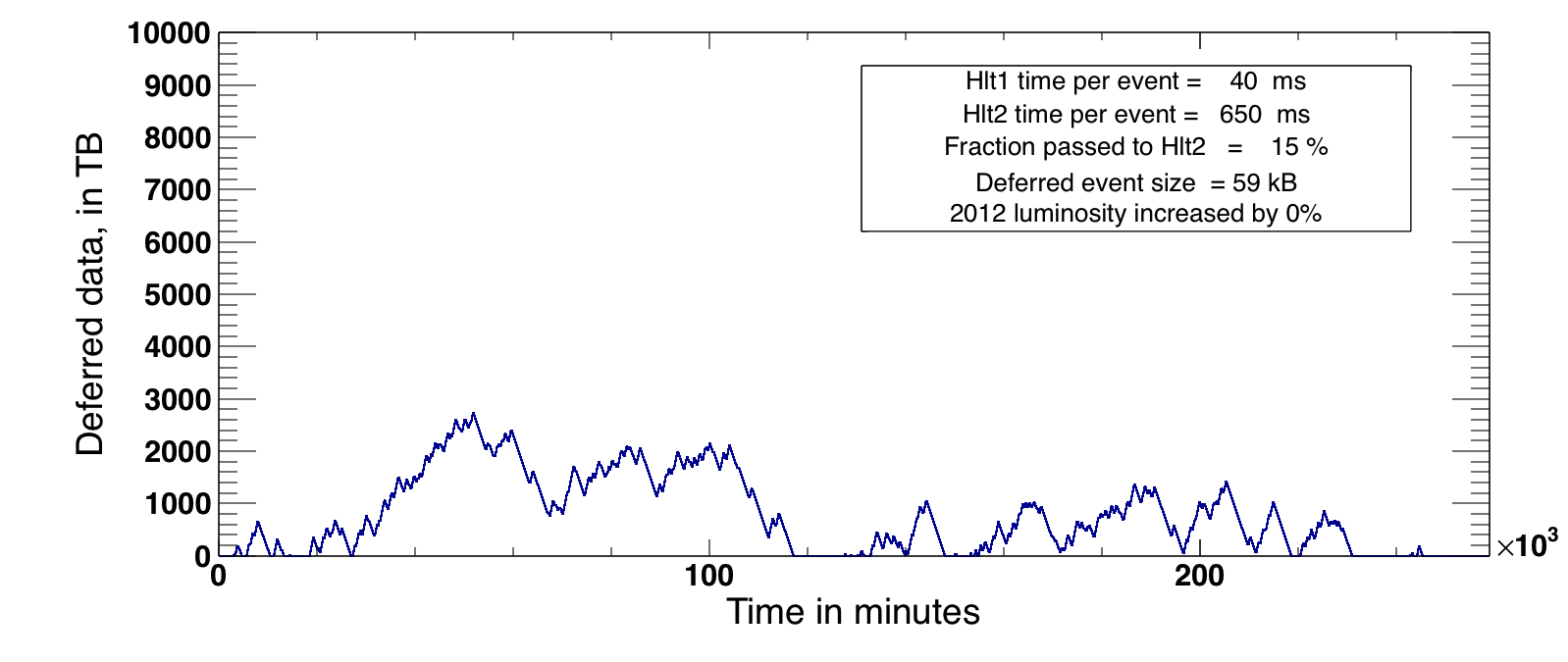}
\includegraphics[width=0.9\linewidth]{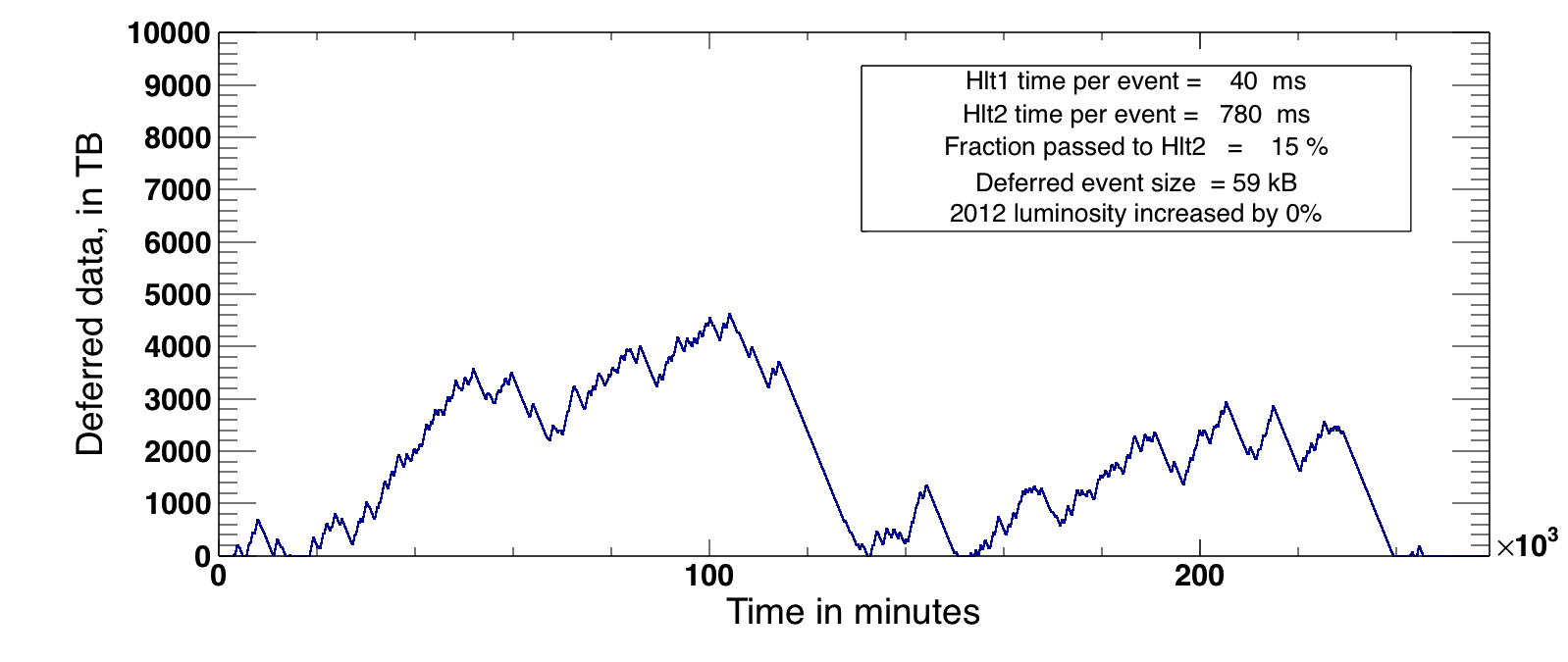}
\includegraphics[width=0.9\linewidth]{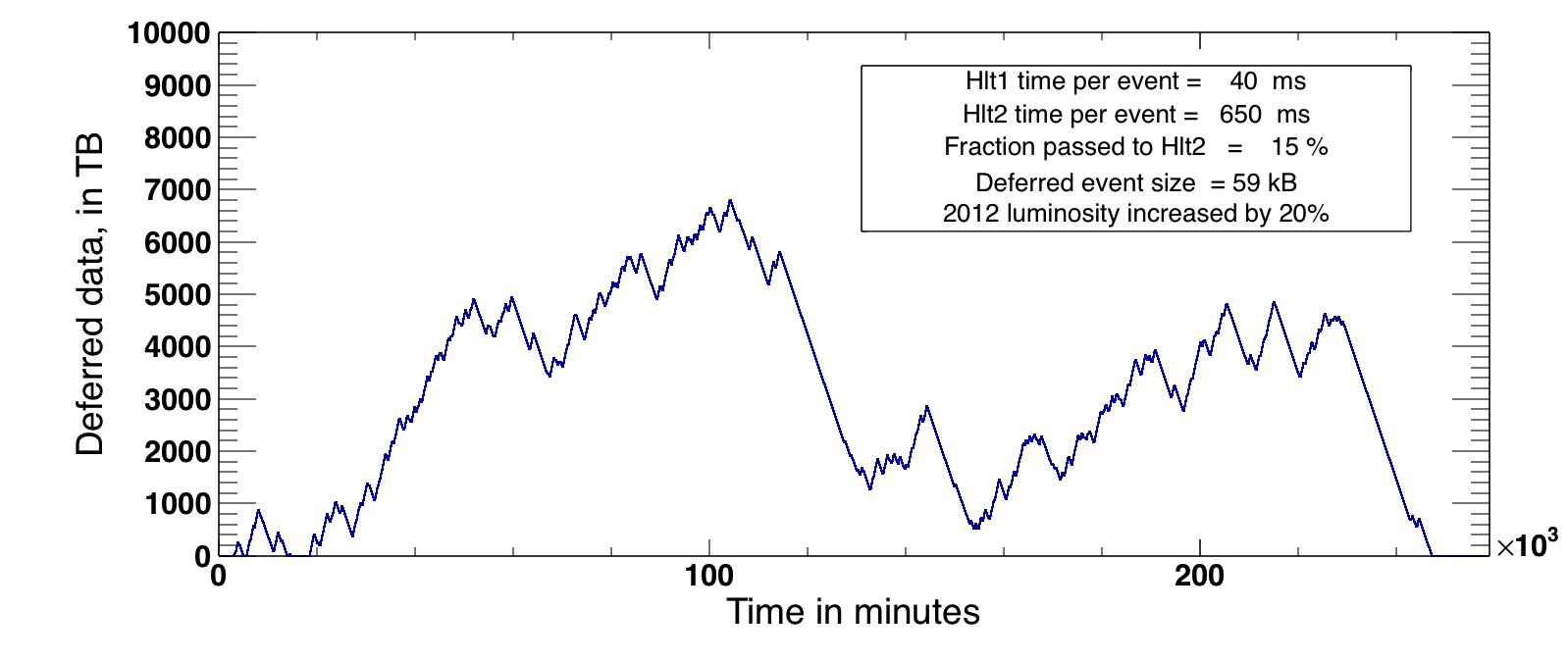}
\caption{Scenarios of disk buffer usage in Run~2, all of which assume that a bunch crossing saved by HLT1 is 59~kB and that HLT1 propagates 150~kHz of bunch crossings to HLT2.
From top to bottom: the baseline scenario, a scenario where the HLT2 reconstruction time is slower by 20\%, a scenario where the LHC provides 20\% more luminosity. We can see
that even though the nominal scenario looks safe and never fills much more than half the disk buffer, small errors in our assumptions can quickly lead to the disk buffer
filling up during a nominal year of datataking.}
\label{fig:diskbufferrun2}
\end{figure}

There were of course reoptimizations of many other parts of the LHCb reconstruction between Runs~1~and~2, and many of those brought important improvements
in the speed of the reconstruction without any decrease in physics performance and contributed to the feasibility of the real-time analysis. What remained
was to optimize the structure of the cascade buffer, in particular the fraction of bunch crossings which could be processed in HLT2, based on the estimates
of how long HLT1 and HLT2 would take to process each bunch crossing. Because the LHC energy would change from 8~TeV at the end of Run~1 to 13~TeV at the start
of Run~2, it was not easy to extrapolate the timing from Run~1 data; instead a mixture of studies based on Run~2 simulation, Run~1 data-simulation agreement,
and studies of how Run~1 timing varied with the number of $pp$ interactions per bunch crossing or the multiplicity of each $pp$ interaction were used. From these 
HLT1 was estimated to take around 40~ms per bunch crossing, while HLT2 was estimated to take around 650~ms per bunch crossing. This led to the disk buffer scenarios
shown in Figure~\ref{fig:diskbufferrun2}, where the nominal disk buffer usage is compared to two possible contingencies. 
This led to a final important operational optimization of the cascade buffer. While the hard disks installed in the server
nodes had a capacity of 10~PB in total, they were used in a mirrored configuration in order to avoid data losses which halved this effective capacity to 5~PB.
As can be seen from the extrapolation plots this would mean that, if the LHC replicated its 2012 fill structure, we could come uncomfortably close to filling up
this disk buffer, in particular if the HLT2 estimate was too optimistic. This was a particular worry because we had never previously gotten these estimates right after a shutdown,
and because we anticipated that the LHC would not only move to a new energy but also to a new bunch filling structure for Run~2. For these reasons, once we had gained 
operational experience with the system in 2015\footnote{Using the disks mirrored in 2015 was a compromise based on the fact that we didn't anticipate the LHC to reach
full luminosity particularly quickly at the new energy, and we wanted to reduce the stress on the operations team.}, the disks in the nodes were unmirrored,
doubling the disk buffer size and making the risk that it would fill up negligible. In practice, we got the HLT time estimates almost exactly right at the start of
2015 datataking, and there was never any real danger of exceeding the allocated budget or any need to retune things in a hurry. And although they did increase the operational workload somewhat, 
the unmirrored disks led to only a percent-level loss of data, most
of it due to human error when testing new system configurations rather than due to disk failures.

\chapter{Il nous faut une proc\'{e}dure: persisting and validating the data}
\label{chpt:uneprocedure}
\pagestyle{myheadings}
\markboth{\bf Il nous faut une proc\'{e}dure}{\bf Il nous faut une proc\'{e}dure}

\begin{flushright}
\noindent {\it On lache pas on s'approche, \linebreak
du but on sera proche \linebreak
Sur disque ou sur porche, \linebreak
c'est grav\'{e} dans la roche} \linebreak
\linebreak
-- Sniper, Grav\'{e} dans la roche
\end{flushright}

The reoptimization of LHCb's reconstruction and spliting of the HLT made real-time analysis possible, but implementing it required
a separate infrastructure to be put in place which would enable the HLT to scaleably persist the data required by analysts to long term storage.
It also required an infrastructure to access this data and transform it into the ROOT format which analysts could read, while remaining flexible
enough to accomodate different requirements of different analyses. In addition, a validation
of this data had to be set up, in order to make sure that all the data required by analysis was indeed being saved at the HLT level, since
real-time analysis in principle left no room for mistakes to be corrected later on. Here I will briefly describe the design of how the
data were to be persisted, and show how a scaleable design allowed the real-time analysis framework to eventually persist not only the signal candidates,
but also derived information about the rest of the bunch crossing. I will also discuss how the system was commissioned and validated in early 2015 datataking,
and the special role played in this validation by the calibration data streams.

\section*{Persisting and resurrecting the data}

The HLT software always had the ability to record trigger-level information to the raw format sent by the HLT to permanent storage. Durning Run~1,
this information consisted of two types of objects: ``decision reports'', which held information about which trigger selections had caused the
bunch crossing to be saved; and ``selection reports'', which held information about the trigger-level objects (tracks, neutrals, composite particles)
built by these selections. This information enabled analysts to measure biases which the trigger selections introduced by matching
their analysis-level signal candidates to the trigger-level objects which had caused the bunch crossing to be saved. Each bunch crossing could be divided
into three categories:
\begin{itemize}
\item Trigger on signal (TOS), in which the trigger would have saved the bunch crossing even if all objects not associated with the signal candidate
were to be removed;
\item Trigger independently of signal (TIS), in which the trigger would have saved the bunch crossing even if all objects associated with the signal
candidate were to be removed;
\item Triggered on both (TOB), those bunch crossings which are neither TIS nor TOS, in other words bunch crossings for which the trigger needed
both some of the information about the signal candidate and some of the information about other objects in order to make a positive decision.
\end{itemize}
This categorization naturally leads to a data-driven tag-and-probe method for measuring trigger efficiencies, proposed and developed before datataking began~\cite{HernandoMorata:2008zz}.
The efficiency for TOS
bunch crossings can be obtained by starting from the unbiased TIS sample\footnote{The TIS sample is not of course entirely unbiased, because frequently the
trigger-level objects which caused the bunch crossing to be saved will be associated to other products of the signal's QCD fragmentation chain, and
therefore be kinematically correlated with the signal. A common example is a trigger which selects \Pqb-hadrons, which if it does not select a bunch
crossing because of the signal \Pqb-hadron candidate will most often select the bunch crossing because of the decay products of the other \Pqb-hadron
produced in the same $b\bar{b}$ fragmentation chain. This means that the kinematics of the signal candidates in the TIS sample will not match the kinematics
of the signal candidates at production, and the integrated trigger efficiency given by this method will be biased.
For this reason, as with any other tag-and-probe method, the efficiencies must be obtained differentially
in the momentum and pseudorapidity of the signal candidates.} and calculating
\begin{equation*}
\epsilon_{\textrm{TOS}} = \frac{N_{\textrm{TOS and TIS}}}{N_{\textrm{TIS}}}
\end{equation*} 
Because of the peculiarity of working off the trigger's efficiency plateau noted in Chapter~\ref{chpt:unrealtime}, and the consequent need to validate
and correct the data-simulation agreement of trigger efficiencies at the few percent level, variants of this kind of efficiency determination have been used in
many LHCb analyses.

\begin{figure}
\centering
\includegraphics[width=1.0\linewidth]{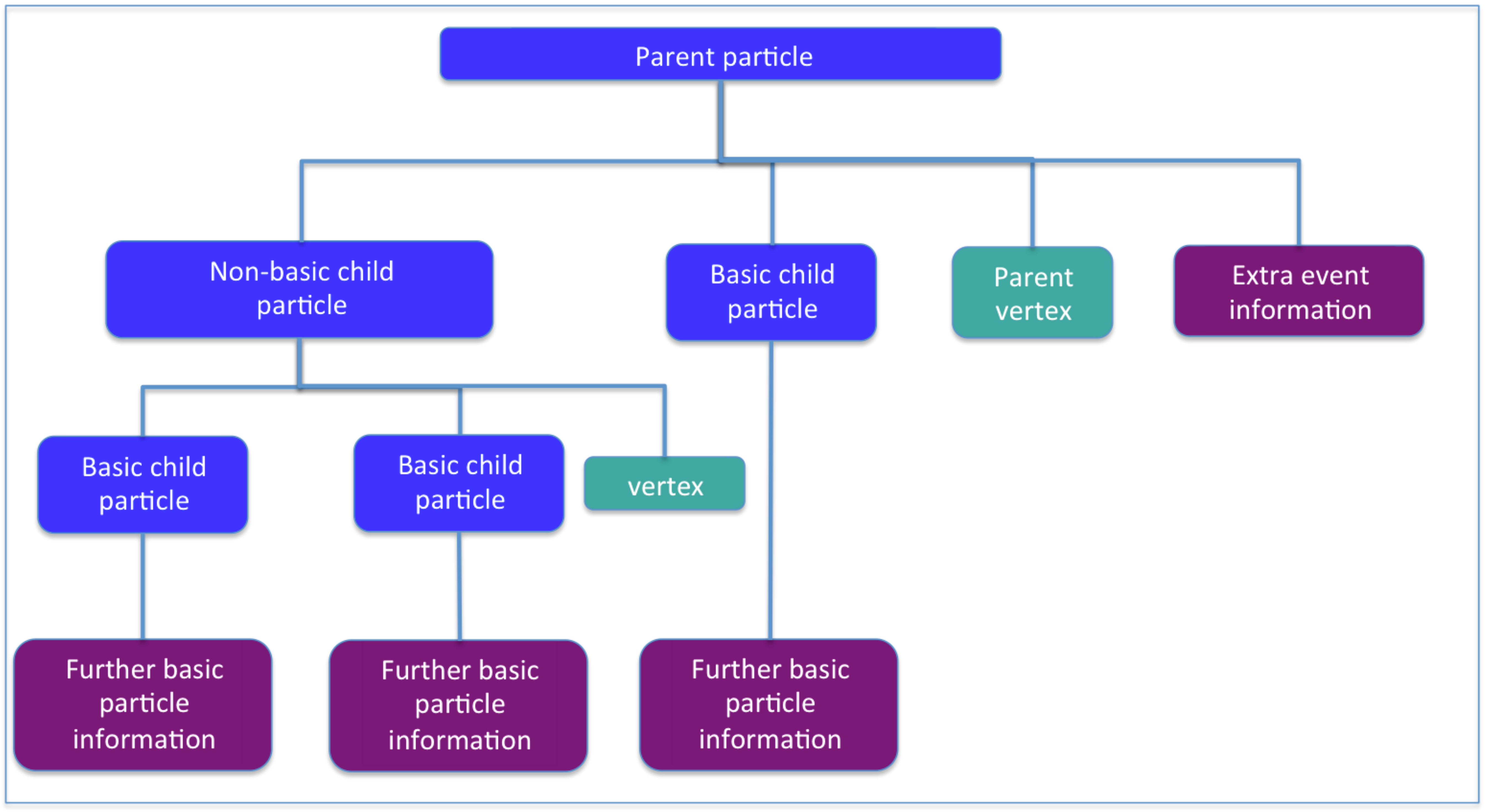}
\caption{Diagrammatic representation of the way real-time analysis data are persisted, reproduced from~\cite{Aaij:2016rxn}. This diagram could for example represent
a $B^-\to D^0 (\to K^- \pi^+) \pi^-$ decay chain, with the $B^-$ as the parent particle and the $D^0$ as the non-basic child particle. ``Extra event information''
allows the flexible storing of other reconstructed particles or complex objects such as particle isolation criteria which are associated to the signal candidate,
as explained in the next section.}
\label{fig:teslastructchain}
\end{figure}

It was clear from the beginning of the real-time analysis development that the HLT would have to continue to write raw files to storage because the limited
personpower and strict time constraint on the project (to be ready for the start of 2015 datataking) made it unfeasible to rework the HLT so that it wrote 
files in a ROOT format. At the same time, the compressed format which had been used to store the selection reports would need some adapting in order to cope
with the much greater volume of data required by real-time analysis. The main goal was to develop a flexible structure which could store analysis-level data
in several levels, each corresponding to one particle in the signal decay chain, and for this structure to allow an arbitrary amount of analysis level information
to be associated with each of these particles. The developed structure is shown in Figure~\ref{fig:teslastructchain}. There is a basic set of quantities which is
stored by default for each type of particle (stable charged, composite, neutral), and analysts can define additional information to be stored when writing
the real-time analysis module. In order for the offline analysis software to be able
to decode the raw information in this structure later, it would have to know the structure itself, that is to say it would have to know which piece of physics information
was stored in which data location. In addition, the structure could evolve over time, for example if an analysis became more sophisticated or analysts realized that they
needed some additional information to reduce an important systematic. For these reasons, the structure definitions are versioned, and the version used is written
to a header associated with every saved bunch crossing; the offline analysis software can later on read these headers and access the correct raw bank definition on a per-bunch-crossing
basis, thus ensuring that all data taken can be processed with a single version of the offline analysis software and at the same time enabling the analyses to evolve flexibly over time.
As we will now see, this structured approach also smoothly solved the problem of storing additional information which was not directly part of the signal candidate.

\section*{Saving global information in real-time analysis}

As touched on in Chapter~\ref{chpt:acunningplan}, the real-time analysis framework is most naturally suited to signals which can be fully reconstructed in real-time,
minimizing the amount of information which needs to be saved.  There is, however,
a specific difficulty for using the real-time analysis for studies which require information about the rest of the bunch crossing, for example spectroscopy studies;
analyses which rely on track isolation information; or analyses which infer the production flavour of a neutral meson through information about other particles
produced in the same QCD fragmentation chain. This difficulty will be illustrated here using the example of a spectroscopy analysis, 
however the same logic is applicable to any other information not directly connected to the signal candidate selected by real-time analysis.

To understand the difficulty, it is worth restating the logic behind real time analysis. There are signals which we wish to analyze which the LHC produces
at such a high rate that if we were to store the full information about the bunch crossings in which these signals were produced, the signal data rate would
overwhelm our storage capacities. However analysing this data does not require information about the entire bunch crossing in which they were produced, only
about the signal candidates and their decay products. Therefore if the detector can be fully reconstructed in real-time, and the signal candidate selected,
it is possible to only save this signal candidate, reducing the storage requirements by an order of magnitude or more, and thus greatly increasing the statistical
reach of the analyses in question.

There are two critical aspects to this logic. Firstly, in order for the signal candidate to be fully reconstructible it must first of all be possible to fully
define it, that is to say, we must know that we wish to study such-and-such a hadron decaying to such-and-such a collection of tracks or neutral particles.
And secondly, it must be possible to select this signal with a high purity, ideally so high that the average number of signal candidates per selected bunch crossing
is 1. Otherwise, much of the space saving benefit of real-time analysis is lost. The problem with spectroscopy analyses is that they break both of these assumptions.
An important aspect of spectroscopy is to look for new particles, and in particular new excited and possibly unforeseen exotic hadrons which will improve our understanding
of QCD; LHCb's recent measurements of tetraquarks~\cite{Aaij:2015eva} and pentaquarks~\cite{Aaij:2015tga} are good examples of this principle in action. 
A better understanding of excited \Pqc-hadrons may also be critical for the study of \Pqb-hadron decays, for example in measurements of the CKM angle $\gamma$ or in tests
of lepton universality in semileptonic \Pqb-hadron decays.
And secondly, excited hadrons are not long-lived, which means that all the particles produced in the same $pp$ collision form a natural and to some extent irreducible background.
Still, real-time analysis is critical to heavy-flavour spectroscopy, and particularly \Pqc-hadron spectroscopy, because all excited \Pqc-hadrons decay into the ground-state
long-lived \Pqc-hadrons. To maximize the signals available for spectroscopy, we will need to reconstruct these long-lived \Pqc-hadrons in their Cabibbo-favoured
decay modes. Since it was already shown in Chapter~\ref{chpt:haystackofneedles} that even the Cabibbo-suppressed \Pqc-hadron signals would saturate the available storage space
in a traditional analysis strategy, there is no alternative but to find a way to perform these studies in real-time.

\begin{figure}
\centering
\includegraphics[width=1.0\linewidth]{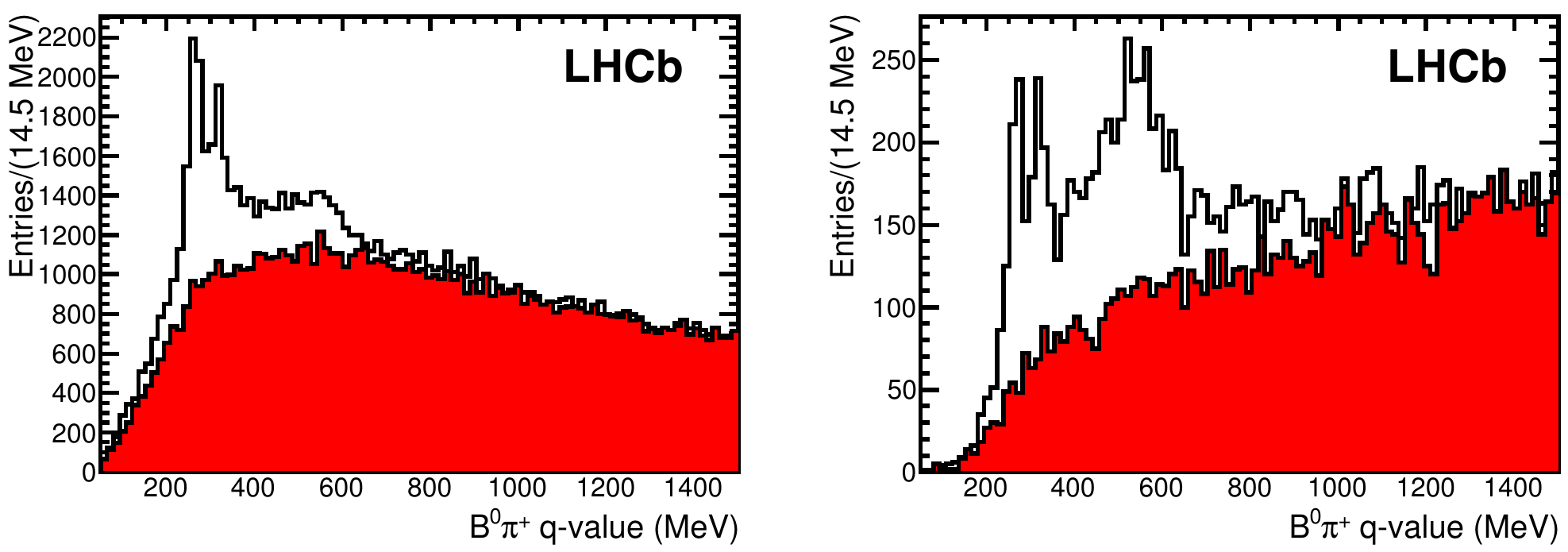}
\caption{Invariant mass of the (open histogram) right-sign and (red) wrong-sign $B^\pm \pi^\mp$ combinations, for all combinations with companion $p_{\textrm{T}} > 500$~MeV on the left,
and with this requirement tightened to 2000~MeV on the right. Reproduced from~\cite{Aaij:2015qla}.}
\label{fig:bststmass}
\end{figure}

It is of course possible to reduce this problem to a manageable level by making particularly stringent requirements on particle identification or transverse momentum
for the companion particle or the excited hadron itself; the former can be well motivated by a desire to avoid fake exotic candidates caused by the misidentification
of more mundane signals, while the latter is clearly motivated by the fact that the excited \Pqb- or \Pqc-hadron will take a larger fraction of the momentum than 
other particles produced in the same fragmentation chain. An example can be seen in Figure~\ref{fig:bststmass}, taken from an LHCb analysis of doubly-excited \Pqb-mesons.\footnote{This
analysis is chosen because I was one of the principal analysts, but a similar point could have been made using any number of other published LHCb spectroscopy analyses.}
Both the signal to background and the candidate multiplicity per bunch crossing significantly decrease when applying a stricter requirement on the companion transverse
momentum. At the same time, the signal yield drops dramatically. For this reason the analysis actually uses all candidates passing the looser transverse momentum requirement,
and simply splits them into more and less pure subsamples according to companion transverse momentum to improve the precision on the measured signal parameters.

Fortunately, the flexible data format described in the previous section provides a natural solution to this issue. The real-time analysis selections must be defined
for the subset of Cabibbo-favoured decays of long-lived ground state hadrons, which are known and small in number. Subsequently, additional companion particles which pass
certain preselection requirements can be added as ``extra event information'' to form the real-time analysis object which is saved to storage. As the raw detector information is not saved,
and as any particles which are not produced in the same $pp$ bunch crossing can be discarded, most of the space saving is maintained. At the same time, there is no proliferation
of real-time analyses or selections to be maintained, and if additional (or more complex) companion particles become interesting later, they can simply be added as a new version
of the raw bank encoding in a way which is analysable together with the data recorded earlier. And of course the same structure can be used for any other information, such as isolation,
mentioned earlier. 

\section*{Commissioning and validating the 2015 data}

As we saw in Chapter~\ref{chpt:acunningplan}, the shift to real-time analysis placed specific additional burdens on the monitoring and validation of LHCb's data. In the most general terms,
the amount of validation which must be performed on the data, and the speed with which this validation must deliver a verdict, depends on the extent to which any problems are recoverable.
Most HEP experiments have an infrastructure for monitoring in real-time, but it is set up to spot genuinely irrecoverable problems, for example broken detector components or incorrect
datataking conditions, before these affect too much of the data. Beyond this, the assumption is that problems which are found by a centralised data quality 
can mostly be recovered after the fact by recalibrating, re-reconstructing, or generally reprocessing this data. Offline data quality checks therefore generally 
happen without any particular urgency, on the timescale of days or even weeks after the data has been taken.
Indeed these data quality checks serve to spot problems with the offline detector alignment, calibration, and reconstruction as much as to spot problems with the data itself.
They also serve to pre-warn analysts that certain data may have problems, however least
in LHCb, nothing prevents an analyst from using this data if they can show that there is no relevant bias for their measurement. From this perspective, after the online monitoring which 
spots problems with the detector itself, the really detailed data quality and validation in LHCb is performed by physics analyses as they go through the internal collaboration review.

Because such a procedure relies on our ability to go back and reprocess or recalibrate the data, it is inherently unsuited to validating real-time analysis. Of course we could all design
an ideal commissioning system for real-time analysis: each group of real-time analysts should provide a list of quantities to be measured using their data sample, the algorithms with which to perform the
measurements, and reference values which these measured quantities should be compared against. The data coming out of the real-time analysis system should then be fed into these
monitoring algorithms and warnings and alarms sounded when the measured quantities go too far from the reference values. However it was clear from the start that such a solution
was not possible in 2015, because the people required to implement it simply did not exist. For this reason, a large part of the solution for commissioning and validating the data involved
a temporary safety blanket: the raw data which the real-time analysis should have thrown away was archived until we gained more confidence in the system. The question nevertheless
remained how to validate the very first data taken in 2015 quickly enough in order to allow the output of real-time analysis (without any cheated offline reprocessing) to be used
to produce physics measurements.

\begin{figure}
\centering
\includegraphics[width=0.48\linewidth]{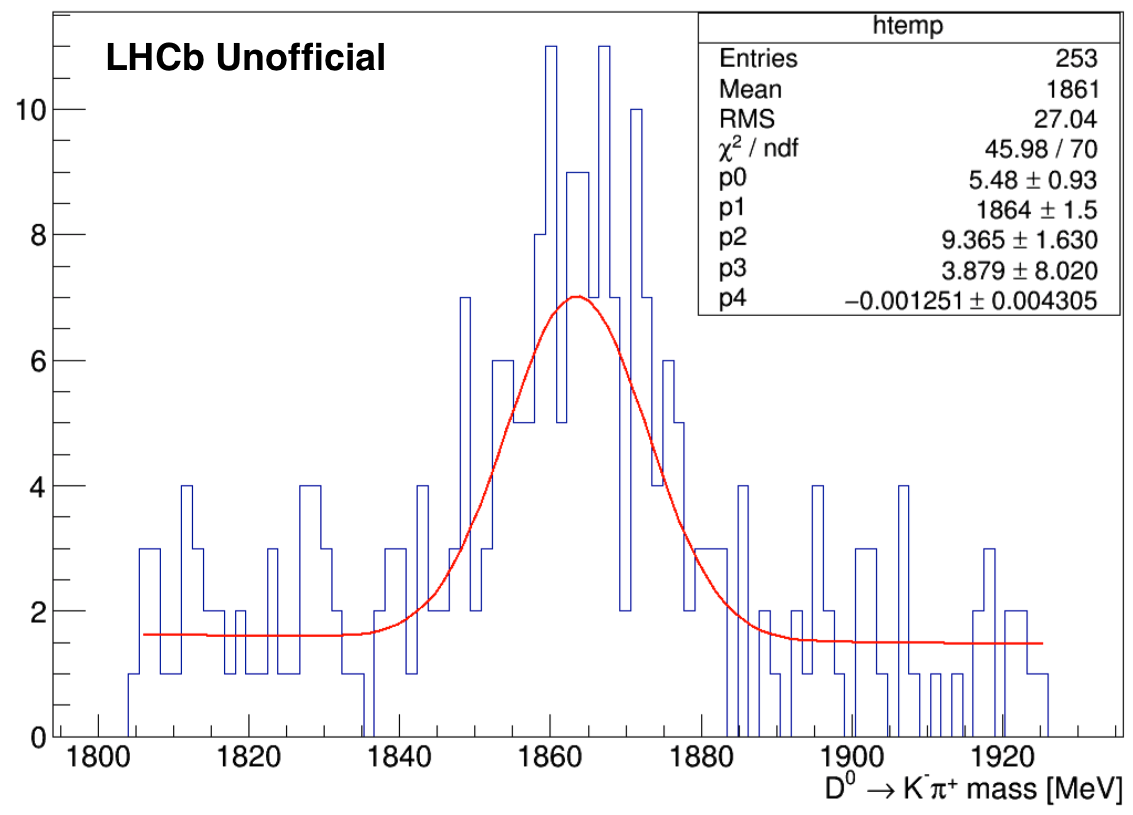}
\caption{Invariant mass of the $D^0\to K^-\pi^+$ signal taken on June 4$^{\textrm{th}}$ 2015, as recorded by the LHCb online monitoring. The signal shape is a single
Gaussian with mean p1 and width p2, the background is a linear function with slope p4.}
\label{fig:d2kpionlinemonit}
\end{figure}

\begin{figure}
\centering
\includegraphics[width=0.96\linewidth]{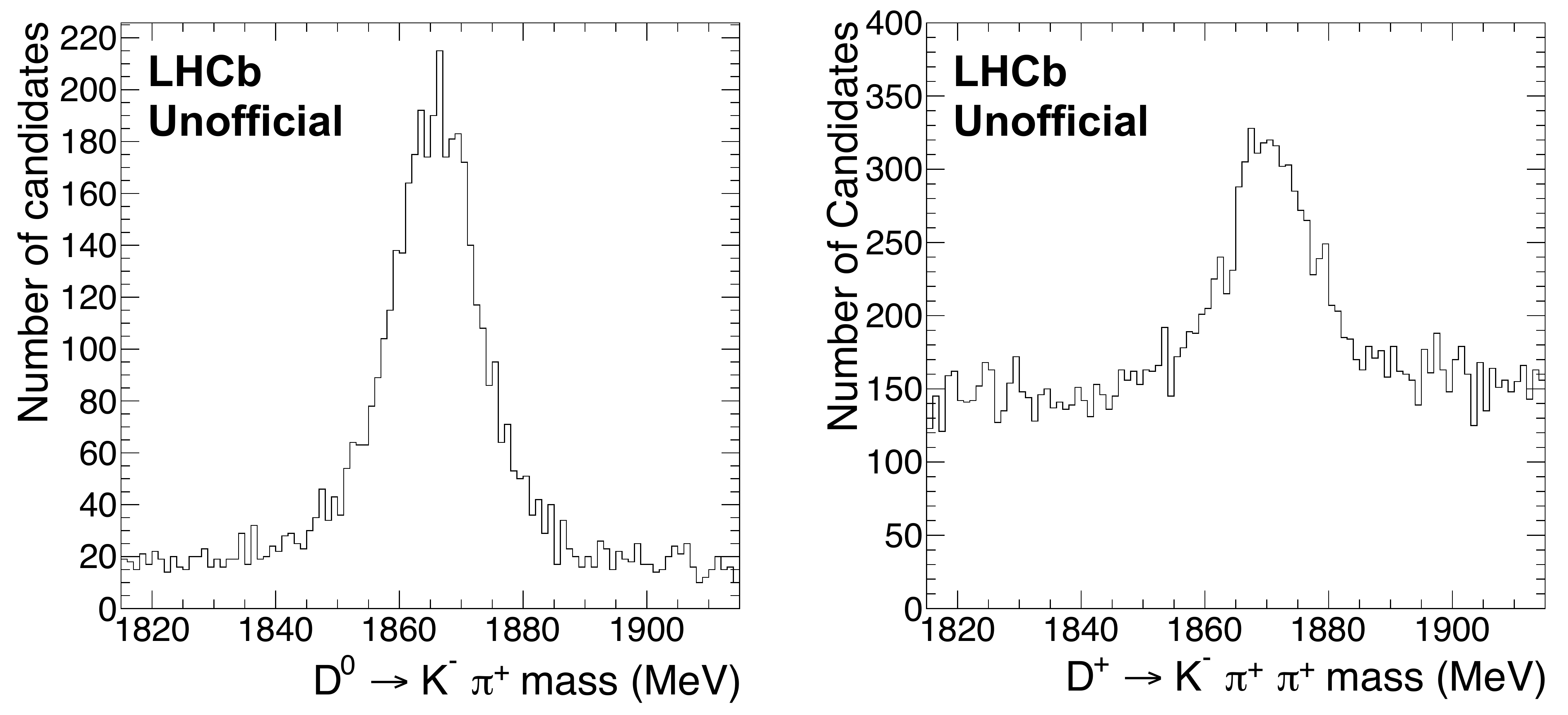}
\caption{Invariant mass of the (left) $D^0\to K^-\pi^+$ and (right) $D^+\to K^-\pi^+\pi^+$ signal taken on June 13$^{\textrm{th}}$ 2015, as processed manually for run 156923.}
\label{fig:d2kpihltmoni}
\end{figure}

\begin{figure}
\centering
\includegraphics[width=0.48\linewidth]{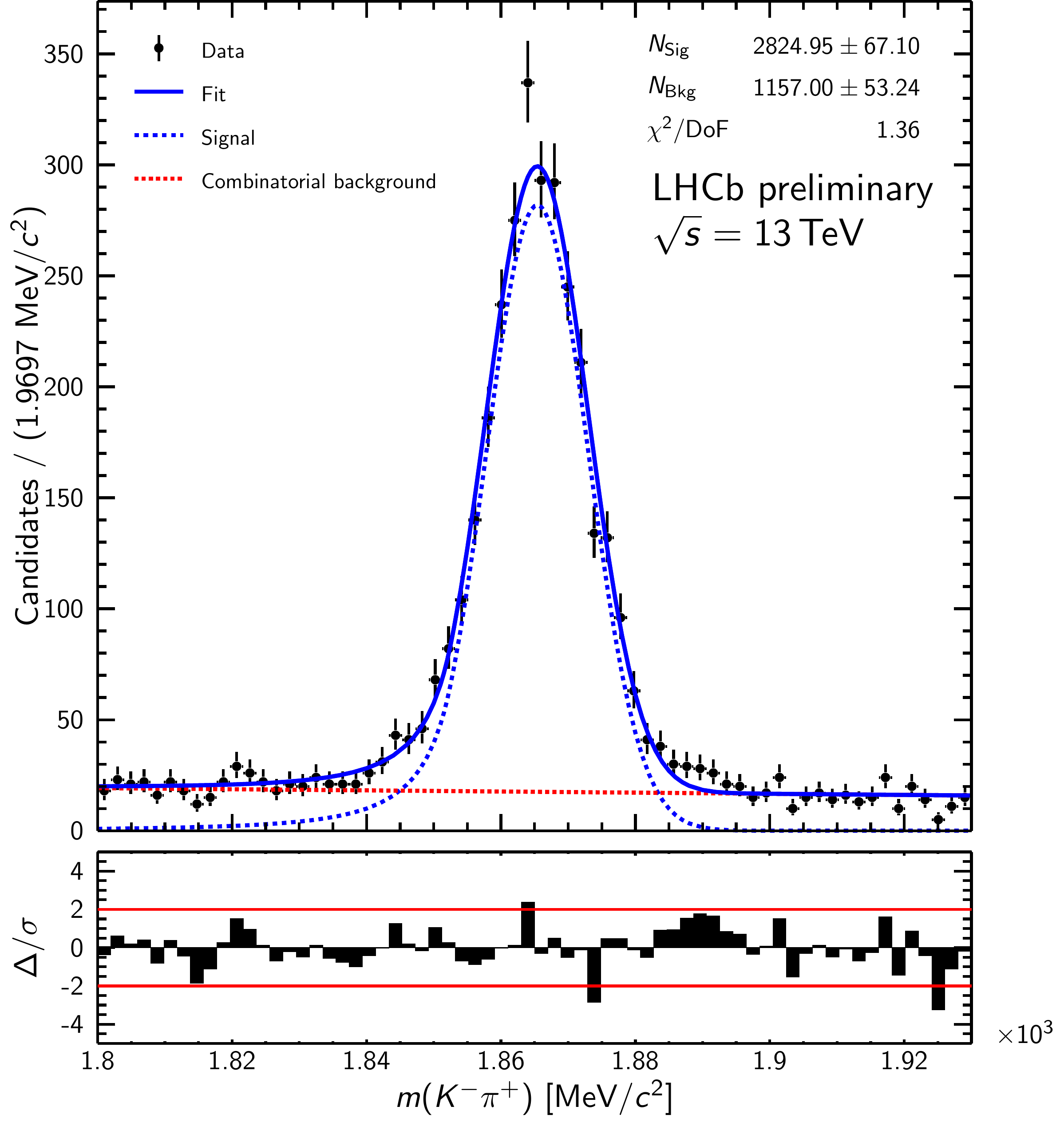}
\includegraphics[width=0.48\linewidth]{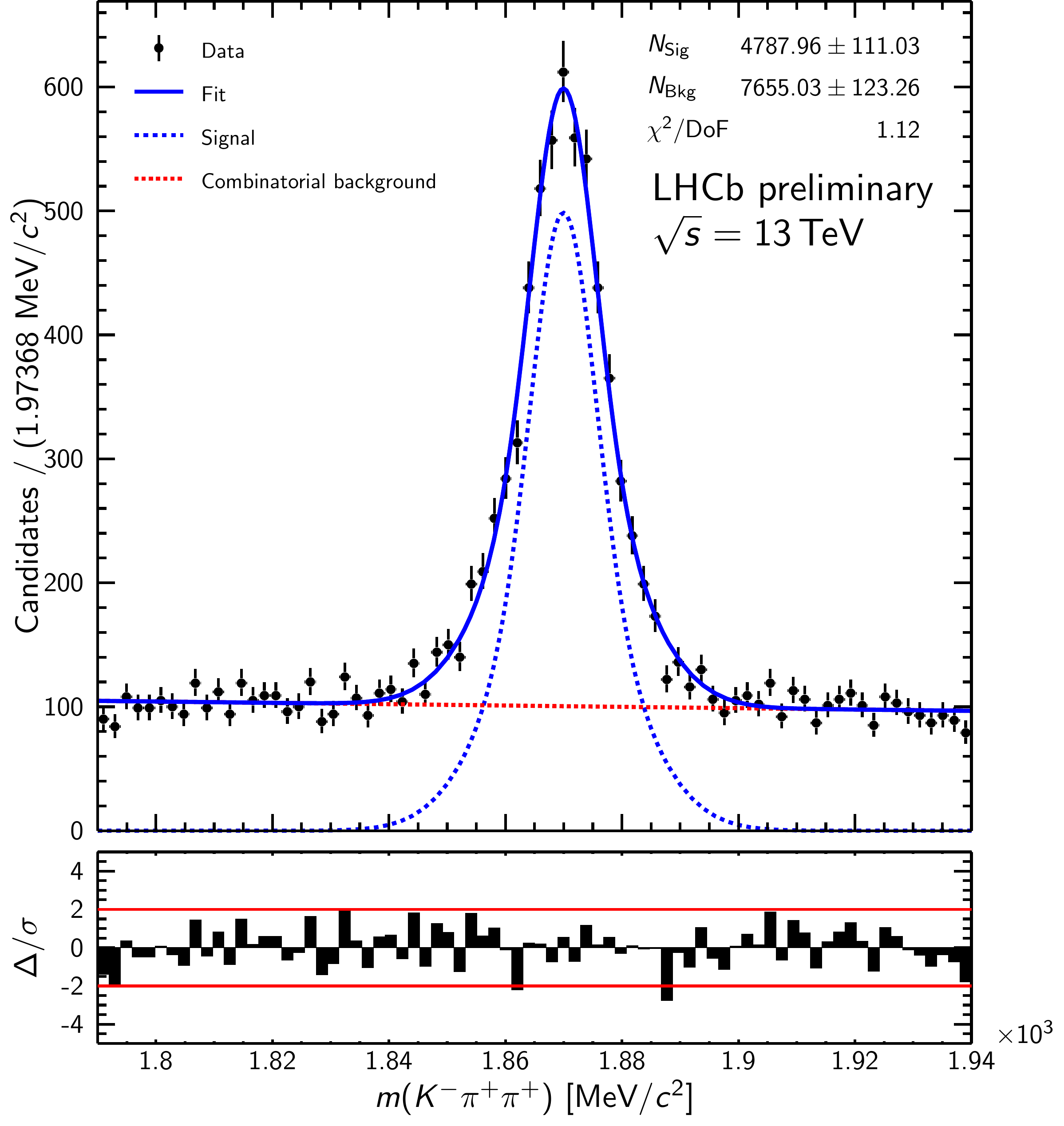}
\caption{Invariant mass of the (left) $D^0\to K^-\pi^+$ and (right) $D^+ \to K^- \pi^+\pi^+$ signals in early 2015 data.}
\label{fig:charmmass2015em}
\end{figure}

The solution was, with one exception, a hodge-podge of pragmatic checks which placed a disproportionate burden on a small number of technical experts and which we knew couldn't scale into a long-term solution.
Our objective was consciously to do whatever was necessary to give people confidence in the system and avoid a psychologically damaging failure which might have harmed the acceptance of real-time analysis
in the experiment. The exception was the detector alignment and calibration where, as already mentioned in Chapter~\ref{chpt:acunningplan}, a scaleable automatic monitoring was in fact
implemented and commissioned with the very first colliding bunches, thanks largely to a superhuman effort by a team of a few experts who all but lived in the control room
during this time. This work has been documented in conference proceedings~\cite{Dujany:2015lxd,Borghi:2017hfp} and I will not discuss it further here. On the HLT side, we had developed
a series of monitoring plots for signal peaks which were abundant enough to be reconstructed in the HLT already during Run~1 ($D^0\to K^-\pi^+$, $J/\psi\to\mu^+\mu^-$, $\phi\to K^+ K^-$).
These were improved a bit for the start of Run~2, and in particular an automated fit to their mass distributions which existed in prototype form in Run~1 was improved to
work reliably. In addition, the $D^0\to K^-\pi^+$ signal plot was connected to the output of the calibration stream which would feed the detector alignment, thus allowing us
to immediately spot any problems with this particularly critical data. 
Figure~\ref{fig:d2kpionlinemonit} shows a $D^0\to K^-\pi^+$ seen in one of the first fills of 2015, taken on June 4$^{\textrm{th}}$. I saved this plot not for its particularly
large signal peak, but because of the quality of the reconstruction: after a two year shutdown, it was a relief to instantly see a clear $D^0$ mass peak in the right place and with a resolution
within around $15\%$ of the Run~1 optimum. In addition to these checks, trigger experts performed other manual verification of the very first data while it was buffered on the trigger farm disks, 
running HLT2 by hand on subsets of the data, verifying that the expected trigger lines were in place and outputing candidates, and checking mass peaks with large statistics and a wider variety of signals
than possible in the online monitoring. Figure~\ref{fig:d2kpihltmoni} shows charm mass peaks from data taken on June~13 in run 156923 and produced manually by myself in this way.  
Once these checks cleared, the data were sent offline for further checks by the early measurement analysts, and Figure~\ref{fig:charmmass2015em} shows charm signals in the real-time analysis data produced by the
charm cross-section analysts and approved on June 16th, less than two weeks after the first 2015 stable beams in LHCb.

\begin{figure}
\centering
\includegraphics[width=1.0\linewidth]{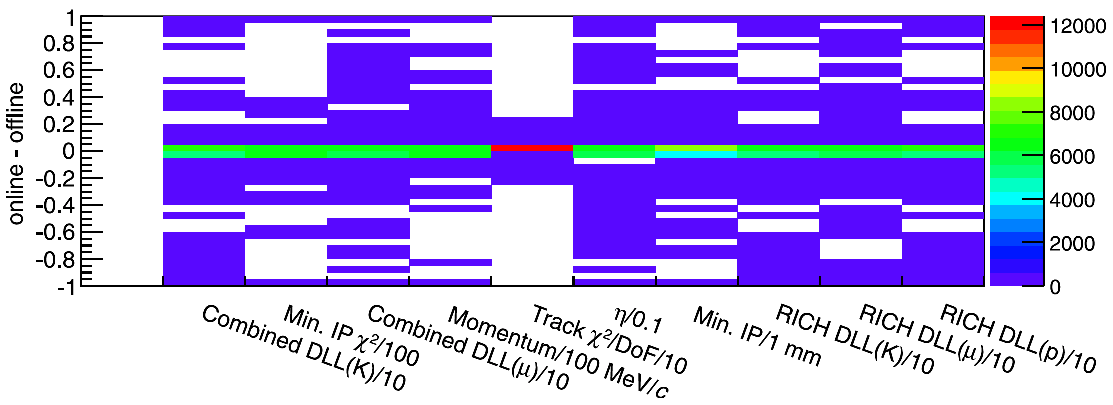}
\caption{Candidate-by-candidate comparison of real-time and offline analysis quantities for $D^0\to K^-\pi^+$ candidates. The y-axis unit for each quantity is given in the corresponding x-axis
bin label. Reproduced from~\cite{Aaij:2015qla}.}
\label{fig:lucio}
\end{figure}

These kinds of checks answered one crucial question related to real-time analysis: we were clearly capable of aligning and calibrating a detector well enough and quickly enough to
select clean signals in our data.\footnote{Of course, a lot of this has to do with the intrinsic quality and reliability of the LHCb hardware; however, I reserve the right to be proud
of the fact that unlike in 2011 or 2012, we started 2015 datataking by seeing signals instead of seeing memory leaks or a reconstruction which couldn't fit into the assigned budget.} What they
didn't answer was whether we could reliably compute the efficiency of our reconstruction and selection, the other crucial component in any physics measurement. They also didn't answer whether
or not the TURBO framework was reliably saving all information which the analyses might need, since it is possible to reconstruct and save the correct signal mass but, for example, get all the 
particle identification information wrong. Here the calibration streams, which selected the signals listed earlier in Table~\ref{table:tagandprobe}, played a crucial role because
they were designed from the start to save both the real-time analysis information \textbf{and} the traditional offline analysis information. This meant that they could be used to produce
not only like-for-like comparisons of the real-time and offline efficiencies, but that we could compare on a candidate-by-candidate level the information given to the analysts by the offline
and real-time analysis software. And example of such a comparison is shown in Figure~\ref{fig:lucio}. As we can see the agreement was not perfect, however this was not caused by any intrinsic difference in
reconstruction quality or even physics between the real-time and offline analysis frameworks, but rather was almost entirely due to different data compressions in the two cases, which in
turn led to floating point differences in the quantites seen by analysts.\footnote{There were, of course, specific problems found which were not caused by data compression. Some variables which should
have been saved in the real-time analysis were not there, others had a slightly different meaning than the one intended. But these problems, while not fully ironed out until 2017, 
were fairly minor and never led to any demonstrable bias in physics analyses.} There were, however, no macroscopic problems found with the real-time analysis, which allowed us to proceed with the
measurement of charm hadron cross-sections with the early 2015 data. 

\chapter{It is not all relative: measuring $\sigma_{c\bar{c}}$ in real time}
\label{chpt:charmxsec}
\pagestyle{myheadings}
\markboth{\bf It is not all relative}{\bf It is not all relative}

\begin{flushright}
\noindent {\it And I thank you for those items that you sent me \linebreak
The monkey and the plywood violin \linebreak
I practiced every night, now I'm ready \linebreak
First we take Manhattan, then we take Berlin \linebreak}
\linebreak
-- Leonard Cohen, First we take Manhattan 
\end{flushright}

Because of the associated risks, in particular the throwing away of raw detector hits
and reconstructed objects not associated to a signal candidate, real-time analysis had to be commissioned
as quickly as possible with Run~2 data. The most reliable way of commissioning a new analysis
framework is of course to publish some physics results with it, so that's what we set out to do with the
first data collected in 2015. The aim was not simply to perform an analysis quickly after taking data in 2015, but
to take the real-time analysis logic to its logical conclusion: the entire analysis was to be planned,
performed, and reviewed using simulation before any data were taken, the outcome of these studies used to design
an HLT1 trigger and HLT2 real-time selections, and the data once taken would simply be used to fill in the final
result tables and plots, and to assign the final values for any data-driven systematic uncertainties. A crucial part of
this was identifying a dedicated trigger expert who would be responsible for the implementation and validation of the early
measurement real-time selections, working in close contact with all the relevant analysts for a year prior to datataking. In order for
this to be a real commissioning, the analyses had to be publishable even with the very limited luminosity available
in the very first LHC fills of 2015; as the LHC had changed its centre-of-mass energy to 13~TeV, production measurements
naturally satisfied this criterion. Two sets of measurements were performed, one of $J/\psi$ production and one of
charm hadron production; I directly participated in the latter and will describe it here.

\section*{Physics motivation and measurement strategy}
A precise understanding of charm hadron production, and its dependence on the transverse momentum and rapidity of the charm
hadron, is important for a number of reasons. The primary motivation for measuring them at LHCb is that the detector's fiducial volume,
$2<\eta <4.5$ and $0 < p_{\textrm{T}} < 15$~GeV, allows us to probe values of the parton momentum fraction $x$ as low as $10^{-4}$, a region
in which gluon parton density functions were only known to around 30\%~\cite{Gauld:2015yia,Zenaiev:2015rfa}. Apart from being a fundamental test of QCD hadronization models, and specific
MC generator tunings for the LHC~\cite{Gauld:2015yia,Cacciari:2015fta,Kniehl:2012ti,Kniehl:2004fy,Kniehl:2005de,Kniehl:2005ej,Kneesch:2007ey,Kniehl:2009ar,Cacciari:1998it,Cacciari:2003zu,Cacciari:2005uk,Cacciari:2012ny,Zenaiev:2015rfa}, 
charm hadrons produced directly in $pp$ collisions are one of the dominant backgrounds in the search for $H\to c\bar{c}$
and rare $W$ and $Z$ boson decays~\cite{LHCb:2016yxg}. In addition, a measurement at a centre-of-mass energy of $13$~TeV and in LHCb's fiducial volume
allows LHCb to constrain the production of atmospheric high energy neutrinos from cosmic-ray induced charm hadron production, which form
an important background to IceCube and other similar detectors searching for anomalous high-energy neutrinos from astrophysical sources~\cite{Gauld:2015yia,Bhattacharya:2015jpa}. The 13~TeV
dataset allows cosmic rays up to 90~PeV to be probed, greatly extending the range of LHCb's previous 7 TeV measurement~\cite{Aaij:2013mga}. 

The production of $D^0$, $D^\pm$, $D^{*\pm}$ and $D^\pm_s$ mesons is measured within the above fiducial volume, in bins of charm meson transverse
momentum and rapidity which will be presented later. The mesons are reconstructed in the following decay modes: $D^0\to K^-\pi^+$,
$D^+\to K^-\pi^+\pi^+$, $D^+_s\to \phi(1020) (\to K^- K^+) \pi^+$, and $D^{*+}\to D^0 (\to K^-\pi^+) \pi^+$. No attempt is made
to distinguish whether these charm mesons are produced directly in the fragmentation process or in the decay of an excited charm resonance,
however charm mesons produced directly in the $pp$ collision are separated from charm mesons produced in the decays of \Pqb-mesons. The production
of $\Lambda_c$ baryons was left for a future measurement, because the yields were expected to be significantly smaller, the purity worse due to the
much shorter (compared to charm mesons) $\Lambda_c$ lifetime, and because proton efficiencies are generally harder to calibrate than kaon or pion ones
as touched on in Chapter~\ref{chpt:acunningplan}. Similarly, the production of $D^{*0}$ and $D^{*0}_s$ mesons, which would have required the reconstruction
of low momentum photons, was left for future measurements.\footnote{Observing the signal would not have been hard, but understanding the reconstruction of low momentum
photons in a calorimeter which had been annealing for two years prior to being reexposed to the LHC's radiation doses would have been an interesting problem.} 

\section*{Dataset and simulation}
The dataset used for this measurement was collected by LHCb between July 7th and 14th 2015 and corresponds to an integrated luminosity of $4.98\pm0.19$~pb$^{-1}$.
The data were collected with a hardware trigger which saved a randomly selected fraction of collisions. This choice was made to avoid having to understand
the efficiency of LHCb's calorimeter trigger, a particularly difficult problem for relatively low-energy hadrons because of the frequent overlap of particle showers in the hadronic
calorimeter. The precise fraction of selected bunch crossings varied with the LHC filling scheme which evolved to collide increasing numbers of bunches during the week of datataking.
Apart from this, the detector was aligned, calibrated, and reconstructed in real-time as described in the previous chapters. A selection of interesting
bunch crossings based on a partial reconstruction of the signal candidates was performed in HLT1, while the entire signal candidates were built in HLT2 and propagated
to the analysts using the described TURBO machinery. 

Simulated detector samples were generated in the usual LHCb fashion, with a specific PYTHIA tune~\cite{LHCb-PROC-2010-056} and using EvtGen~\cite{Lange:2001uf} to decay
the hadronic particles with final-state radiation described by PHOTOS~\cite{Golonka:2005pn}. The simulated samples were provided in advance of datataking
and used to optimize the entire analysis chain presented below, and the analysis methodology was reviewed and signed off based on these simulation prior
to any data being taken, with the data used to fill in the results and validate that everything behaved as expected.
\section*{HLT1 selection}
The HLT1 selection was used to reduce the input rate from the output of the random hardware trigger to the roughly $100$~\kHz which could be safely buffered
to disk in the HLT computing cluster. Although the rate of the random hardware trigger varied, the HLT1 selection was tuned to give the correct rejection for the highest
anticipated random trigger rate of around $660$~kHz; no attempt was made to run with a looser configuration in the earlier LHC filling schemes which had fewer colliding bunches. Indeed HLT1 was 
generally tuned for simplicity rather than absolute performance, since the measurement was expected to be systematics limited. 

\begin{table}
  \caption[HLT1 optimization]{%
    Efficiencies and rate the different considered HLT1 triggers on the relevant signal signal samples. The rate
    was evaluated with respect to 13~TeV minimum bias simulated events, while the efficiency was evaluated with respect
    to a preliminary version of the full selection for each signal. In the top row ``R'' refers to the reconstruction
    and preselection of charged particles, ``Rect.'' to a rectangular or ``cut-based'' selection, and ``BDT'' to the
    optimized multivariate selection. The 1~Track~Rect. baseline selection is fully described in Table~\ref{table:1track}.
  }
  \label{table:hlt1optim}
  \begin{center}
  \begin{tabular}{lcccc}
    \toprule
    & R & 1~Track~Rect. & 1\& 2~T~Rect. & 1\& 2~Track~BDT \\
    \midrule
    Signal (down) / Rate (right)     & 666~kHz     & 99~kHz & 98~kHz & 98~kHz  \\
    \midrule
    $D^0\to K^-\pi^+$ & 97.7\% & 89.5\% & 91.8\% & 93.0\% \\ 
    $D^0\to K^-\pi^+\pi^-\pi^+$ & 97.9\% & 91.3\% & 95.2\% & 96.0\% \\
    $D^+\to K^-\pi^+\pi^+$ & 98.2\% & 85.1\% & 87.5\% & 91.9\% \\
    $D^+_s\to \phi(1020) \pi^+$ & 96.3\% & 81.8\% & 87.6\% & 89.9\% \\
    \bottomrule
  \end{tabular}
  \end{center}
\end{table}

Four different versions of the HLT1 trigger were considered and optimized. Two were based on a single high-transverse-momentum track displaced from the primary vertex, and two
were based on a pair of high-transverse-momentum tracks forming a displaced secondary vertex. For each of these 1-track and 2-track schemes, a simple selection based on
rectangular cuts and a more sophisticated boosted decision tree selection were considered. The performance of these options was evaluated on different simulated signal
samples and the results of this comparison are shown in Table~\ref{table:hlt1optim}. As can be seen, the BDT-based 2-track HLT1 trigger gave the best performance, however the gains were rather marginal considering
the fact that it was the most complicated of the four options. For this reason, the decision was made to use the rectangular cut based 1-track trigger, whose selection criteria
are listed in Table~\ref{table:1track}. 

\begin{table}
  \caption[HLT1 \texttt{1Track} selection]{%
    Requirements made on the track that fires the HLT1 \texttt{1Track} 
    trigger line. The impact parameter (IP) $\chi^2$ is a measure of a charged
    particle's displacement from the primary vertex, calculated
    as the difference in the $\chi^2$ of the primary vertex fit when the charged particle
    is or is not included in it.
  }
  \label{table:1track}
  \begin{center}
  \begin{tabular}{lcc}
    \toprule
    Particle                   & Variable     & Cut value          \\
    \midrule
    \multirow{5}{*}{Any track} & p$_{\textrm{T}}$          & $> 0.8$~GeV \\
                               & p        & $> 3$~GeV   \\
                               & Track $\chi^2/\textrm{ndof}$ & $< 3$              \\
                               & IP $\chi^2$     & $> 10$             \\
                               & VELO hits   & $> 9$              \\
    \bottomrule
  \end{tabular}
  \end{center}
\end{table}

\begin{figure}
\centering
\includegraphics[width=0.7\linewidth]{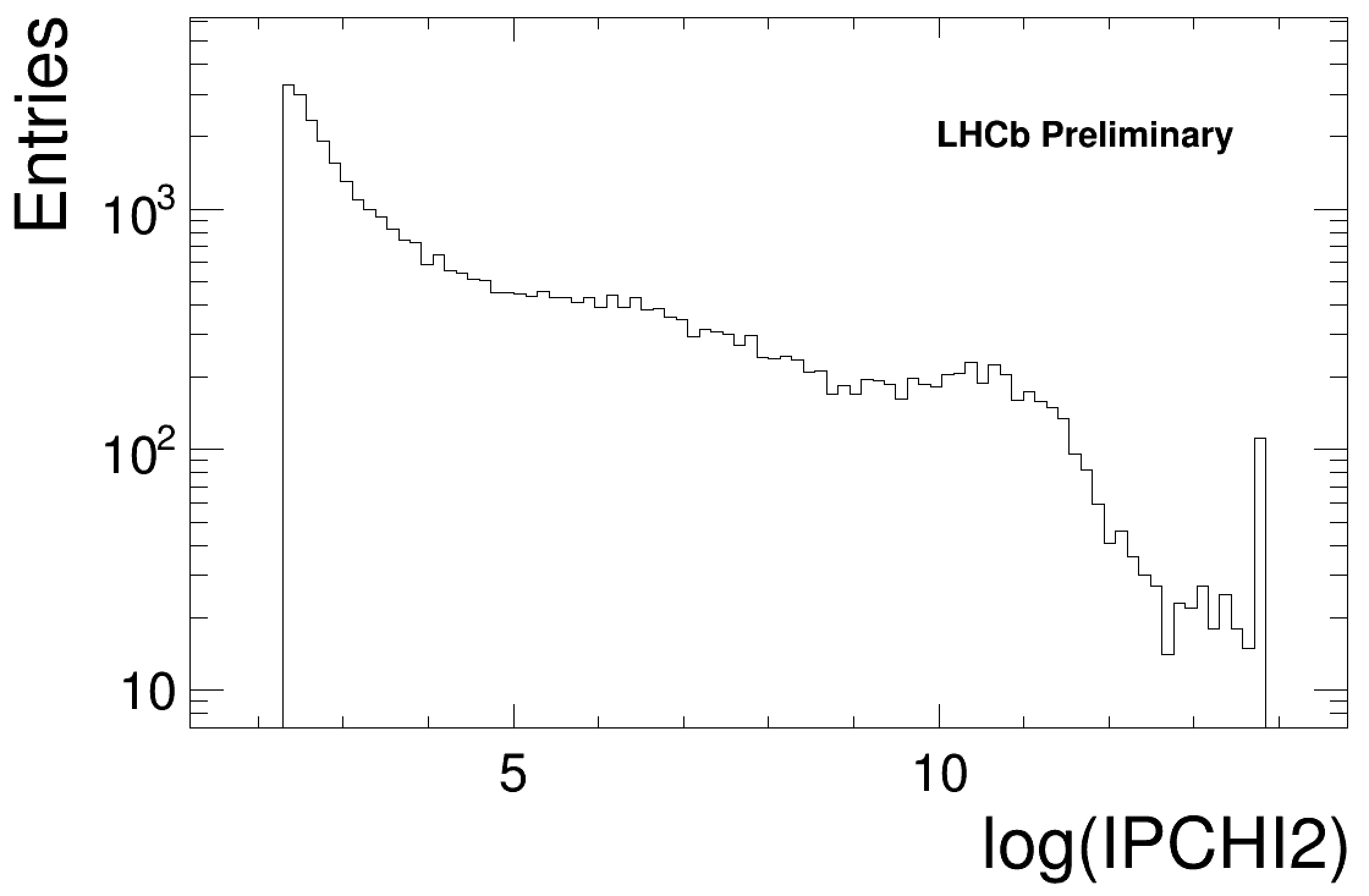}
\caption{The logarithm of the impact parameter $\chi^2$ of the tracks passing all other baseline 1 Track trigger requirements.}
\label{fig:higeipchi2tracks}
\end{figure}

A particular background for
this trigger, exacerbated by the relatively loose requirement on transverse momentum, were found to be decays of $K^0_S$ mesons, which left tracks with extreme displacement
as seen in Figure~\ref{fig:higeipchi2tracks}. These could have been removed either by imposing a requirement on the maximum particle displacement from the PV (as opposed to the usual requirement
on the minimal displacement) or by requiring a minimal number of hits in the vertex detector, since most $K^0_S$ mesons in question decayed near the final few VELO stations.
I decided to go with the minimal hits requirement, in part due to a worry about understanding the decay-time acceptance of a requirement on maximal displacement, and in part
because this requirement had proved to be useful at rejecting fake tracks in Run~1. As we will see in the next chapter, that perhaps wasn't the most inspired choice. 

\section*{HLT2 selection}
Unlike HLT1, which had access only to information about primary vertices and a subset of high-transverse-momentum tracks, HLT2 executed the full detector reconstruction
and had access to all information which would have been available in any traditional analysis performed during Run~1. The objective of HLT2 was to select signal candidates
with a reasonable purity, using a minimum of selection criteria which were known to have a high signal efficiency from experience with similar Run~1 analyses. The requirement
of reasonable purity was not simply about writing a small number of signal candidates to disk, but also about not writing too many candidates for any given bunch crossing, since
in a real-time analysis the size of the output data are no longer proportional to the number of bunch crossings we save but rather to the number of signal candidates.

This task was made particularly simple by the availability of full-quality particle identification information in HLT2. First of all,
we could apply less stringent requirements on transverse momenta and displacement from the primary vertex, preserving efficiency. And secondly, we had to build a much smaller
number of candidates, since each track would be considered only once, either as a pion or as a kaon, according to the particle identification information. This makes a big
difference for three and four-body final states in particular, and again allows other criteria to be made looser. Because of this and because of experience with such selections
gained in the analysis of Run~1 data, there was no particular need to optimize the HLT2 selections, which are listed in Tables~\ref{table:sel:trigger:dztohh}~to~\ref{table:sel:trigger:commondstp}. 
The purity achieved by these selections was already seen in Figure~\ref{fig:charmmass2015em} in the previous chapter.

\begin{table}
  \caption[\DzToKpi \hlttwo selection]{%
    Selection requirements made in the \DzToKpi \hlttwo trigger line.
    The track \chisq criterion is applied in the reconstruction and listed here 
    for completeness.
    The DIRA requirement, $\cos\theta > 0.99985$, is approximately equivalent 
    to a cut of $\theta < 17.32$~$\milli\radian$.
  }
  \label{table:sel:trigger:dztohh}
  \begin{center}
  \begin{tabular}{lcc}
    \toprule
    Particle                     & Variable                                    & Cut value                                                  \\
    \midrule
    \multirow{4}{*}{\pipm, \Kpm} & \pT                                         & $> 250$~$\MeV$                                           \\
                                 & \ptot                                       & $\ptot > 2$~$\GeV$                                               \\
                                 & Track \chisq                                & $< 3$                                                             \\
                                 & \ipchisq                                    & $> 16$                                                    \\
    \midrule
    \pipm                        & \dllkpi                                     & $< 5$                                                      \\
    \midrule
    \Kpm                         & \dllkpi                                     & $> 5$                                                          \\
    \midrule
    \multirow{5}{*}{\Dz}         & $m(\Km\pip)$                                & $1784 \MeV < m < 1944 \MeV$                    \\
                                 & \Kpm to \pipm DOCA                          & $< 0.1$~mm                                               \\
                                 & Vertex fit \chisq                           & $< 10$                                                       \\
                                 & Direction angle                             & $> 0.99985$                                                       \\
                                 & Vertex displacement                         & $VD\chisq > 49$                                              \\
    \bottomrule
  \end{tabular}
  \end{center}
\end{table}

\begin{table}
  \caption[$\decay{\Dp}{h^{-}h^{+}h^{+}}$ \hlttwo selection]{%
    Selection requirements made in the three-body \hlttwo lines.
    The lines are split into \Dp (\DpToKpipi, \DpToKKpi) and \Dsp (\DsToKKpi) 
    decays according to the mass window.
    Cuts of the form $x > x_{1},\, x_{2},\, x_{3}$ require that all particles 
    satisfy $x > x_{1}$, at least two satisfy $x > x_{2}$, and at least one 
    satisfies $x > x_{3}$.
    The DIRA requirement, $\cos\theta > 0.9994$, is approximately equivalent to 
    a cut of $\theta < 34.64$~$\milli\radian$.
  }
  \label{table:sel:trigger:dptohhh}
  \begin{center}
  \begin{tabular}{lcc}
    \toprule
    Particle                     & Variable             & Cut value                                                   \\
    \midrule
    \multirow{4}{*}{\pipm, \Kpm} & \pT                  & $> 200,\, 400,\, 1000$~$\MeV$                            \\
                                 & \ptot                & $\ptot > 2$~$\GeV$                                                 \\
                                 & Track \chisq         & $< 3$                                                                    \\
                                 & \ipchisq             & $> 4,\, 10,\, 50$                                             \\
    \midrule
    \pipm                        & \dllkpi              & $< 5$                                                   \\
    \midrule
    \Kpm                         & \dllkpi              & $> 5$                                               \\
    \midrule
    \Dp                          & $m(h^{-}h^{+}h^{+})$ & $1789 \MeV < m < 1949 \MeV$                 \\
    \midrule
    \Dsp                         & $m(h^{-}h^{+}h^{+})$ & $1889 \MeV < m < 2049 \MeV$               \\                                
    \midrule
    \multirow{3}{*}{\Dp/\Dsp}    & Vertex fit \chisq    & $< 25$                                                       \\
                                 & Direction angle      & $> 0.9994$                                                         \\
                                 & Vertex displacement  & $VD\chisq > 16$ \text{AND} $\tau > 0.150$~$\pico\second$\\
    \bottomrule
  \end{tabular}
  \end{center}
\end{table}

\begin{table}
  \caption[Common \Dstp \hlttwo selection]{%
    Common selection requirements made on \hlttwo \Dstp candidates.
  }
  \label{table:sel:trigger:commondstp}
  \begin{center}
  \begin{tabular}{lcc}
    \toprule
    Particle                     & Variable                                                    & Cut value                                                  \\
    \midrule
    \multirow{2}{*}{\pipm}       & \pT                                                         & $> 100$~$\MeV$                                               \\
                                 & Track \chisq                                                & $< 3$                                                                  \\
    \midrule
    \multirow{2}{*}{\Dstp}       & $m(\Dstp)-m(\Dz)$                                           & $130 \MeV < m < 160 \MeV$                    \\
                                 & Vertex fit \chisq                                           & $< 25$                                                  \\
    \bottomrule
  \end{tabular}
  \end{center}
\end{table}

\section*{Offline selection}
The selections applied in HLT2 were almost optimal for the analysis as a whole, however for the $D^+$ and $D^+_s$ candidates it was found that slightly tightening the kinematic and displacement
criteria offline gave a noticable improvement in purity and analysis sensitivity, particularly in the corners of the detector acceptance where the signal purity was worst. There
was no special tuning of these criteria performed, they were tightened by eye to some reasonable round numbers looking at the data. The $D^0$ and $D^{*\pm}$ selection criteria
applied in HLT2 were found to be already good enough for the analysis and were not modified.
\section*{Efficiencies and data-simulation corrections}
The selection efficiency factorizes into several components: the efficiency for the decay to occur in the LHCb geometric acceptance, the efficiency for the
tracks to be reconstructed, and the efficiency for the decay to be selected. These efficiencies are, with two exceptions, evaluated independently using fully simulated signal samples,
and multiplied in order to obtain the overall signal efficiency in each bin of transverse momentum and rapidity. 

The efficiency of the particle identification criteria is particularly sensitive to the specific dattaking conditions because of the gaseous RICH detectors, and
is obtained not with simulation but using dedicated tag-and-probe calibration samples given in Table~\ref{table:tagandprobe}. The kinematic and geometric distributions of the signal and calibration samples
are not perfectly aligned, and must therefore be corrected to avoid introducing a bias into the measurement. The correction is applied in the particle transverse momentum,
pseudorapidity, and the detector occupancy, with the distributions of these quantites in the  signal and calibration samples obtained using the sPlot~\cite{Pivk:2004ty} background subtraction
technique. Notice that the $D^{*+}\to D^0 (\to K^-\pi^+) \pi^+$ signal sample has some overlap with the tag-and-probe  $D^{*+}\to D^0 (\to K^-\pi^+) \pi^+$ sample, although
not a complete overlap as the HLT2 selection criteria were different in the two cases. Nevertheless this sample can be said to ``self calibrate'', and this is properly taken into account in
the analysis. 

\begin{figure}
\centering
\includegraphics[width=0.96\linewidth]{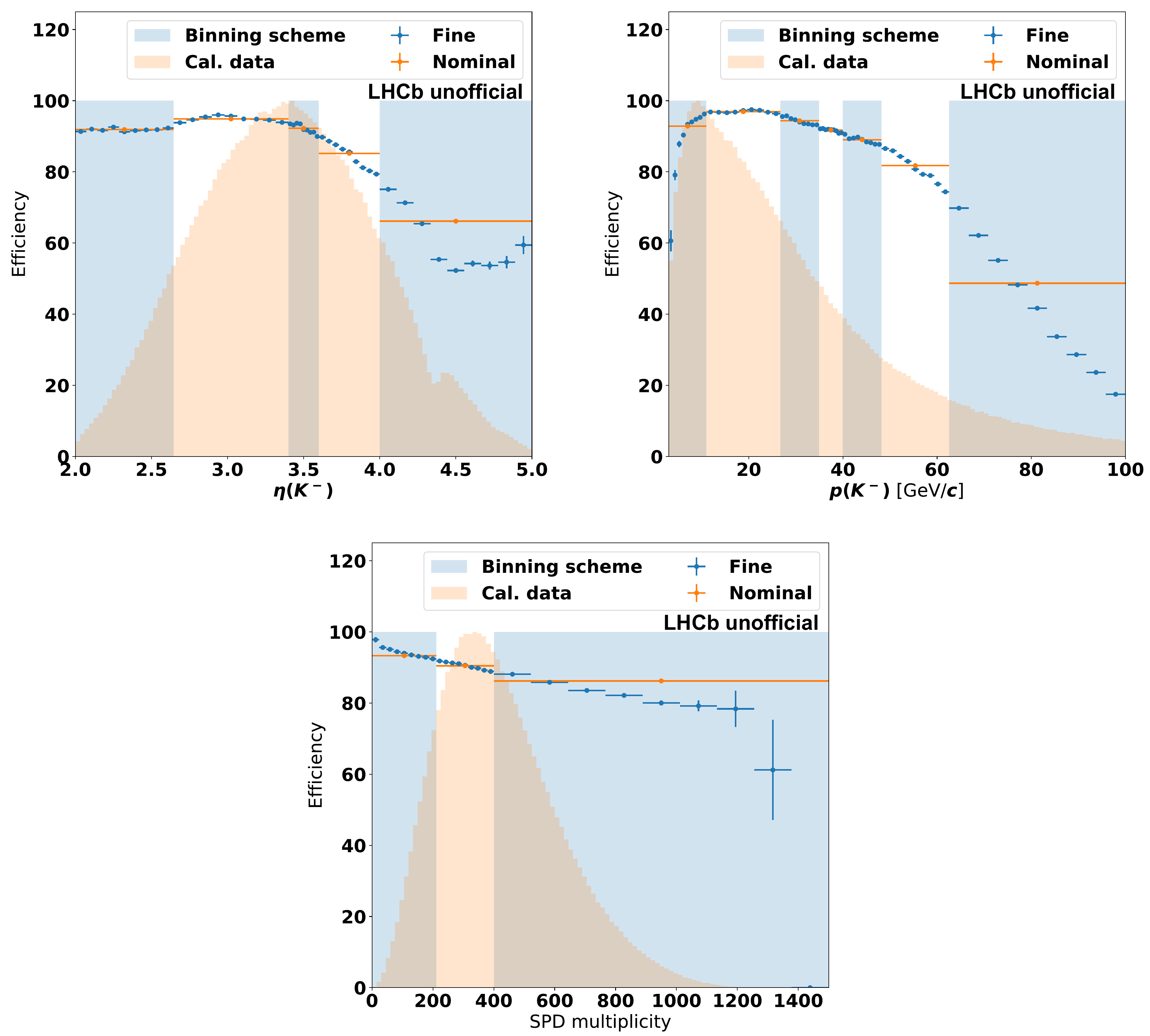}
\caption{The particle identification efficiency for kaons as a function of (top left) pseudorapidity, (top right) momentum, and (bottom) event occupancy. The distribution of the calibration
sample in each quantity is shown as a shaded orange area, the blue points give the efficiency in a fine binning which would be optimal if reweighting in only that single dimension, while
the orange points show the optimized three-dimensional binning used in the analysis.} 
\label{fig:pidbinsyst}
\end{figure}

The correction binning size was optimized using toy studies to minimise associated biases. In particular, because the
kinematic distributions of the signal and calibration samples differ, a binning which is too coarse will mean that the bin-average of the calibration sample does not
accurately reflect the bin-average of the signal sample. On the other hand, a binning which is too fine will introduce large statistical uncertainties and may result in empty
bins in sparsely populated areas, which are also undesirable.  
Examples of the optimized calibration and signal sample distributions are given in Figure~\ref{fig:pidbinsyst}. As can be seen, the choice of a three-dimensional binning scheme imposes
bin widths which are larger than would be optimal in any given dimension, but the systematic uncertainty which would have been associated with performing a one-dimensional correction
would have been larger than the systematic caused by this coarse binning.

\begin{figure}
\centering
\includegraphics[width=1.0\linewidth]{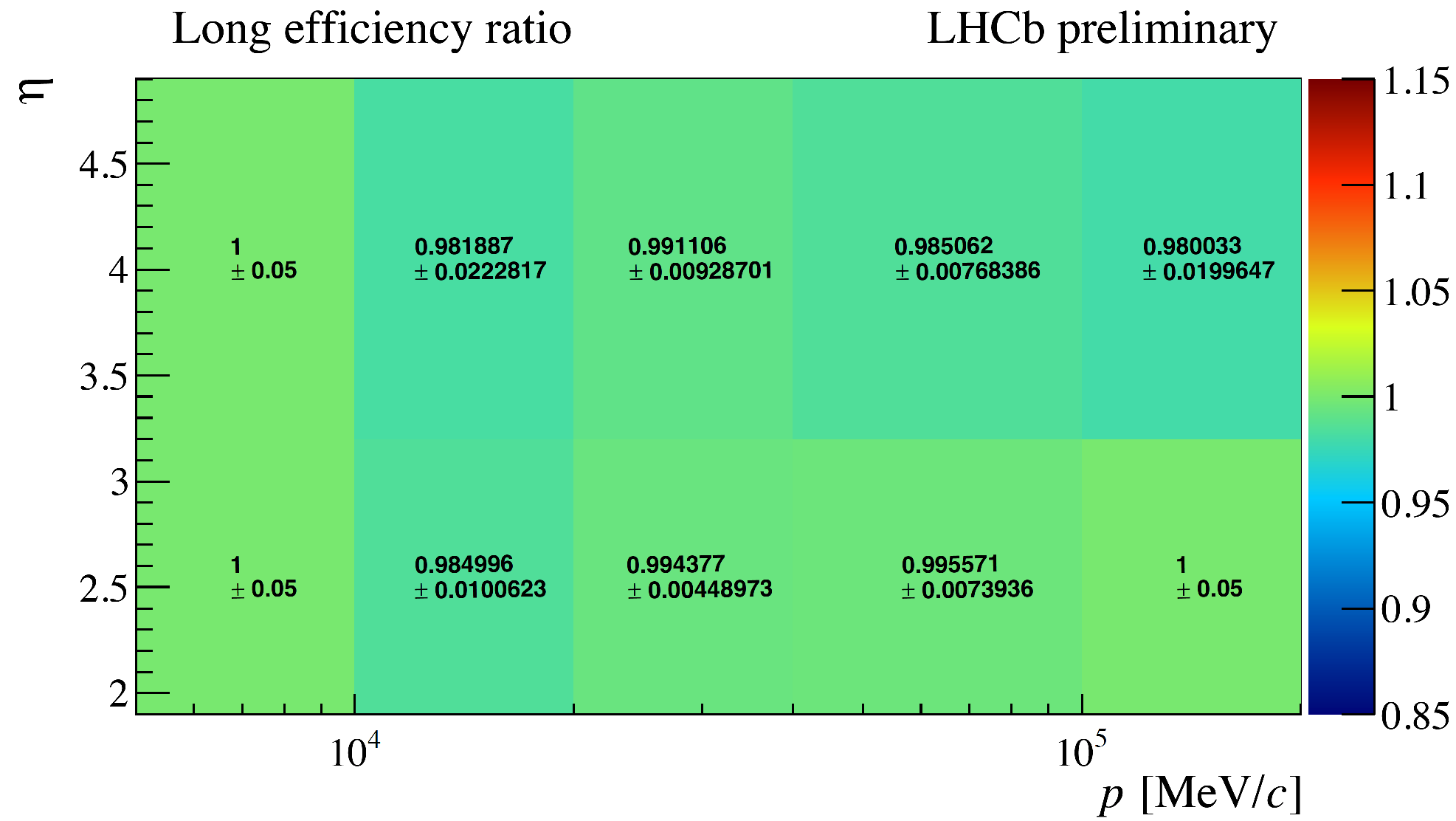}
\caption{The per-track efficiency correction factor, in percent, as a function of transverse momentum and pseudorapidity. The quoted uncertainties are a combination of
statistical uncertainties due to the limited calibration sample sizes and systematic uncertainties due to the finite accurary of the
calibration method when compared to the truth-level efficiency on simulated samples. The coarse binning is an indication of the limited
size of the tag-and-probe calibration samples available. Reproduced from~\cite{TrackEffPage}.} 
\label{fig:trefffinal}
\end{figure}

The track reconstruction efficiency is obtained from simulation, but corrected for data-simulation differences observed in the relevant tag-and-probe calibration samples. The correction
table is divided in bins of transverse momentum and pseudorapidity, and shown in Figure~\ref{fig:trefffinal}. This
approach is favoured as systematic uncertainties in the tag-and-probe measurement of reconstruction efficiencies cancel in the data-simulation ratio, in principle allowing the simulation
to be corrected in an unbiased way. In practice, as we will see in the next chapter, this correction introduced a bias into the original measurements which had to be corrected by
later errata. 
 
\section*{Measurement of the signal yields}
In order to obtain the cross-section in a given bin of transverse momentum and rapidity, it is necessary to measure the yield of charm mesons produced directly in
the $pp$ collision in that bin. There are three sources of background which have to be considered: random combinations of charged particles; partially reconstructed
decays of other long-lived particles, for example reconstructing the $K^-\pi^+$ pair from the $B^- \to K^- \pi^+ \pi^-$ decay as $D^0\to K^-\pi^+$; and genuine charm mesons produced
not directly in the $pp$ collision but in the decay of \Pqb-hadrons. The first two sources can be separated from the signal by looking at the invariant mass of the charm meson candidate.
In practical terms, both of these backgrounds have a smooth distribution in the candidate mass, and can be modelled by a single component, henceforth referred to as the ``combinatorial background''. 
The third background however, referred to as ``secondary'', has by definition the same mass shape as the signal, and therefore has to be measured in another way. 

\begin{figure}
\centering
\includegraphics[width=1.0\linewidth]{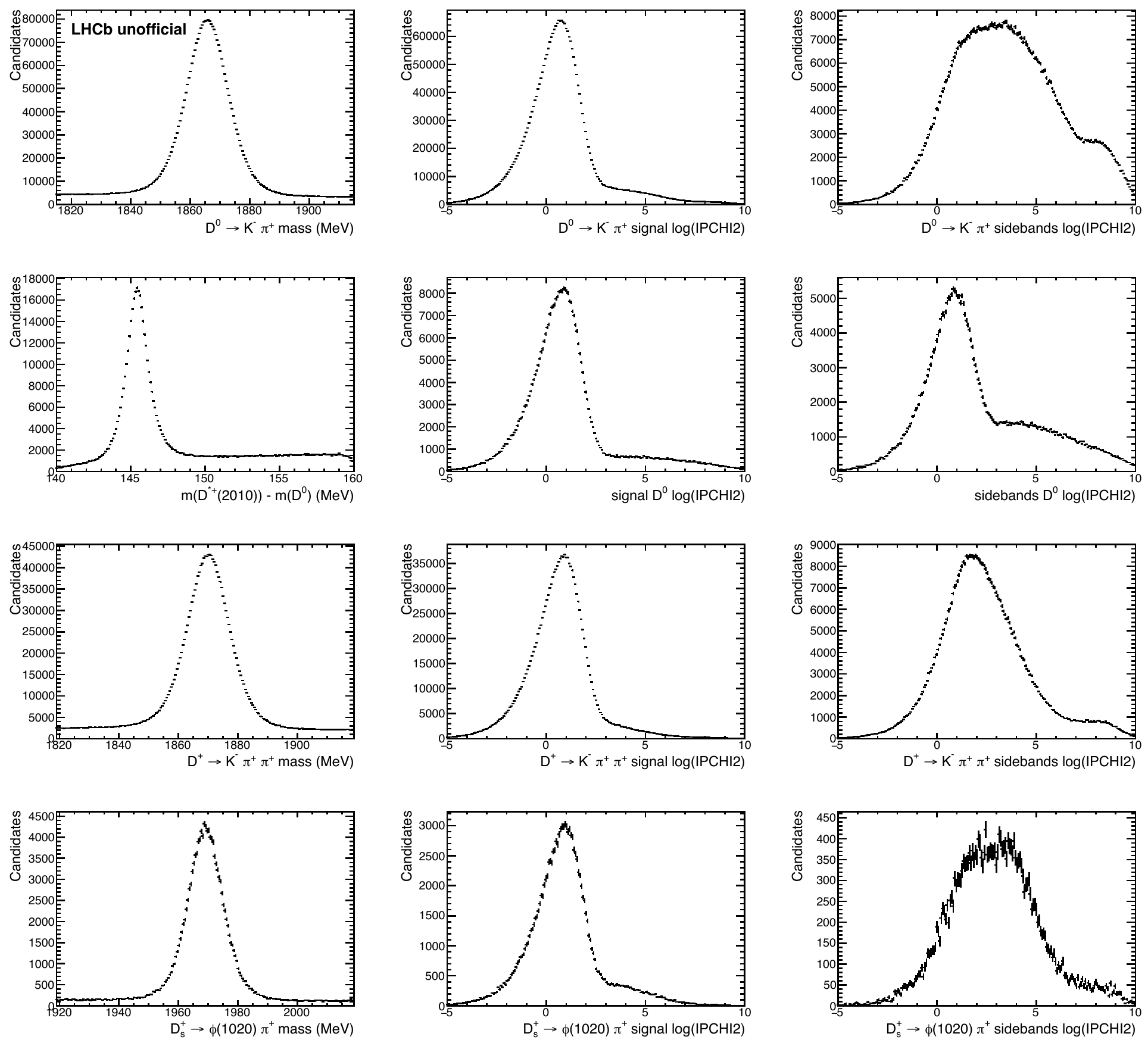}
\caption{The (left) signal mass, (middle) impact parameter $\chi^2$ in the signal region and (right) impact parameter $\chi^2$ in the sideband region for the different
signals analysed. From top to bottom:  $D^0\to K^-\pi^+$, $D^{*+}\to D^0 (\to K^-\pi^+) \pi^+$, $D^+\to K^-\pi^+\pi^+$, and $D^+_s\to \phi(1020) (\to K^- K^+) \pi^+$. 
All plots correspond to the full datasample taken in 2015 for these measurements. Note that
in the case of $D^{*+}\to D^0 (\to K^-\pi^+) \pi^+$, the sideband region is computed in the $D^{*+} - D^0$ mass difference and therefore still contains a significant
number of true $D^0$ mesons originating in the primary proton-proton interaction, so the impact parameter $\chi^2$ looks rather signal like even in the sideband region.  }
\label{fig:signalsnofit}
\end{figure}

Charm mesons which originate in the $pp$ collision can be identified by 
a coincidence between their measured momentum and displacement vectors, where the displacement vector connects the $pp$ collision vertex and the charm meson decay vertex. 
By contrast, charm mesons produced in the decay of a \Pqb-hadron will not have such a coincidence. Therefore we might think to use the angle between the momentum and displacement
vectors as a discriminant, however it is also important to take into account the uncertainty on this angle, particularly in the case of $pp$ collision vertices which produce few
charged particles and are therefore poorly measured. In practice, experience with previous analyses shows that the best discriminating
variable is the logarithm of the impact parameter $\chi^2$ of the charm meson, defined as the change in the $\chi^2$ of the fit to the $pp$ collision vertex when the charm meson
is and isn't included in this fit. This quantity properly takes into account all available information on both the positions of the vertices and their uncertainties.
Charm mesons which come directly from the $pp$ collision vertex have small values of this quantity, whereas charm mesons which come from \Pqb-hadron decays have large values; the
logarithm serves to transform the discriminating variable into a form which is simpler to model.
An example distribution of the mass and impact parameter $\chi^2$ is shown in Figure~\ref{fig:signalsnofit} for $D^0\to K^-\pi^+$ candidates, where the impact parameter $\chi^2$ plot has been split into the
signal and sideband regions in order to better illustrate the shape of the combinatorial background in the impact parameter $\chi^2$. 

\begin{figure}
\centering
\includegraphics[width=0.7\linewidth]{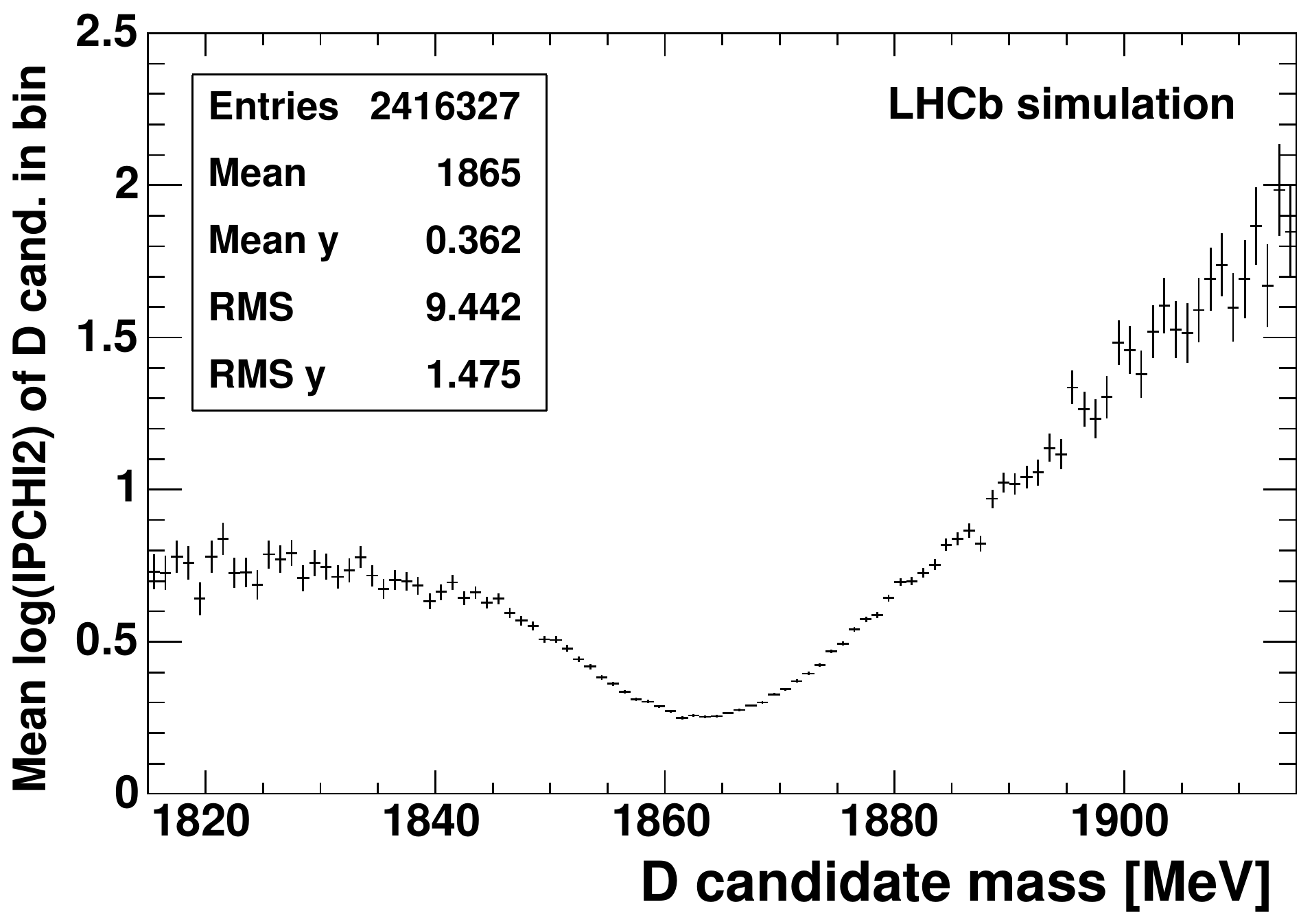}
\caption{The average logarithm of the impact parameter $\chi^2$ of simulated signal $D^0\to K^-\pi^+$ candidates in a given bin of reconstruction $D^0$ mass.}
\label{fig:sigmassipchi2corr}
\end{figure}

\begin{figure}
\centering
\includegraphics[width=1.0\linewidth]{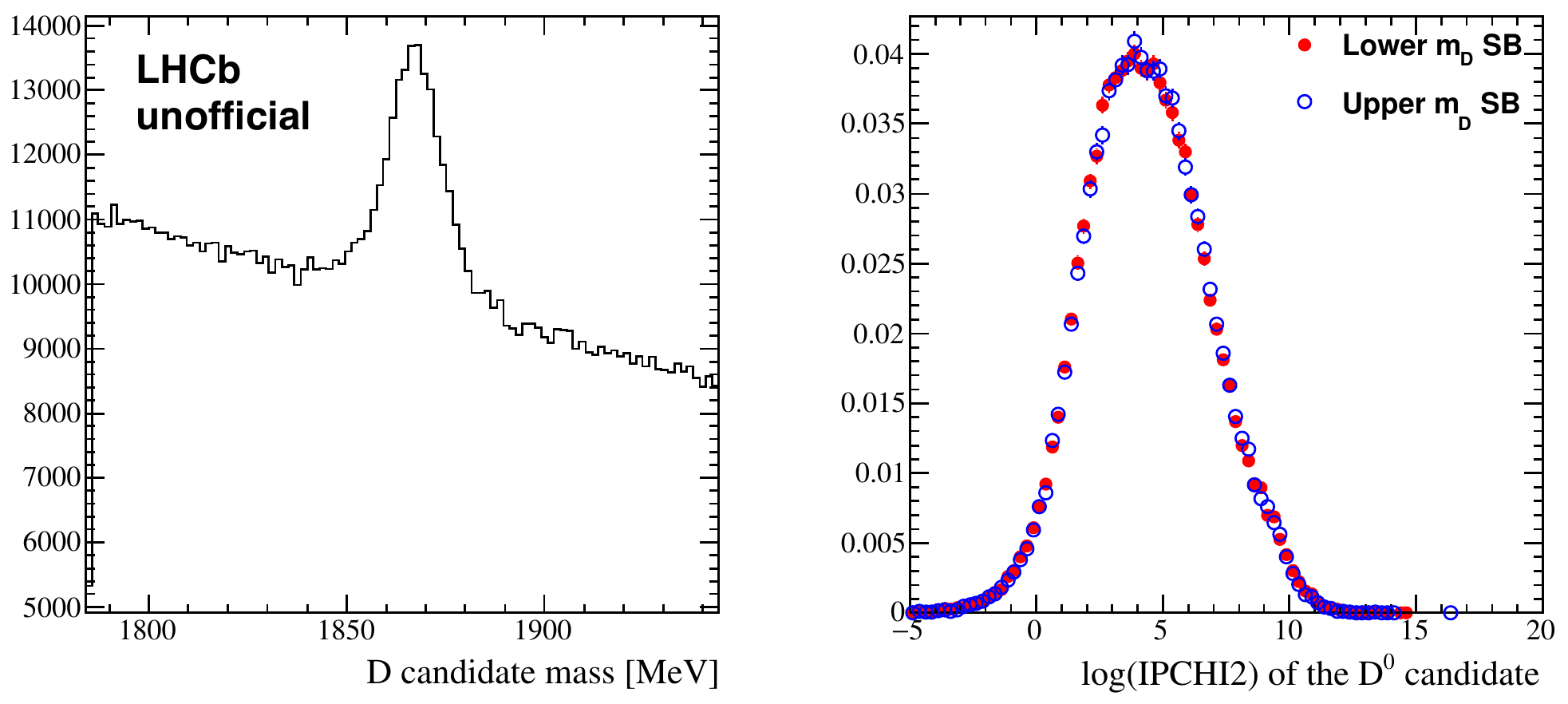}
\caption{The (right) reconstructed mass of 2011 data $D^0\to K^-\pi^+$ candidates and (left) logarithm of the impact parameter $\chi^2$ of these candidates in the lower and upper mass sideband. By contrast with
Figure~\ref{fig:signalsnofit} this plot was made using 2011 data, and was used to fix the analysis procedure prior to datataking. }
\label{fig:bkgmassipchi2corr}
\end{figure}

It is tempting to try and measure the yields of the
three fit components -- signal, combinatorial, and secondary background -- by performing a two-dimensional likelihood fit to these discriminating variables. Such an approach is however
not possible because of correlations between the signal mass and the signal impact parameter $\chi^2$, shown in Figure~\ref{fig:sigmassipchi2corr}. The further the signal candidate mass lies from the
true value, the larger the average impact parameter $\chi^2$ of this candidate will be. The difference between the measured mass and the true mass is essentially driven by two effects: the
error on the measured momentum of each of the decay products, and the error on the reconstructed opening angle between these decay products. Because impact parameter $\chi^2$ is not a signed
quantity, any error on the measured momenta or opening angles will make it larger and therefore introduce this correlation. By contrast, no correlation is observed for the background,
as can be seen in Figure~\ref{fig:bkgmassipchi2corr}, which can be explained by the fact that the background has a randomly distributed impact parameter $\chi^2$ in the first place.

This correlation leads to a two-step procedure for measuring the signal yields. In the first step, the reconstructed mass of the charm meson candidate is fitted to measure the yield of the combinatorial
background in the signal region, defined as $\pm2.5$ mass resolutions either side of the signal mean. In the second step, the distribution of the candidate impact parameter $\chi^2$ in this signal region is
fitted in order to measure the yield of the signal and secondary background components. In this fit, the yield of the combinatorial background is constrained to the value found in the
mass fit, while the shape of the combinatorial background in the impact parameter $\chi^2$ is taken from distributions in two mass sideband regions, one centered 50~MeV above and the other
50~MeV below the signal mean. In the case of the $D^{*\pm}$ measurement, the mass fit is performed in the mass difference of the $D^{*\pm}$ and $D^0$ candidates, called $\Delta m$.

The signal and secondary background distributions in mass are described by the sum of a Gaussian function and a Crystal Ball~\cite{Skwarnicki:1986xj} function which accounts for the tail towards
small reconstructed masses caused by final state radiation. The Gaussian and Crystal Ball are required to have the same mean but allowed to have different widths. The combinatorial 
background is empirically described by a first-order polynomial. The $\Delta m$ signal and secondary background distributions are described by a sum of three Gaussian functions with
common mean and different widths, while the combinatorial background is empirically described by a threshold function fixed to turn on at the nominal pion mass $m = 139.57$. 
The mass fits are performed for each charm meson species simultaneously across all transverse momentum and rapidity bins, with all shape parameters except the tails of the Crystal ball (which
are fixed from fits to simulation) free to vary between the bins. 

\begin{figure}
\centering
\includegraphics[width=0.48\linewidth]{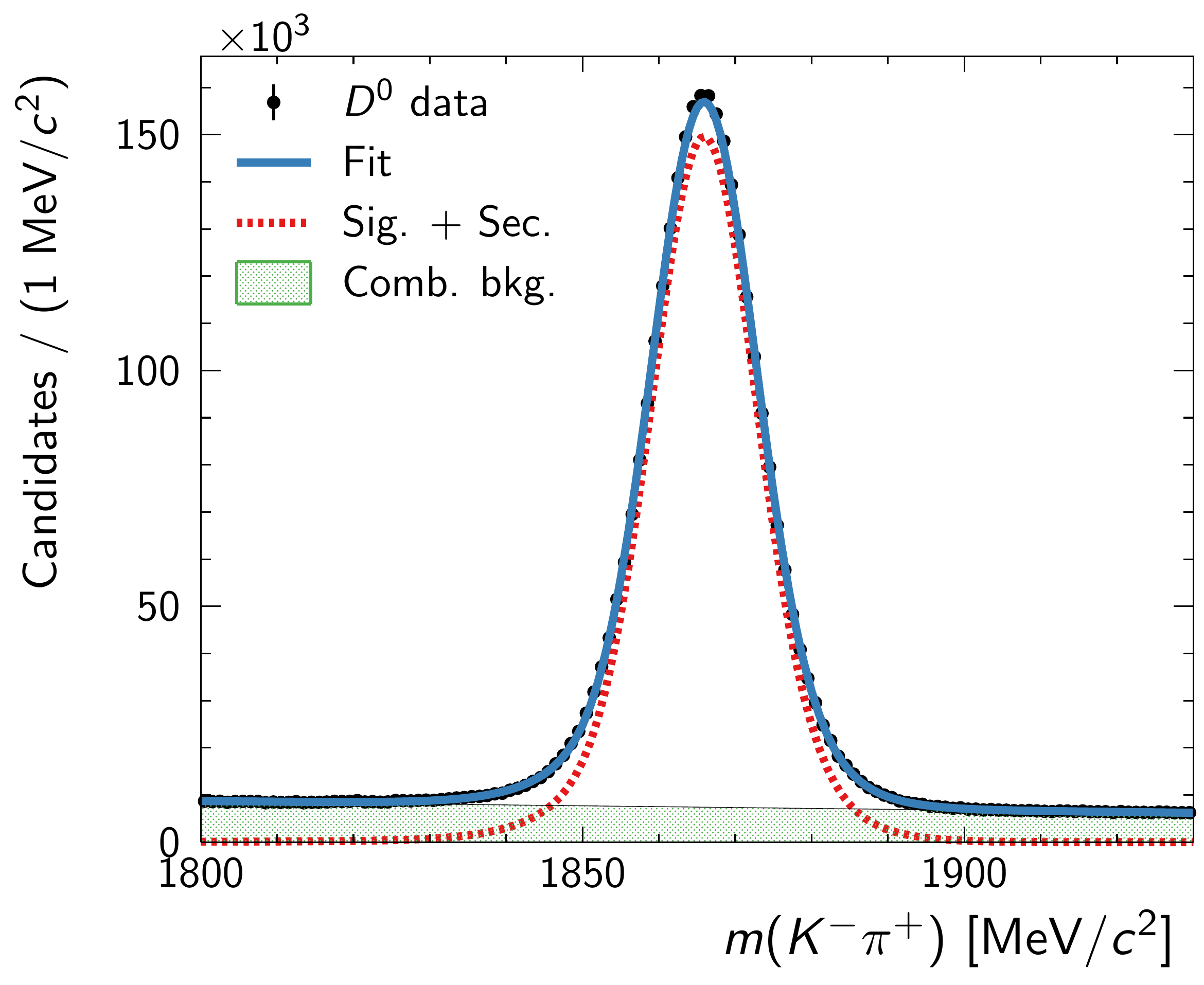}
\includegraphics[width=0.48\linewidth]{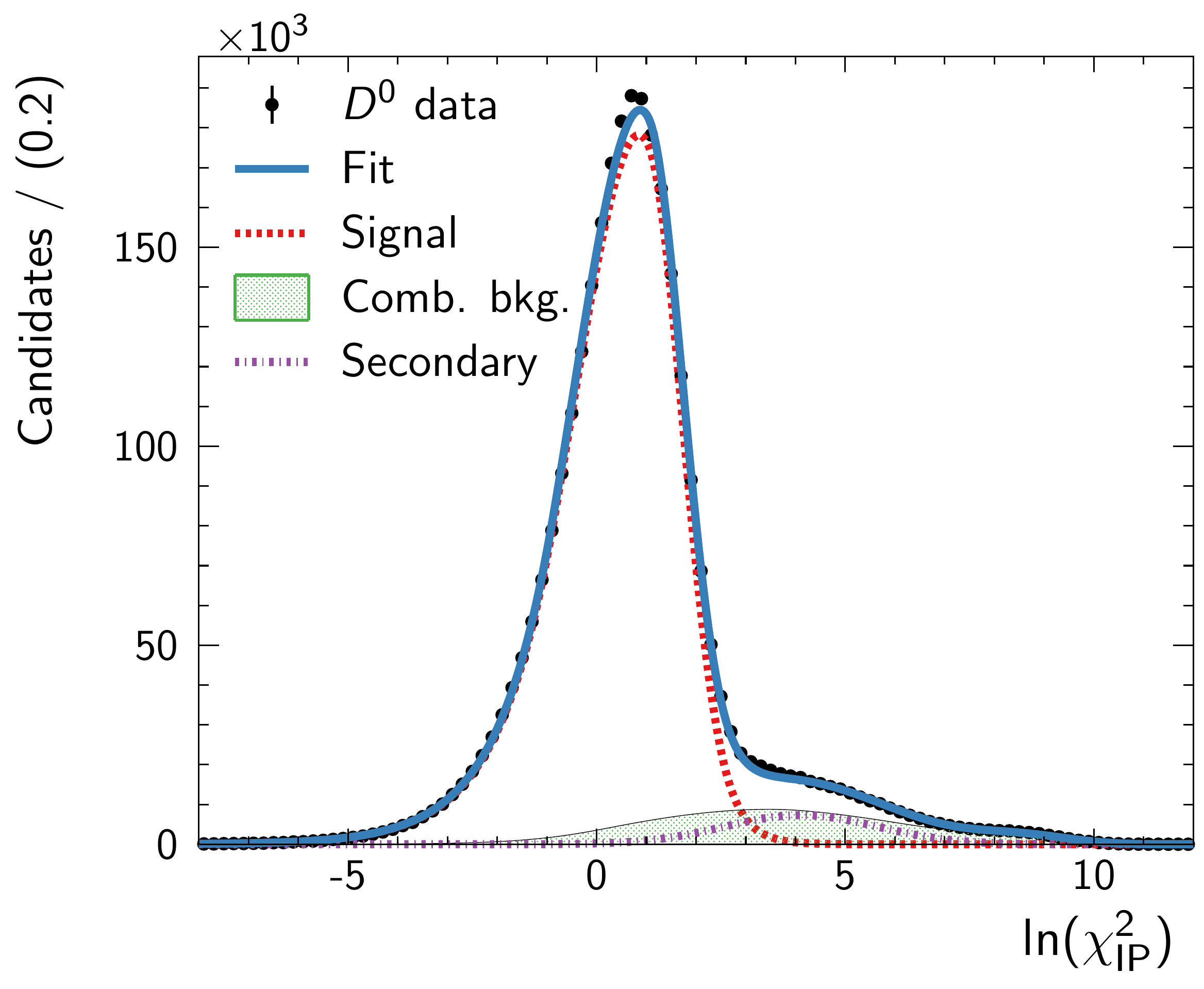}
\caption{The data fit to the (left) reconstructed mass and (right) logarithm of the impact parameter $\chi^2$ of the $D^0\to K^-\pi^+$ candidates. 
Fit components are indicated in the legend.
This plot is integrated over all bins of transverse momentum and pseudorapidity, and the displayed fit result is the sum of the individual fits in each bin.}
\label{fig:fitresultd2kpi}
\end{figure}

\begin{figure}
\centering
\includegraphics[width=0.48\linewidth]{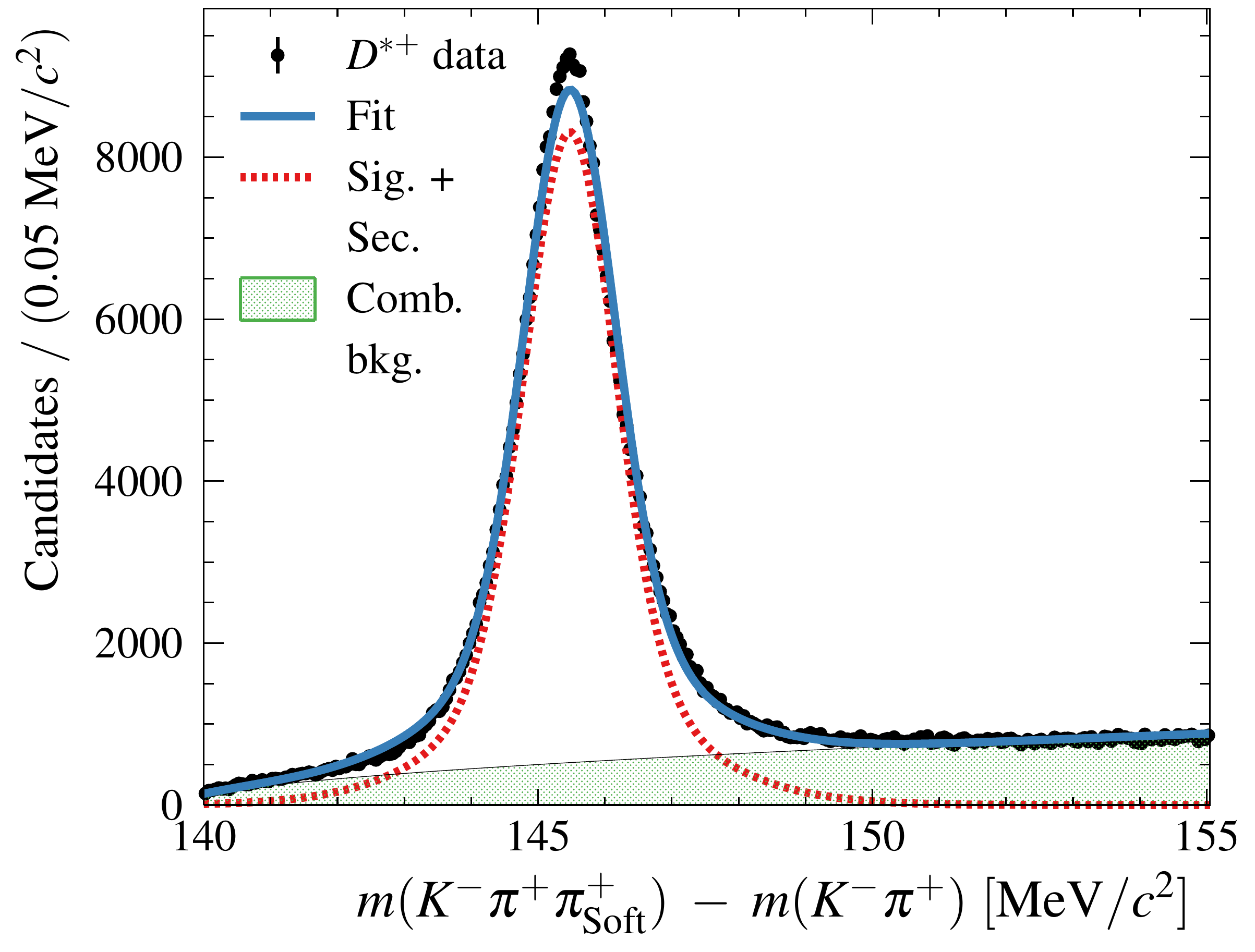}
\includegraphics[width=0.48\linewidth]{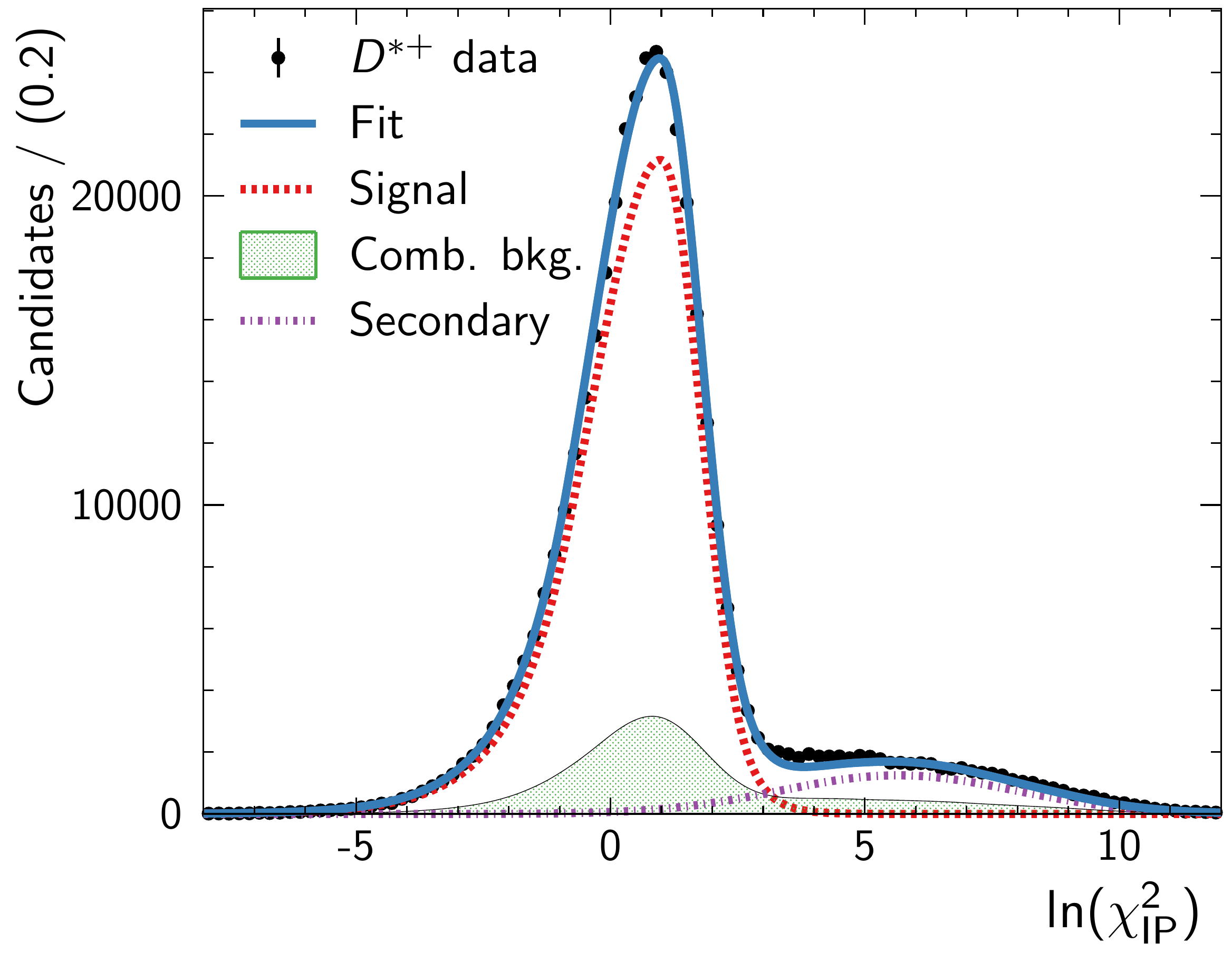}
\caption{The data fit to the $D^{*+}\to D^0 (\to K^-\pi^+) \pi^+$ candidates, performed in the (left) mass difference of the $D^{*\pm}$ and $D^0$ and (right) 
logarithm of the impact parameter $\chi^2$ of the $D^0$. Fit components are indicated in the legend. This plot
is integrated over all bins of transverse momentum and pseudorapidity, and the displayed fit result is the sum of the individual fits in each bin. Unlike
in the other fits, the combinatorial background mostly consists of true $D^0$ mesons combined with a random pion, and for this reason its PDF
looks signal-like in the logarithm of the impact parameter $\chi^2$ of the $D^0$.}
\label{fig:fitresultdst2d0pi}
\end{figure}

\begin{figure}
\centering
\includegraphics[width=0.48\linewidth]{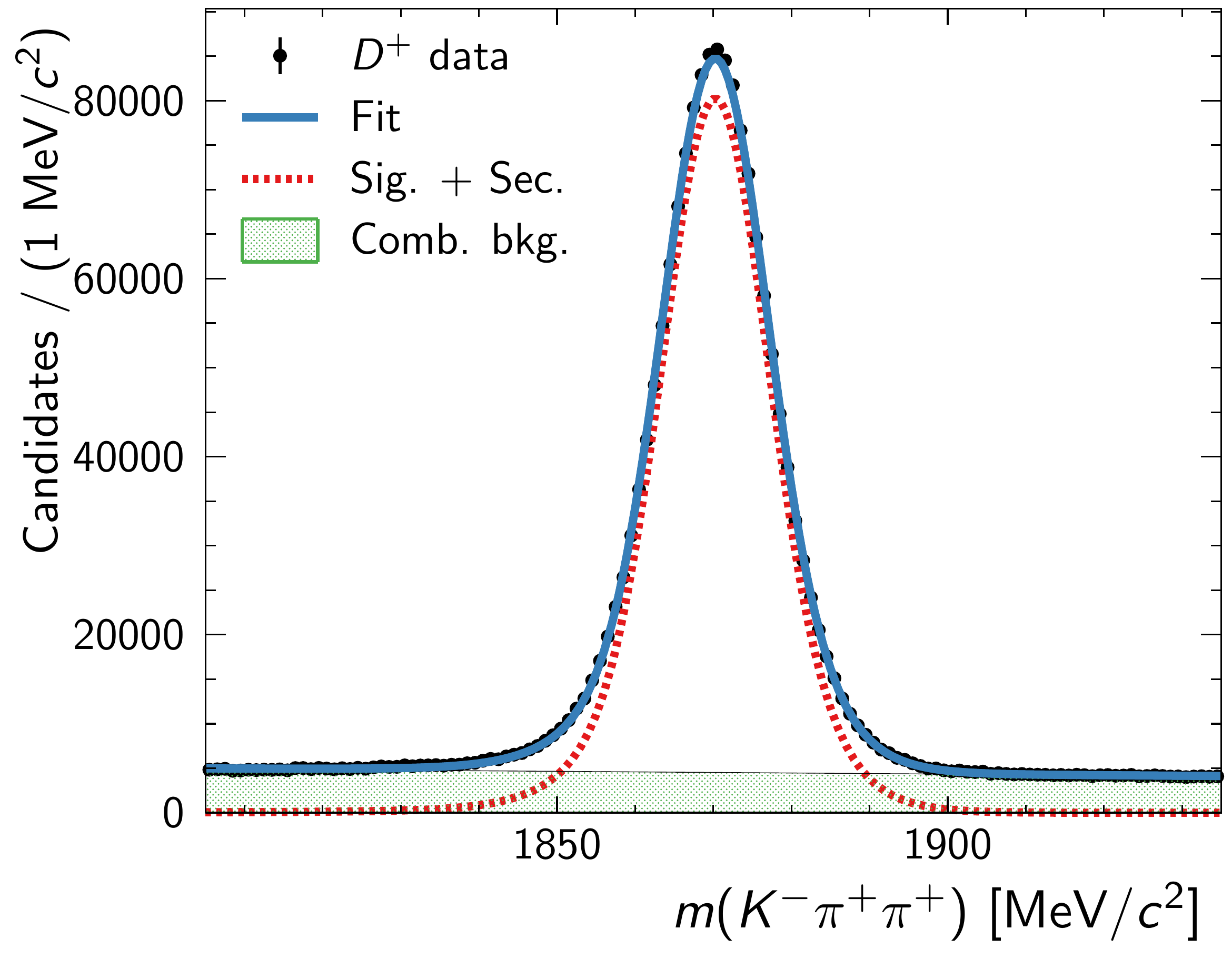}
\includegraphics[width=0.48\linewidth]{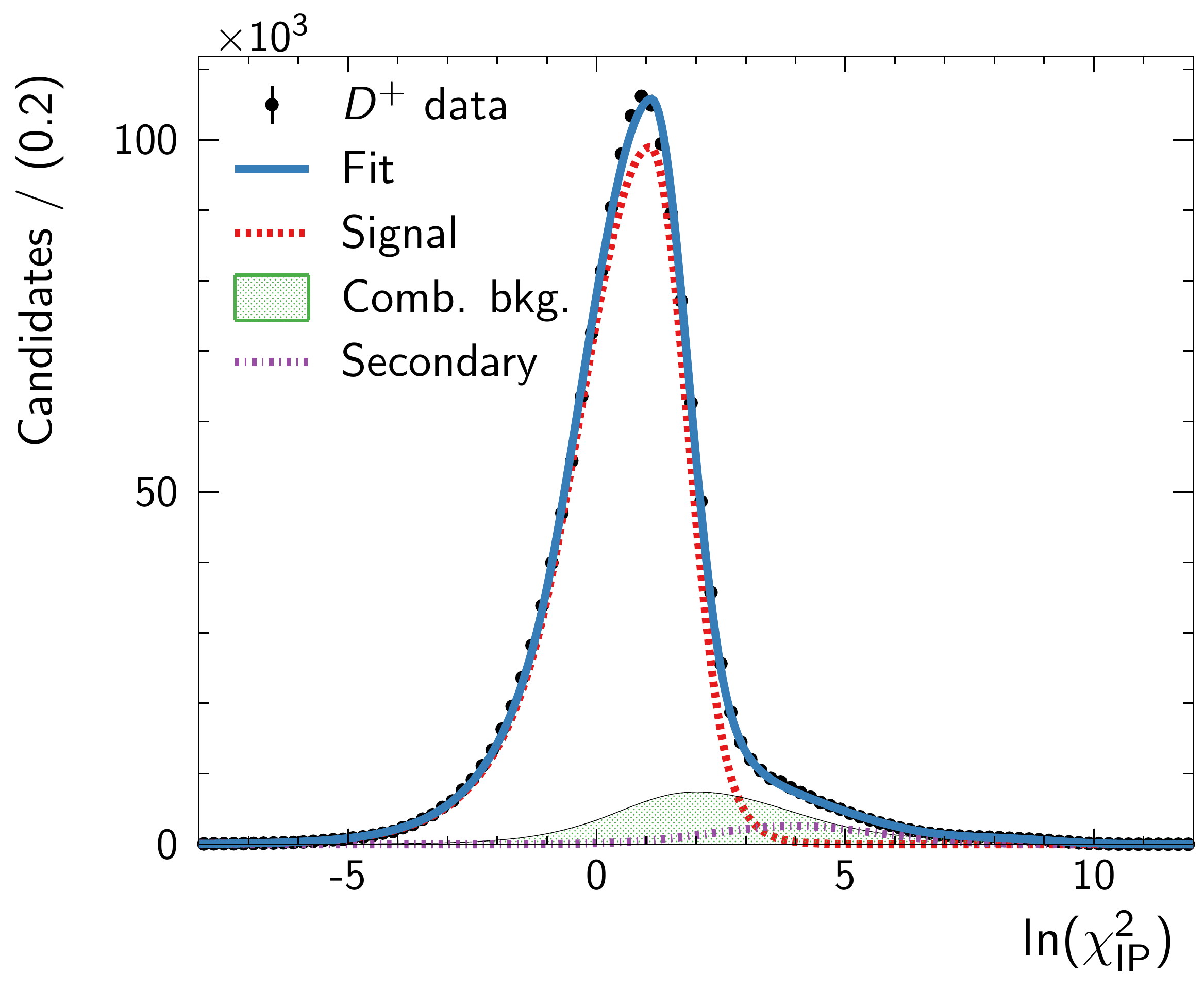}
\caption{The data fit to the (left) reconstructed mass and (right) logarithm of the impact parameter $\chi^2$ of the $D^+\to K^-\pi^+\pi^+$ candidates. 
Fit components are indicated in the legend. This plot
is integrated over all bins of transverse momentum and pseudorapidity, and the displayed fit result is the sum of the individual fits in each bin.}
\label{fig:fitresultd2kpipi}
\end{figure}

\begin{figure}
\centering
\includegraphics[width=0.48\linewidth]{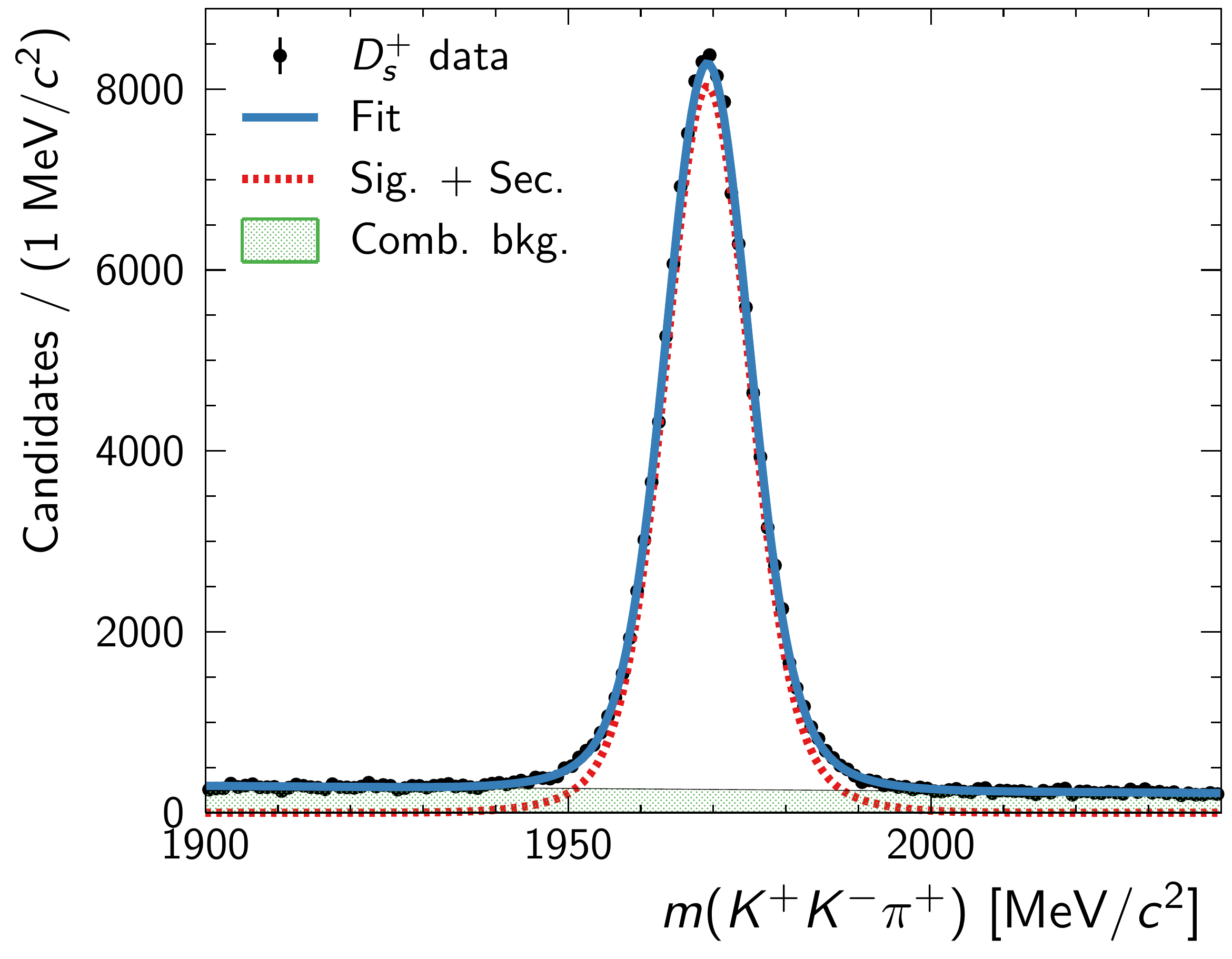}
\includegraphics[width=0.48\linewidth]{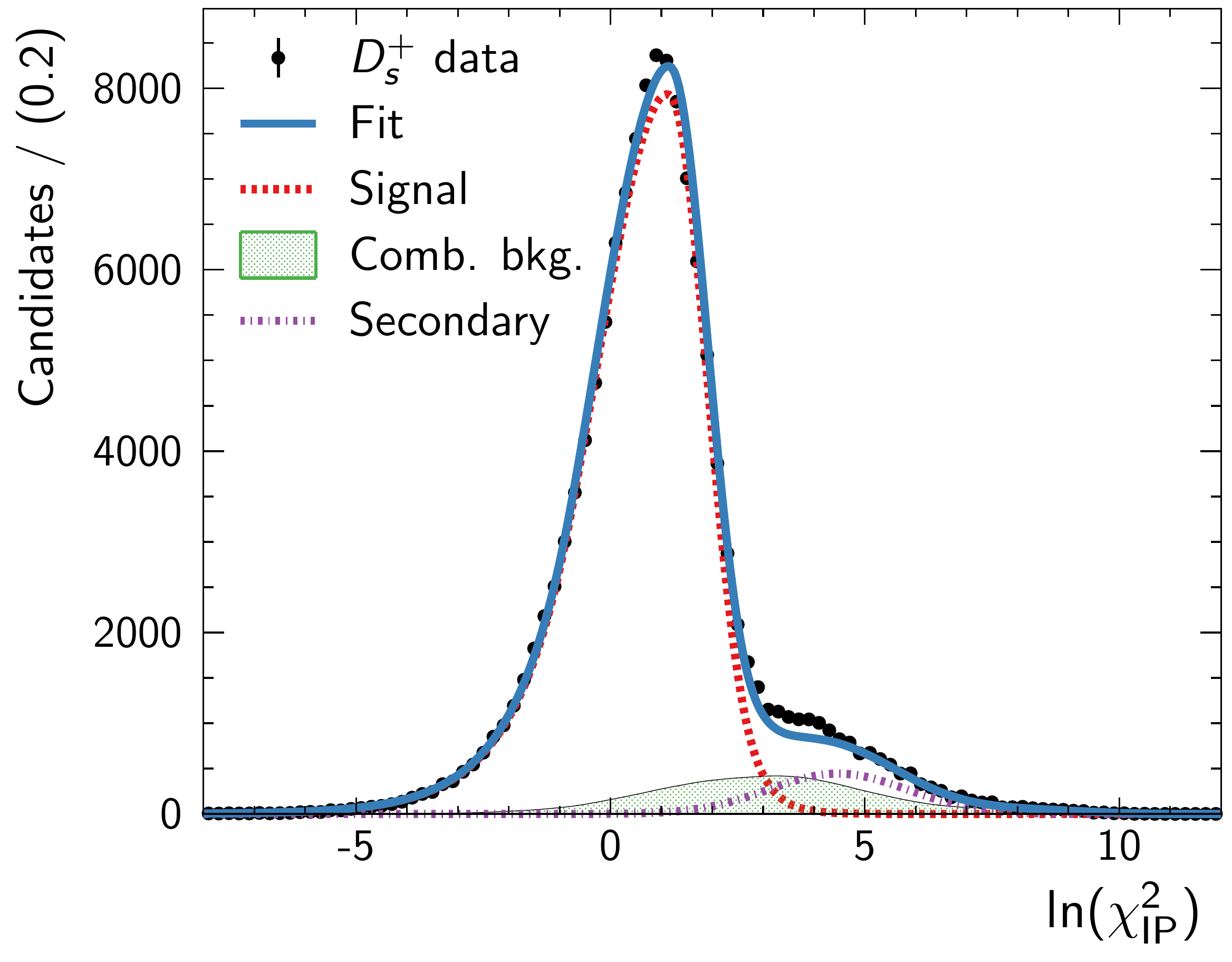}
\caption{The data fit to the (left) reconstructed mass and (right) logarithm of the impact parameter $\chi^2$ of the $D^+_s\to \phi(1020) (\to K^- K^+) \pi^+$ candidates. 
Fit components are indicated in the legend. This plot
is integrated over all bins of transverse momentum and pseudorapidity, and the displayed fit result is the sum of the individual fits in each bin.}
\label{fig:fitresultds2kkpi}
\end{figure}

\begin{figure}
\centering
\includegraphics[width=0.48\linewidth]{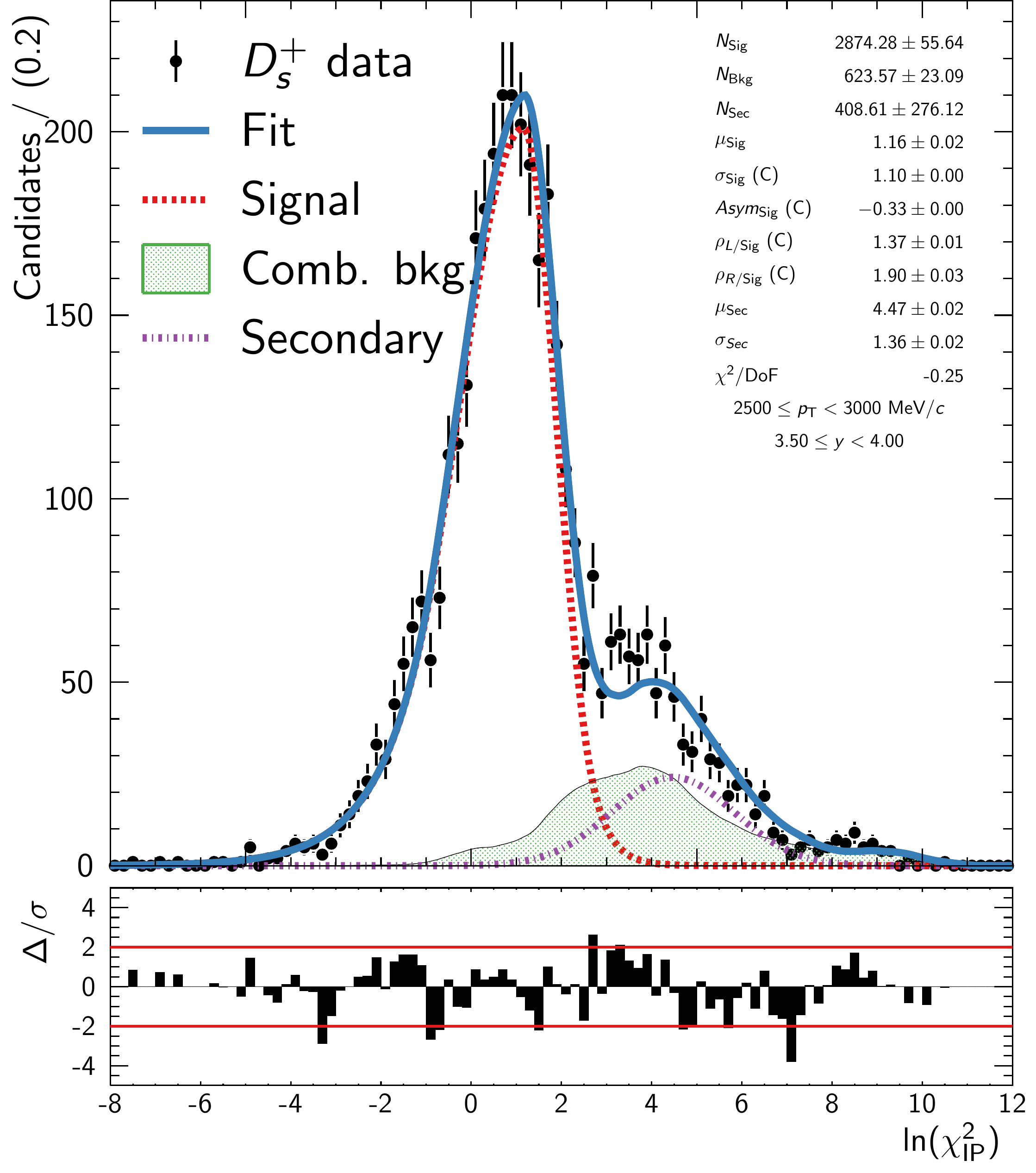}
\includegraphics[width=0.48\linewidth]{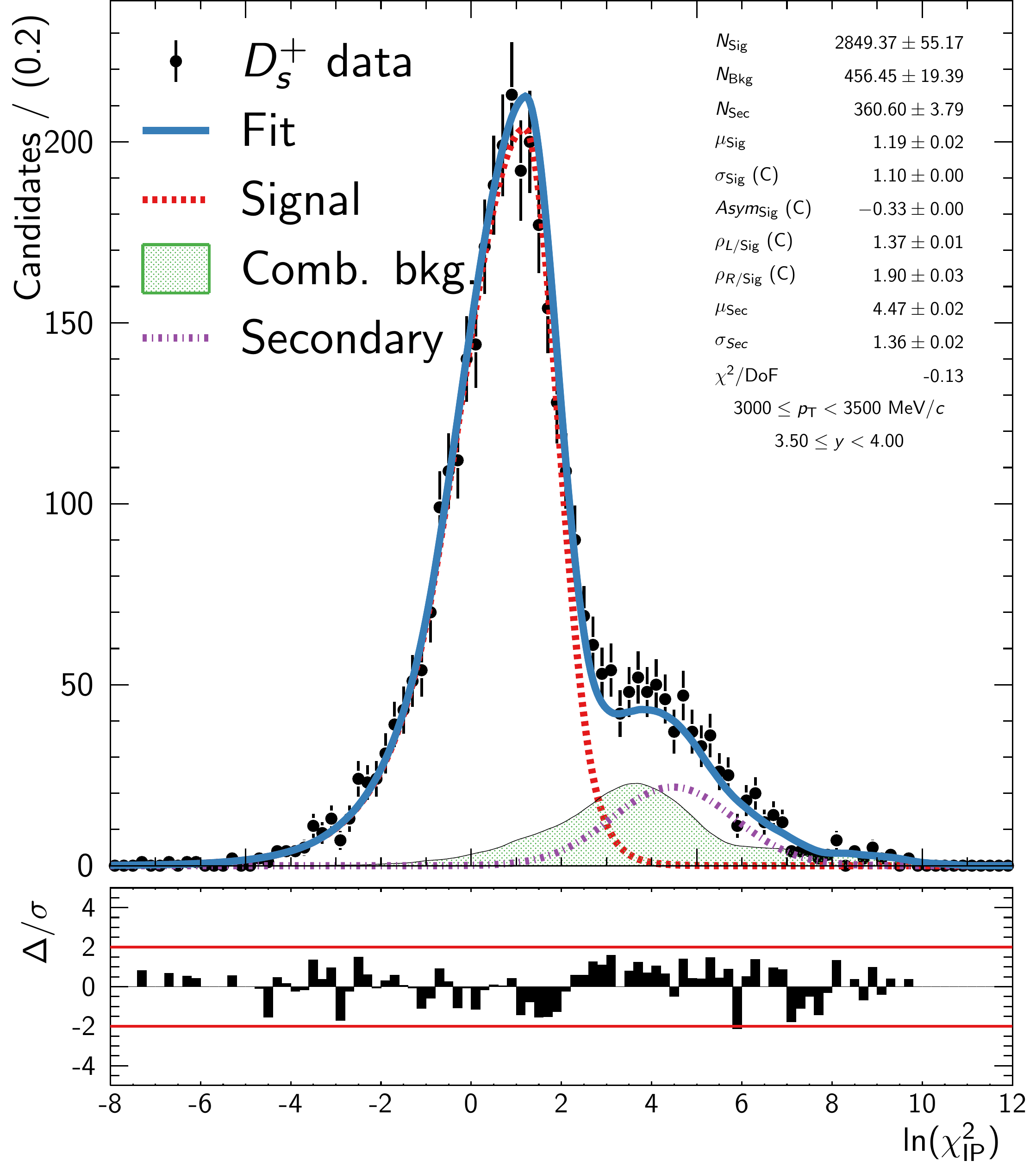}
\includegraphics[width=0.48\linewidth]{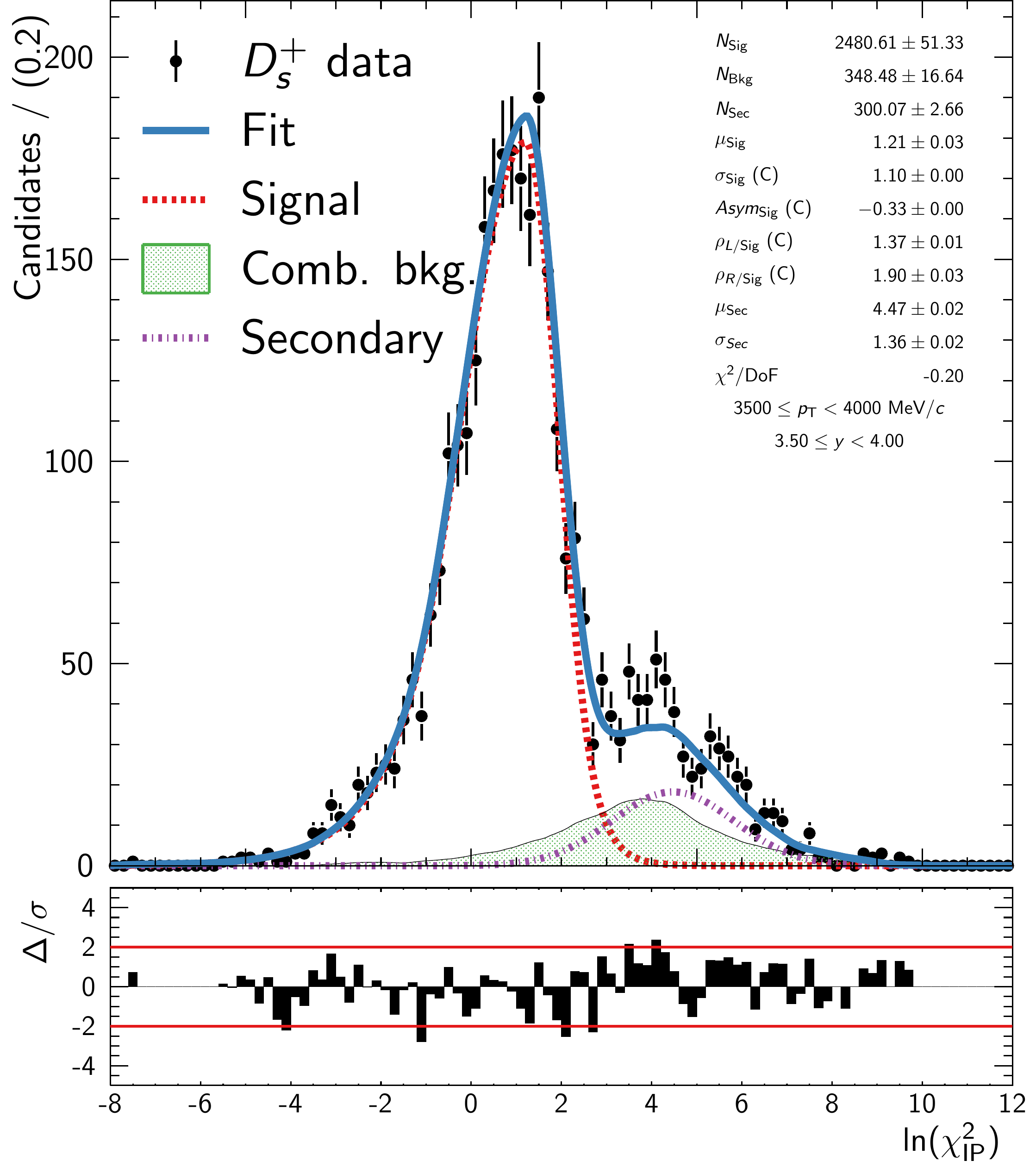}
\includegraphics[width=0.48\linewidth]{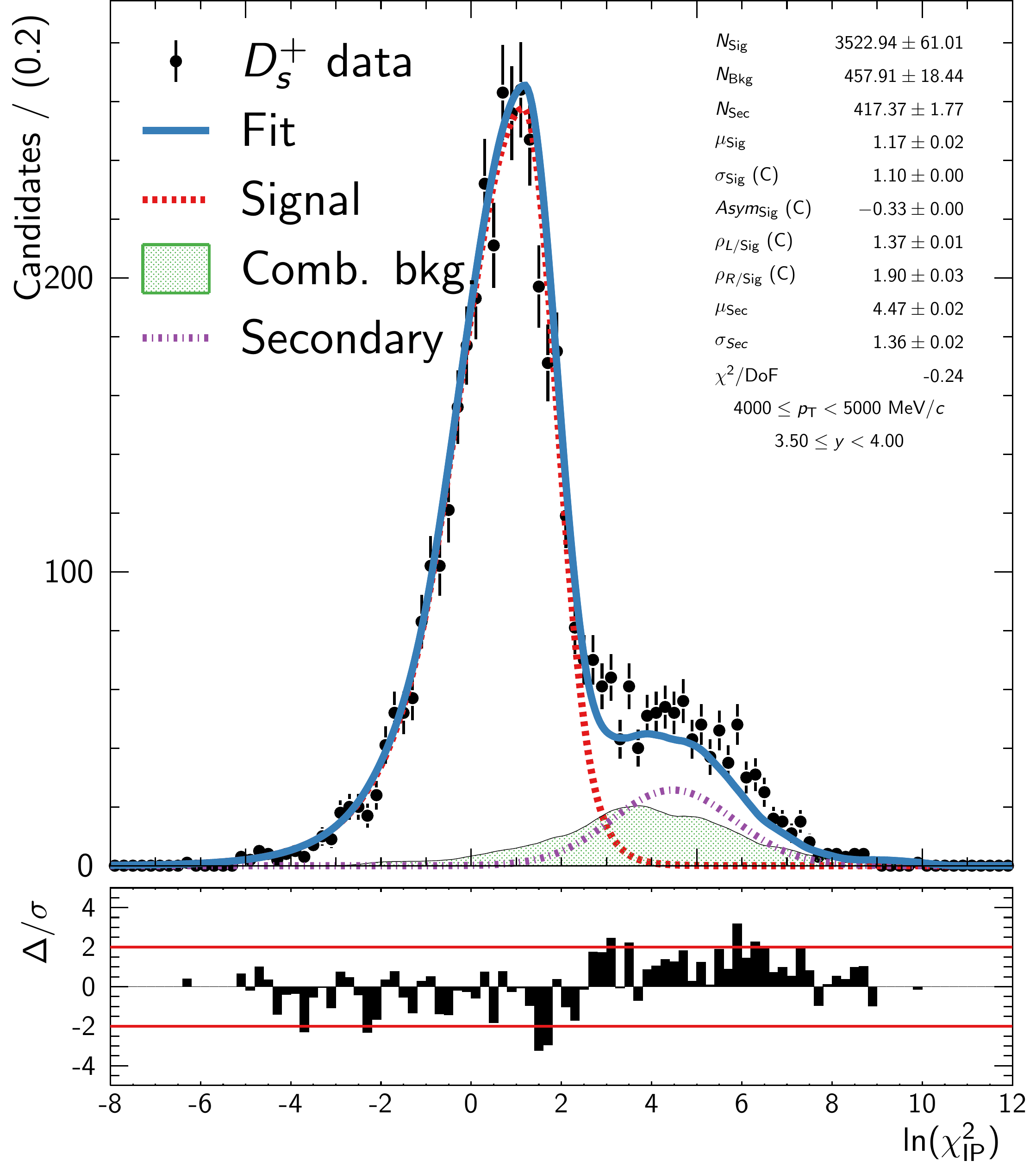}
\caption{The data fit to the logarithm of the impact parameter $\chi^2$ of the $D^+_s\to \phi(1020) (\to K^- K^+) \pi^+$ candidates in four specific
bins of transverse momentum and pseudorapidity. 
Fit components are indicated in the legend. }
\label{fig:fitresultds2kkpiinbins}
\end{figure}

The signal distribution in logarithm of the impact parameter $\chi^2$ is described by a Gaussian distribution with independently varying left- and right-hand widths and exponential tails, 
while the secondary background is described by a Gaussian.
The tail parameters and the asymmetry between the signal Gaussian components is obtained from simulation, while all other shape parameters are free to float in
the fit to the data. The combinatorial background is described by a kernel density estimator which is evaluated using the earlier mentioned mass sidebands. 
The impact parameter $\chi^2$ fits are performed for each charm meson species simultaneously across all transverse momentum and rapidity bins, and all parameters except the 
peak of the signal distribution are shared between the bins. Example mass and impact parameter $\chi^2$ fits are given in Figures~\ref{fig:fitresultd2kpi}~to~\ref{fig:fitresultds2kkpi}
for the sum over all bins. The fit quality is generally good, however some discrepancies between the fit and data can be noted, particularly in the background
impact parameter $\chi^2$ region of the fit to the $D^+_s$ candidates. Figure~\ref{fig:fitresultds2kkpiinbins} shows the results of the fit in four bins of pseudorapidity and transverse momentum
where the effect is most visible. In all cases, we can see that while the fit model clearly doesn't give a good description of this crossover region between the signal,
combinatorial background, and secondary backgrounds, the actual signal component is sufficiently isolated so that it is not sigificantly affected by this mismodelling. 
Similar observations could be made about the mismodelling visible in the fit to the $D^{*+}\to D^0 (\to K^-\pi^+) \pi^+$ candidates.

\section*{Systematic uncertainties}
Systematic uncertainties are computed separately for each charm meson species and each bin of transverse momentum and rapidity. Because
of the way in which they are evaluated, systematic uncertainties can be correlated between bins of transverse momentum and rapidity, between
particle species, or between both. The dominant systematic uncertainties are those on the luminosity, which is fully correlated between
all measurement bins and particle species, and the systematic uncertainty on the track reconstruction efficiency, which is nearly fully correlated
between the measurement bins and particle species. The statistical uncertainty on the selection efficiencies caused by the finite simulated sample sizes used
in the analysis is treated as a systematic and is fully uncorrelated between measurement bins and decay modes. It is the dominant systematic in those bins
with the smallest signal yields, but is otherwise subleading with respect to the track reconstruction and luminosity systematics.

The systematic uncertainty on the track efficiency includes an uncertainty for the finite size of the calibration samples, an uncertainty from
the finite accuracy of the tag-and-probe method when compared to the truth efficiency in simulated events, and an additional uncertainty on the 
material interactions of pions and kaons with the detector. The finite size of the calibration samples and the material interaction uncertainties
are the dominant sources of this systematic, which is between $5-10$\% depending on the particle species and measurement bin. 

The evaluation of selection efficiencies from simulation is calibrated on the data itself, by varying the selection criteria and comparing the
measured signal yields in data with the predicted efficiency variation in simulation. This is done for each selection criterion separately, and the 
sum in quadrature of the differences between the measured and expected efficiency is assigned as a systematic, taking correlations between variables into account.
Particle identification efficiencies are reevaluated using different binnings of the calibration samples, and differences in the measured efficiencies
are assigned as systematic uncertainties. 

Finally, the likelihood fits used to measure the signal yield are repeated with alternative fit models, and a systematic is assigned based on the largest
deviation from the nominal result. A summary of the overall systematic uncertainties is shown in Table~\ref{tab:sys:summary}.

\begin{table}
  \begin{center}
  \caption[Systematic uncertainties summary]{%
    Overview of systematic uncertainties and their values, expressed as relative fractions
    of the cross-section measurements in percent (\%). Uncertainties that are computed bin-by-bin
    are expressed as ranges giving the minimum to maximum values of the bin uncertainties. Ranges for
    the correlations are also given seperately for bins and modes in \%. \\ 
  }
  \label{tab:sys:summary}
  \begin{tabular}{c|c|c|c|c|cc}
    \toprule
                              & \multicolumn{4}{c}{Uncertainties (\%)} & \multicolumn{2}{c}{Correlations}\\
                              & \Dz   & \Dp & \Dsp & \Dstp & bins    &   modes \\
    \midrule                                                                                     
    MC stat.                  &   1-26    & 1-39            & 1-55                 & 1-23             & 0      &     0      \\
    MC modelling                  &   1   & 1            & 0.2                 & 0.9             & 0      &     0       \\
    \midrule                                                                                         
    Fit model                 & 1-6 & 1-5 & 1-2 & 1-2 & 0 & 0 \\
    Tracking                  &   3-10     & 3-14             & 4-14                 & 5-11              & 90-100  & 90-100      \\
    \pid cal.                 &   0-2     & 0-1               & 0-2                  & 0-1               & 0-100   & 0-100          \\
    \pid binning              &   0-44    & 0-10             & 0-20                  & 0-15              & 100     & 100          \\
    $\mathcal{BR}$            &   1.2     & 2.1             & 5.8                  & 1.5              & 100     & 0-95     \\
    \midrule
    Luminosity                &           \multicolumn{4}{c|}{3.9}                                       & 100     & 100      \\

    \bottomrule
  \end{tabular}
  \end{center}
\end{table}

\section*{Results and discussion}
In the paper, four sets of results were presented: the absolute cross sections for each charm meson species in bins of transverse momentum and rapidity,
the integrated cross sections for each charm meson species in the LHCb fiducial volume, ratios of cross-sections between 13 and 7~TeV data for each charm
meson species, and ratios of cross-sections between different charm meson species. The full presentation of these results would require around twenty pages of text
and achieve little except to kill a few trees, so I refer you to the published paper for the full results and restrict myself to a few comments here.

\begin{figure}
\centering
\includegraphics[width=0.9\linewidth]{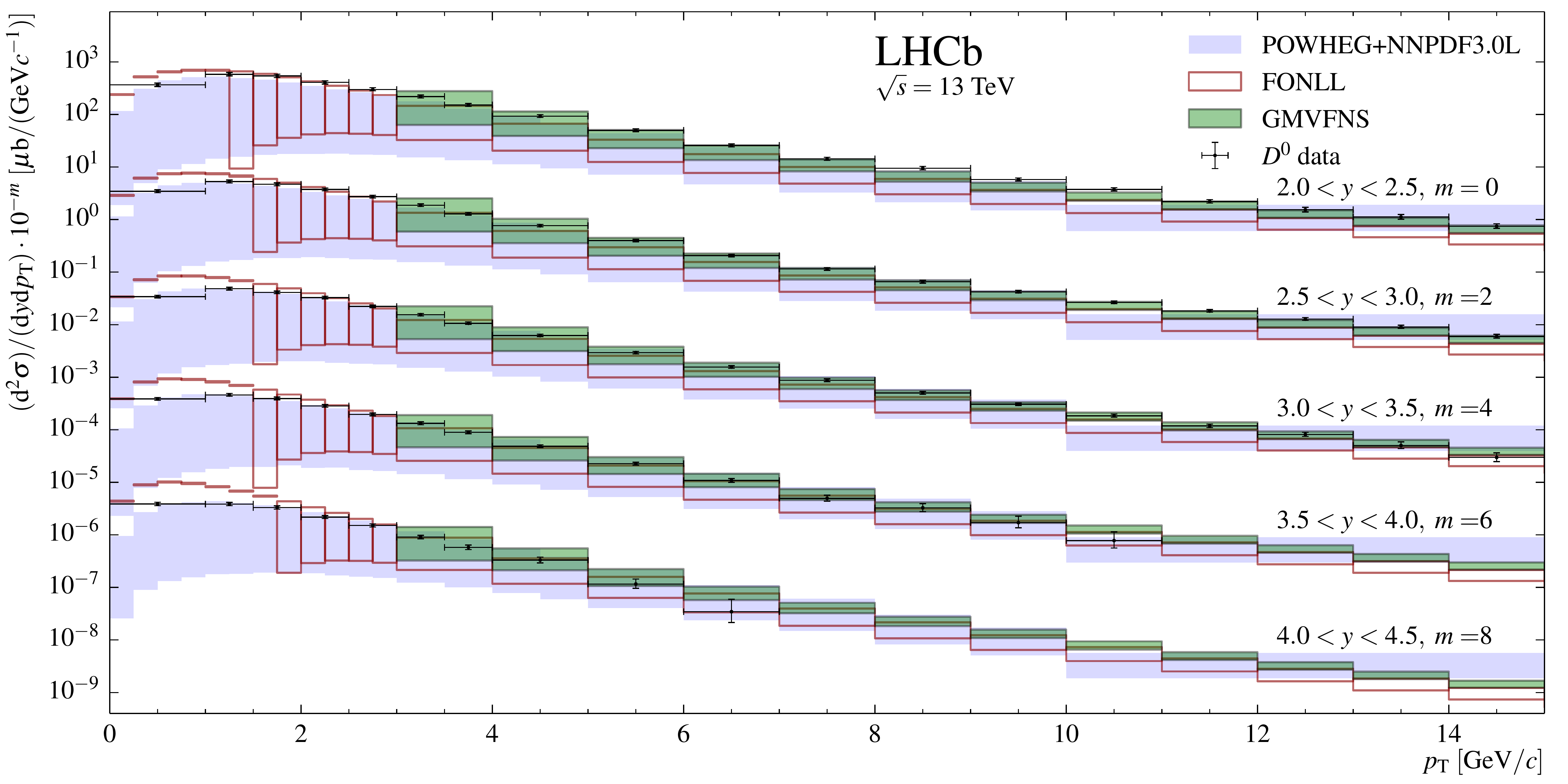}
\includegraphics[width=0.9\linewidth]{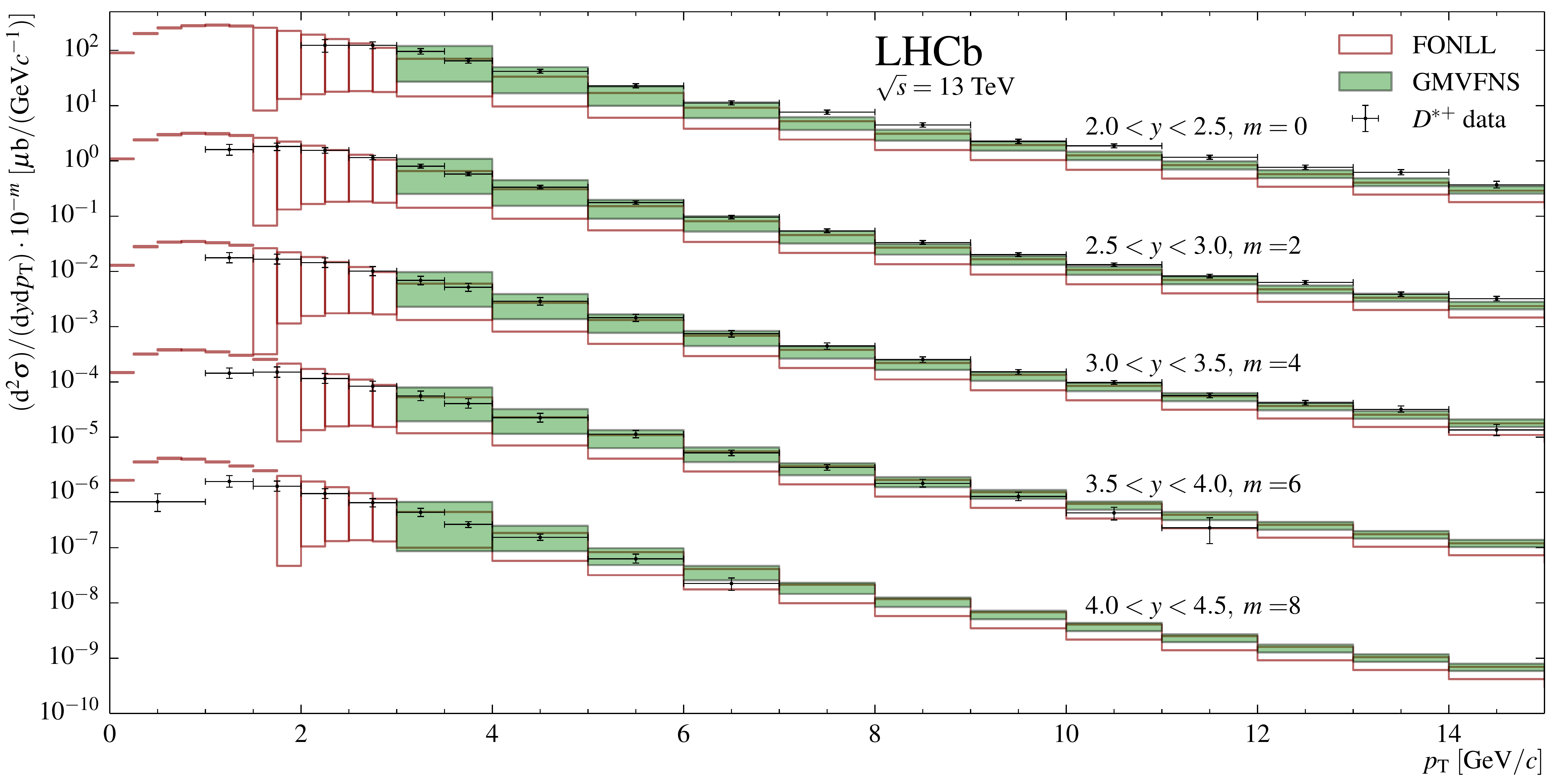}
\caption{Absolute differential cross-sections for the (top) $D^0$ and (bottom) $D^{*+}$ mesons, reproduced from~\cite{Aaij:2015bpa}.
Plot components are indicated in the legend, the abbreviations refer to the available theoretical predictions.
The vertical axis is subdivided into five ranges, one for each of the pseudorapidity bins, as labelled on the right-hand side
of the plot.}
\label{fig:diffxsecddst}
\end{figure}

\begin{figure}
\centering
\includegraphics[width=0.9\linewidth]{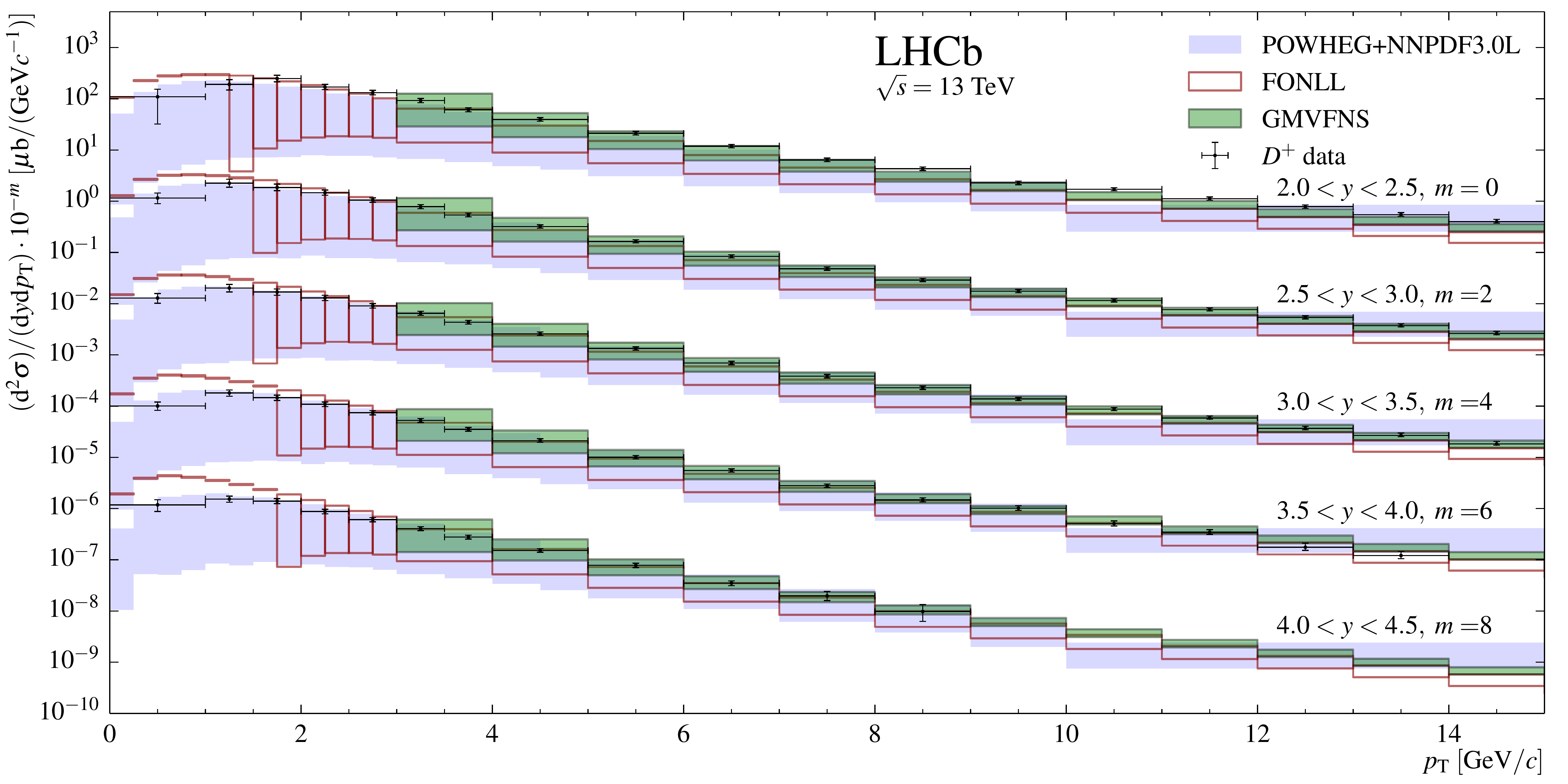}
\includegraphics[width=0.9\linewidth]{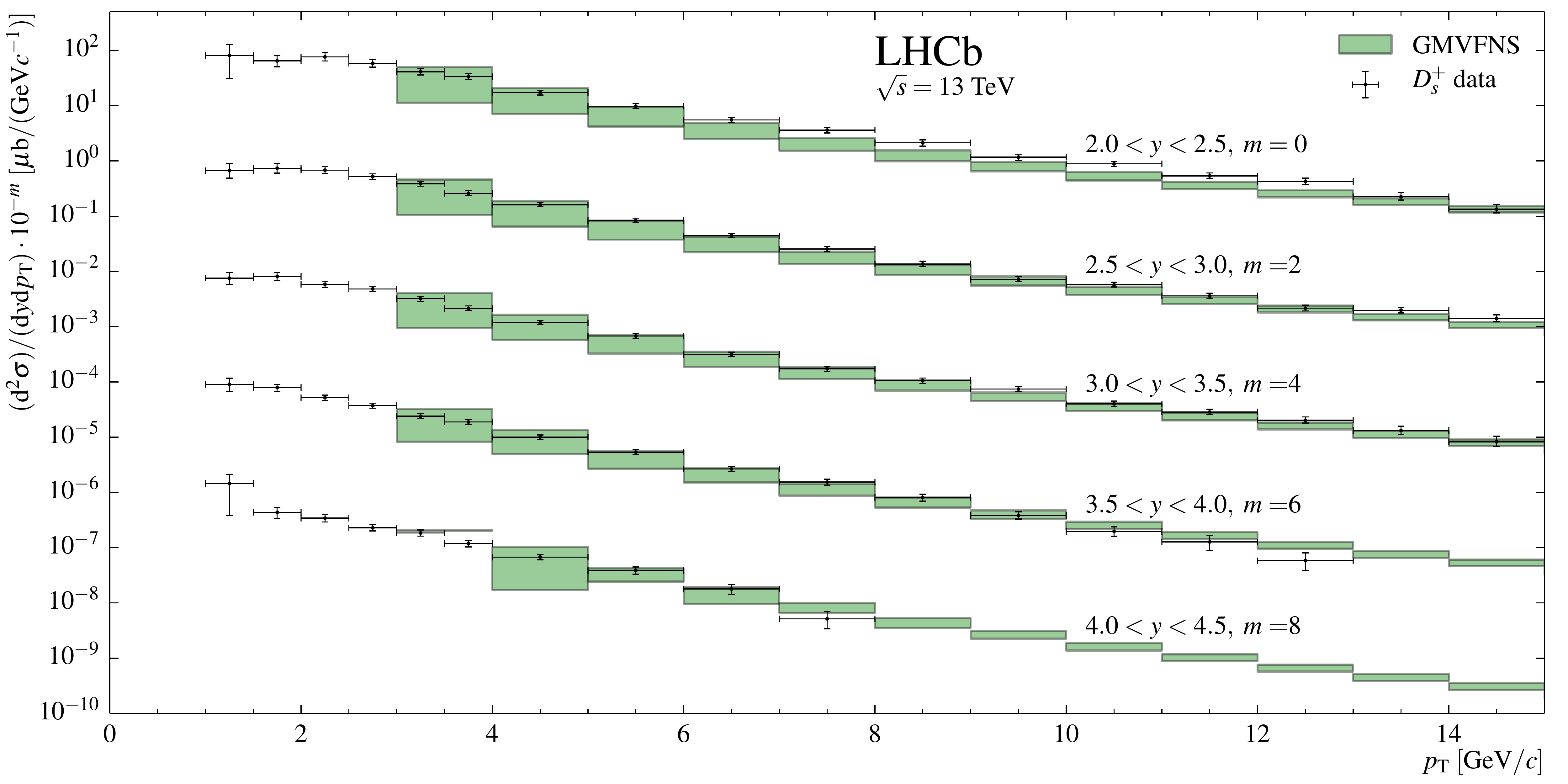}
\caption{Absolute differential cross-sections for the (top) $D^+$ and (bottom) $D^+_s$ mesons, reproduced from~\cite{Aaij:2015bpa}.
Plot components are indicated in the legend, the abbreviations refer to the available theoretical predictions.
The vertical axis is subdivided into five ranges, one for each of the pseudorapidity bins, as labelled on the right-hand side
of the plot.}
\label{fig:diffxsecdpds}
\end{figure}

\begin{figure}
\centering
\includegraphics[width=0.9\linewidth]{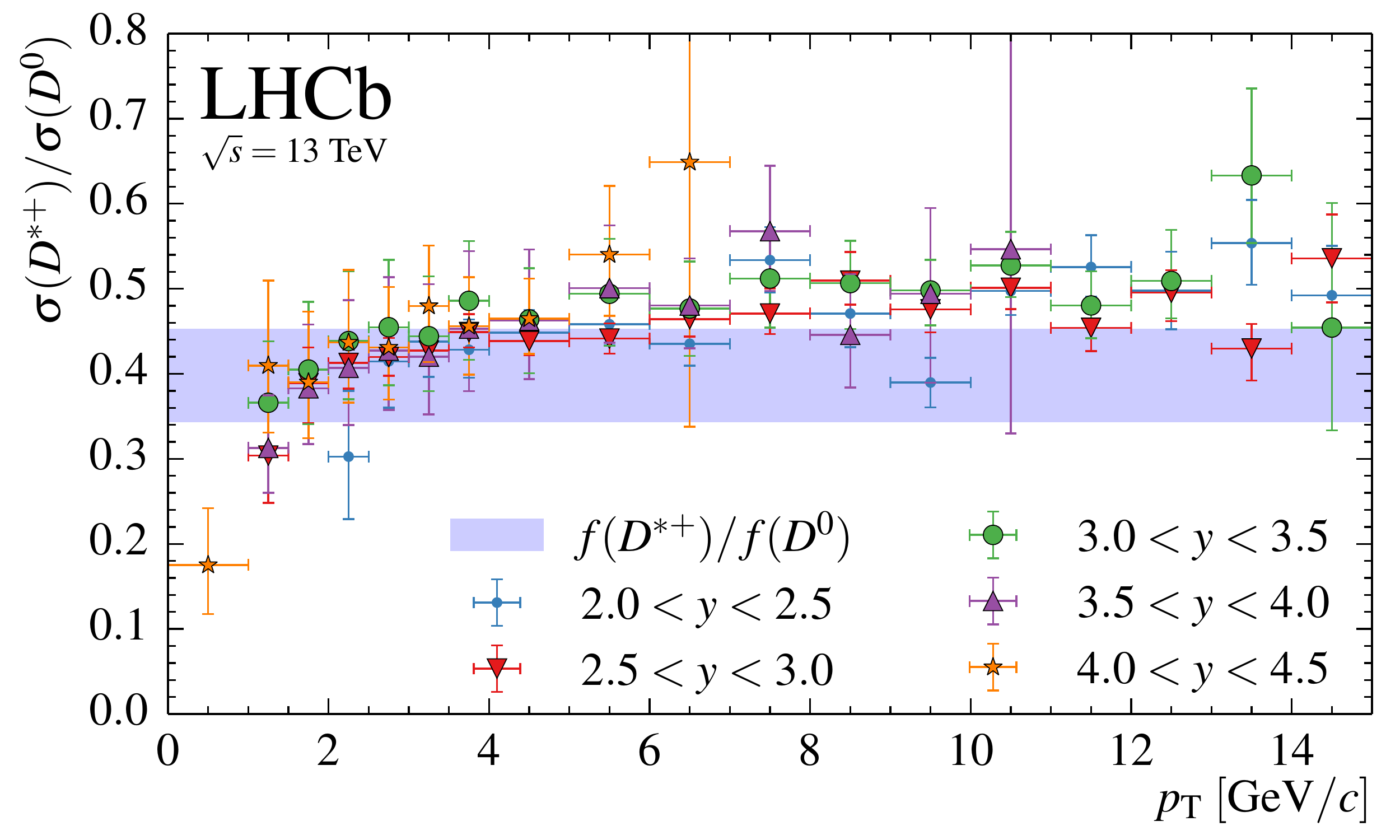}
\includegraphics[width=0.9\linewidth]{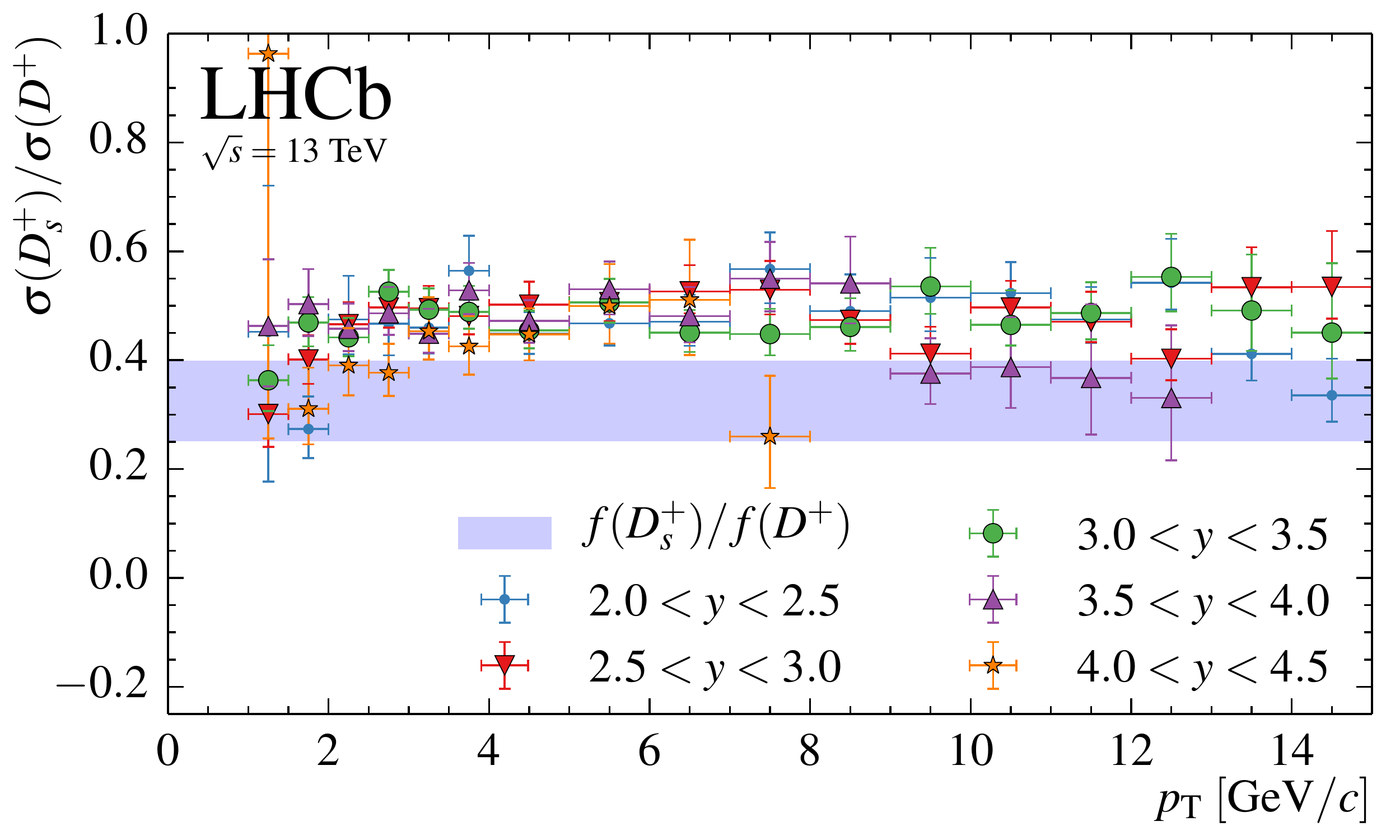}
\caption{Differential cross-sections for the (top) $D^{*+}$ vs. $D^0$ and (bottom) $D^+_s$ vs. $D^+$ mesons, reproduced from~\cite{Aaij:2015bpa}.
Plot components are indicated in the legend, the abbreviations refer to the available theoretical predictions.}
\label{fig:xsecratios}
\end{figure}

The absolute cross-sections are shown in Figures~\ref{fig:diffxsecddst}~and~\ref{fig:diffxsecdpds},
compared to the different available theoretical predictions. In general all of these measurements agree with all the theoretical models within uncertainty.
The ratios of the cross-sections of different particle species are shown in Figure~\ref{fig:xsecratios}, and support the
conclusion that heavier mesons have a harder $\pt$ spectrum. Finally, the integrated cross-section is computed as the sum of the per-bin measurements, with
correlations in the systematic uncertainties taken into account. In bins where the signal yield was insufficient to allow a measurement, a theoretical
prediction is used instead, and the uncertainty on this prediction added to the total as a systematic uncertainty. The integrated results for the $D^0$ and $D^+$
mesons in the fiducial volume $0<\pt <8$~GeV and $2 < y < 4.5$ are
\begin{equation*}
\sigma(D^0) = 2709 \pm 2 \pm 165 \mu b ,
\end{equation*} 
\begin{equation*}
\sigma(D^+) = 1102 \pm 5 \pm 111 \mu b ,
\end{equation*}
where the uncertainties are statistical and systematic, respectively. These can be combined with the charm hadronisation fractions measured at 
$e^+e^-$ colliders~\cite{Artuso:2004pj,Aubert:2002ue} to produce an overall result
\begin{equation*}
\sigma(pp\to c\bar{c}X) = 2369 \pm 3 \pm 152 \pm 118 \mu b ,
\end{equation*}
where the uncertainties are statistical, systematic, and due to the fragmentation functions, respectively. Considering that the LHC inelastic cross-section
has been measured~\cite{PhysRevLett.117.182002} to be $78.1 \pm 2.9$~mb in the full phase space, this means that around 6\% of all LHC inelastic collisions at 13~TeV produce
a charm hadron within the acceptance of the LHCb detector.

\chapter{LHCb repent!: errors in real-time analyses and their implications}
\label{chpt:lhcbrepent}
\pagestyle{myheadings}
\markboth{\bf LHCb repent!}{\bf LHCb repent!}

\begin{flushright}
\noindent {\it 
Houston do we have a problem? Obviously the answer's naw\linebreak
Ball hoggers won't pass the ball, I'mma steal that rock, then pass to y'all\linebreak
}
\linebreak
-- Chamillionaire, Won't let you down (DJ Smallz remix)
\end{flushright}

The results presented in the previous chapter are the final and accurate outcome of the analysis, but
they are not the results which were first presented at the EPS conference in 2015, nor published in the journal
later that year. The published results contained several bugs in the analysis procedure which led to biases
in the measured cross-sections, and required errata to be published. These bugs were not limited to the charm
cross-section analysis, and several other 2015 measurements also required errata to correct biases resulting from
the same underlying source. In this chapter I will explain how these biases occured, what errata were required,
and how while none were really the fault of doing real-time as opposed to traditional analysis, we can nevertheless
learn some valuable lessons from them for future real-time analyses. To avoid any risk of coming across as smart
after the fact, not only was I implicitly responsible for these bugs as one of the analysts, but I was explicitly
tasked with finding the more general bug which was affecting multiple LHCb analyses and miserably failed to do so before
we had published several further papers which went on to require errata. So the following is very much offered in the spirit
of not only criticism, but self-criticism.

\section*{The first, specific, erratum}
\begin{figure}
\centering
\includegraphics[width=0.48\linewidth]{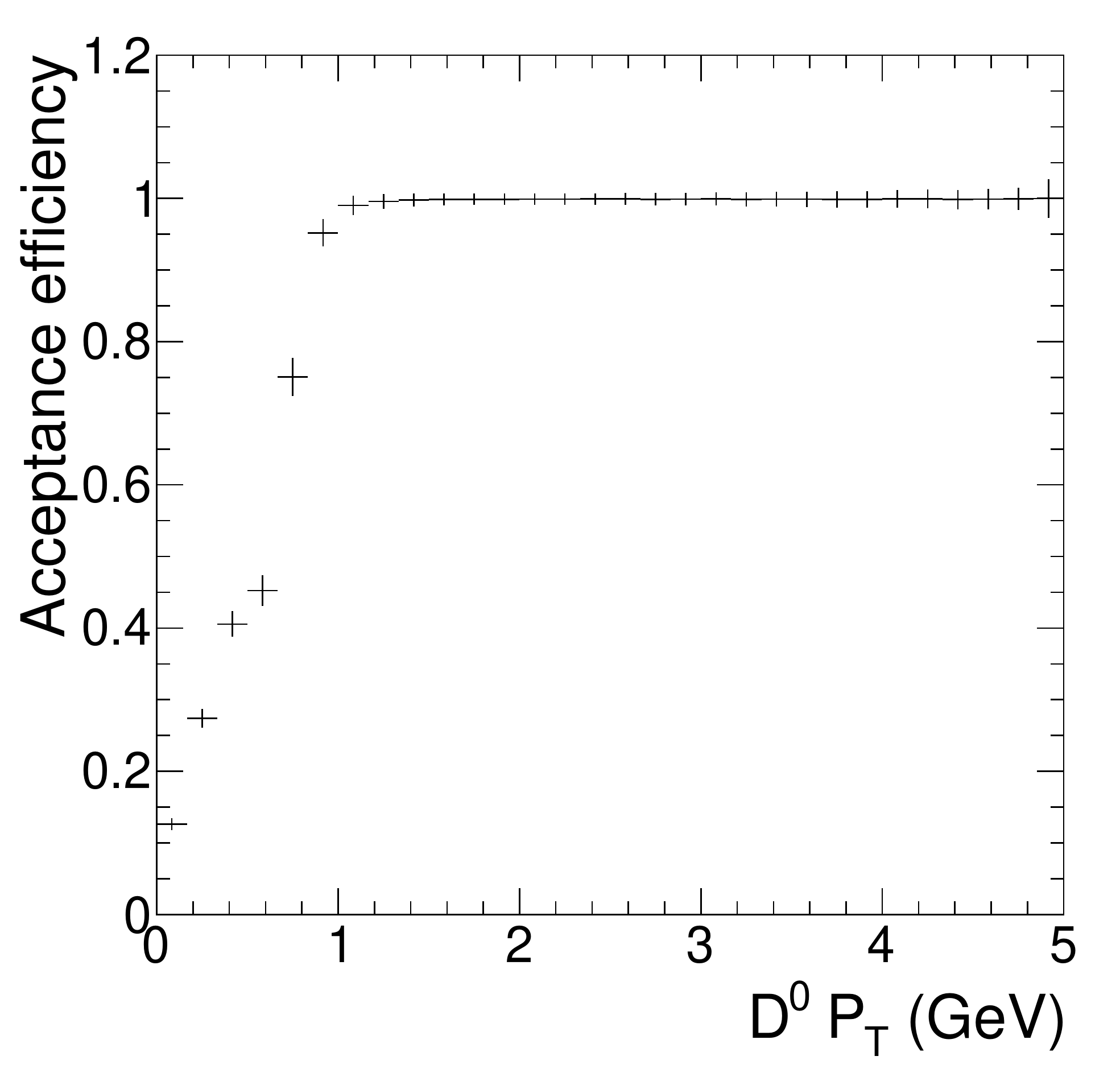}
\includegraphics[width=0.48\linewidth]{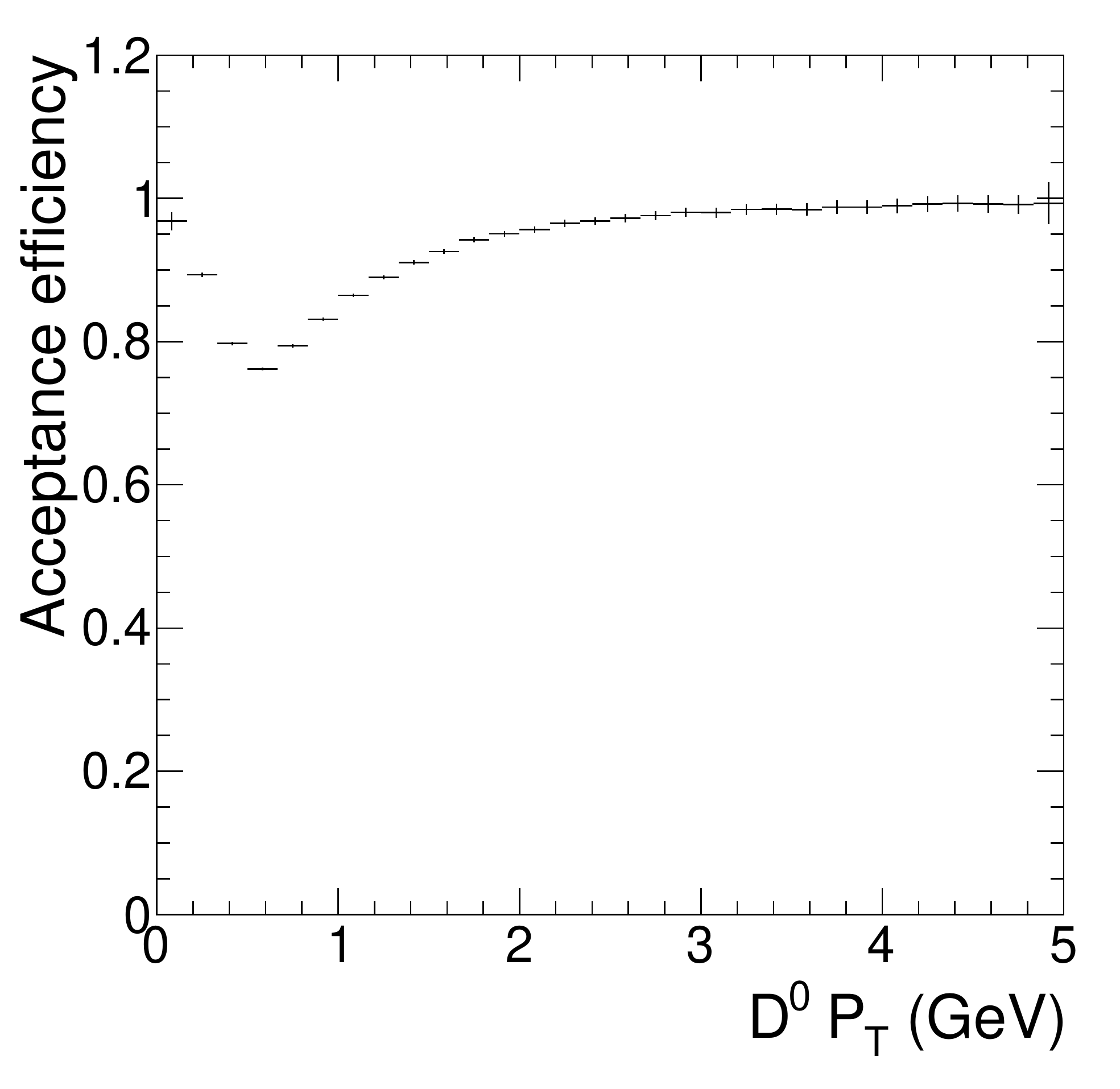}
\caption{Effect of requiring the slow pion from the $D^{*+}\to D^0 \pi^+$ decay chain to be in the LHCb acceptance on the acceptance efficiency for
the $D^0$ decay products. The ratio of the efficiencies with and without the requirement on the slow pion
is plotted as a function of the $D^0$ transverse momentum. In order to make the effect more visible, the $D^0$ is restricted to have (left) a
pseudorapidity of at most $2.5$ or (right) a pseudorapidity of at least $4$.}
\label{fig:dstd0accept}
\end{figure}

The first of the two errata was caused by two bugs in the calculation
of the efficiencies. Both of these bugs affected only the $D^0$ cross-section. The first bug was caused by the way in which
the geometrical efficiency of the LHCb detector was calculated. In order to save time, LHCb generates simulated signal samples
only if the decay products are within the geometrical acceptance of the detector, defined in the broadest sense to be between
10 and 400~mrad in polar angle. The efficiency of this requirement must then be calculated from additional generator-level samples
which are not themselves propagated through the detector simulation. In other to save further time and storage space, only the
$D^{*+}\to D^0 \pi^+$ decay chain was simulated, and it was assumed that effects due to the soft pion cancelled when computing the
$D^0$ efficiencies. Unfortunately this assumption was broken by the fact that the code also required the soft pion from the decay
chain to be in the detector acceptance, leading to a systematic underestimation of the $D^0$ efficiency and hence a systematic
overestimation of the $D^0$ cross-section. The effect was particularly large at low transverse momenta and large rapidities, a region
in which the $D^0$ is anyway at the edge of the detector acceptance. This is illustrated in Figure~\ref{fig:dstd0accept} where the effect has been simulated
using a sample of $D^{*+}\to D^0 \pi^+$ decays made with the RapidSim~\cite{Cowan:2016tnm} package.

The second bug was caused by the fact that, in order to increase
precision and reduce systematic uncertainties, most of the efficiencies (excluding the ones related to particle identification, where
the limited size of the calibration samples imposes specific constraints) are computed using a finer-grained binning than is used when reporting
the differential cross-sections themselves. There was a bug in the way these fine-grained efficiencies were converted to coarser-grained
ones for the $D^0$ sample, caused by a simple overwriting of the correct code by an earlier incorrect version when integrating developments
from a parallel branch of the software. This second effect was a lot smaller than the first one, and was only spotted in the process
of preparing the erratum for the first bug.

In both cases, the analysts conceptually intended to do the right thing, reported doing what they intended, but had actually ended up doing another.
As LHCb, in common with many other large HEP collaborations, reviews analysis notes containing a physics summary of what was done,
these bugs couldn't have been reasonably found in
review, whether the analysis had been done in real-time or traditionally. Indeed, the problems were only spotted when the same
group of analysts proceeded to use their analysis code to measure charm cross-sections at 5~TeV; if the code had not been reused,
it is highly likely that the problem would never have been found. 

\section*{The second, general, erratum}

The second erratum was linked to a much more general, and therefore  important and instructive, problem with how LHCb
corrects its detector efficiencies for data-simulation differences. In the simplest possible terms, it turned out that the methods
which LHCb had used for correcting such differences fundamentally relied on the differences being small in the first place. It is
interesting however to understand how and why this happened in more detail.

LHCb analyses correct for data-simulation differences in various places: particle identification efficiencies are taken directly
from data calibration samples,\footnote{Either by computing the efficiencies in the relevant kinematic ranges from the data calibration
samples, or by resampling the simulated samples with data values of the particle identification features. The former is generally
used for analyses which simply need to know the efficiency of one specific particle identification requirement, e.g. branching ratio
measurements. The latter is generally used for analyses which input multiple particle identification features into a multivariate
classifier and hence do not have a single simple efficiency to compute at the end.} tracking efficiencies are corrected using
data-simulation correction tables, and other simulated features such as particle kinematics or the results of vertex or Kalman fits are generally
reweighted to match data using dedicated control samples. The second erratum was caused by a much larger than usual difference between
data and simulation in the tracking efficiencies. 

Interestingly, and perhaps contrary to what one might expect, the problem was not that the simulation overestimated  the 
tracking efficiencies, but rather that the simulated samples produced at the start of Run~2 dramatically underestimated the tracking
efficiencies, particularly at large track angles. The cause of this effect was an imperfect modelling of the impact of radiation
damage on the hit efficiency in the LHCb vertex detector. The simulation accounts for this effect using a parametrization whose
terms can be thought of as quantifying the amount of radiation damage the detector has sustained. During the shutdown between Run~1 and Run~2 
an additional ``second metal layer'' term was added to this radiation damage modelling, whose parametric form was wrong,
especially for high-angle tracks. The result was that the simulation modelled a too low hit efficiency in the outer regions of the VELO,
and thus a too low tracking efficiency for high-angle tracks. To be complete, I should note that the charm cross-section analysis
was affected by this twice: once in the tracking efficiencies, and once because the trigger documented in the previous chapter
explicitly cut on the number of hits on the VELO track. The former effect applied to other analyses as well,
so I will focus on it here, but the bias in the measured trigger efficiencies actually affected the measured charm-cross sections more severely,
especially at very low pseudorapidities.

\begin{figure}
\centering
\includegraphics[width=0.9\linewidth]{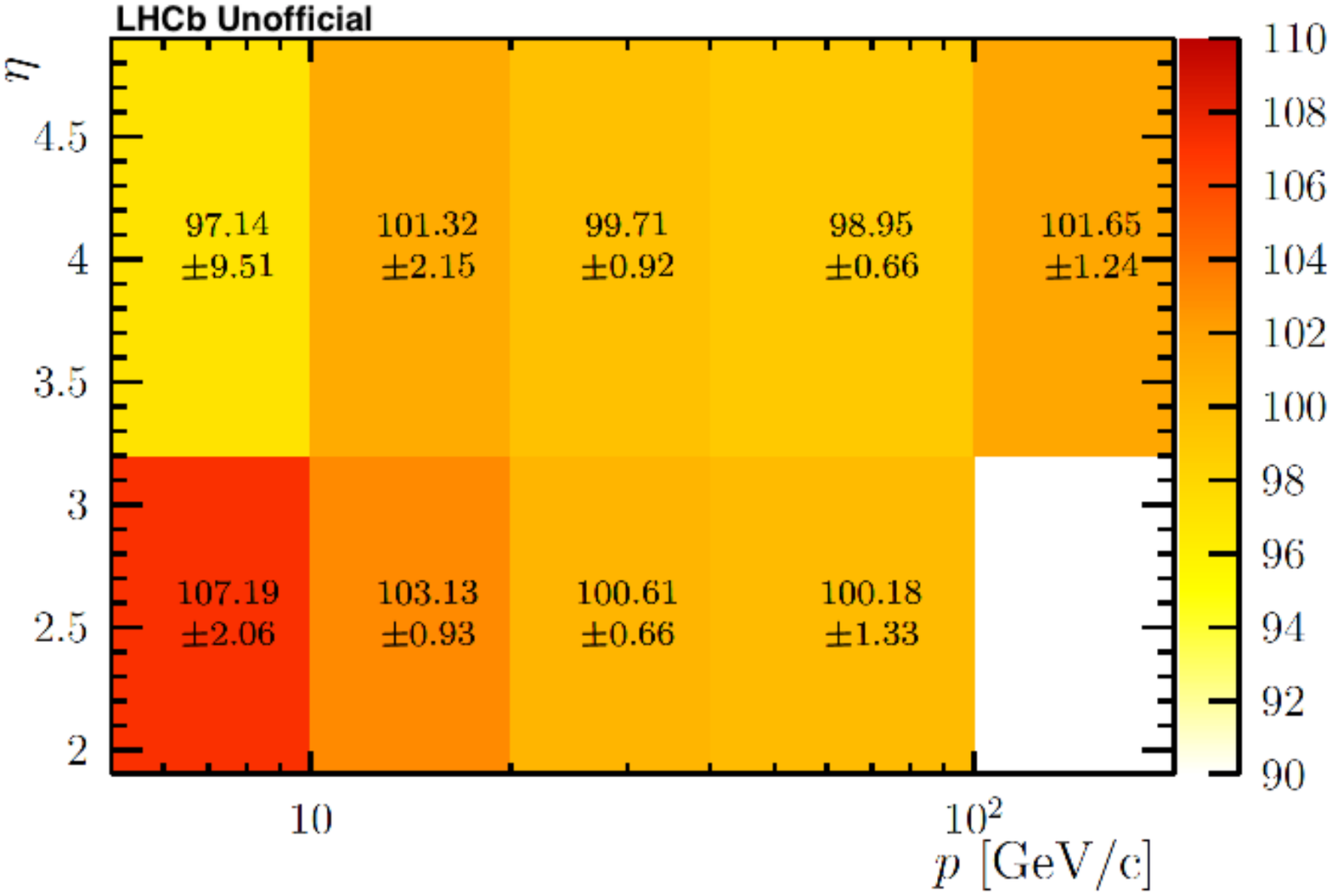}
\caption{The per-track efficiency correction factor, in percent, as seen in early 2015 data. The quoted uncertainties are a combination of
statistical uncertainties due to the limited calibration sample sizes and systematic uncertainties due to the finite accurary of the
calibration method when compared to the truth-level efficiency on simulated samples. The coarse binning is an indication of the limited
size of the tag-and-probe calibration samples available. This plot is reproduced from unpublished analysis documentation and is otherwise
no longer available even internally in LHCb, and no citation is therefore given.} 
\label{fig:treffbroken}
\end{figure}

The discrepancy in the data and simulation tracking efficiencies was seen very early in 2015, 
as the tracking efficiency correction tables used to produce this plot were an integral part of the
calibration samples discussed in Chapter~\ref{chpt:uneprocedure} and the analysis workflow itself. 
The corresponding correction table can be seen in Figure~\ref{fig:treffbroken}, and can be compared to the final table used
for the eventual analysis in Figure~\ref{fig:trefffinal}. The underlying problem, described in the previous paragraph,
was also quickly understood by the subdetector experts, however there was a general assumption throughout the collaboration that
our data-simulation correction tables were designed for exactly such a situation and could be relied on to remove the bias in the analysis.
Why, then, did they fail to do so?

The data-simulation corrections to LHCb's tracking efficiencies are applied as a two-dimensional table in pseudorapidity and transverse
momentum. Each track is located within this table, and the corresponding efficiency ratio between data and simulation is obtained.
The overall data-simulation efficiency ratio for a signal candidate is then the product of the individual data-simulation ratios for 
each of the tracks produced by the signal. The essential assumption behind such a correction procedure is that either
the tracking efficiency correction does not rapidly vary as a function of pseudorapidity
or transverse momentum, or that the calibration samples used to obtain the tracking efficiencies on data have a sufficiently
similar distribution of pseudorapidity and transverse momentum as the signal decay products. Had either of these assumptions been satisfied,
the method would have worked. Unfortunately both were broken:
the calibration samples have significantly different distributions in these variables compared to the signal decay
products, and in early Run~2 the corrections varied rapidly as a function of track pseudorapidity. This problem was made worse by the coarse
pseudorapidity binning in the early 2015 data, necessitated by the limited available tag-and-probe statistics.

\begin{figure}
\centering
\includegraphics[width=0.48\linewidth]{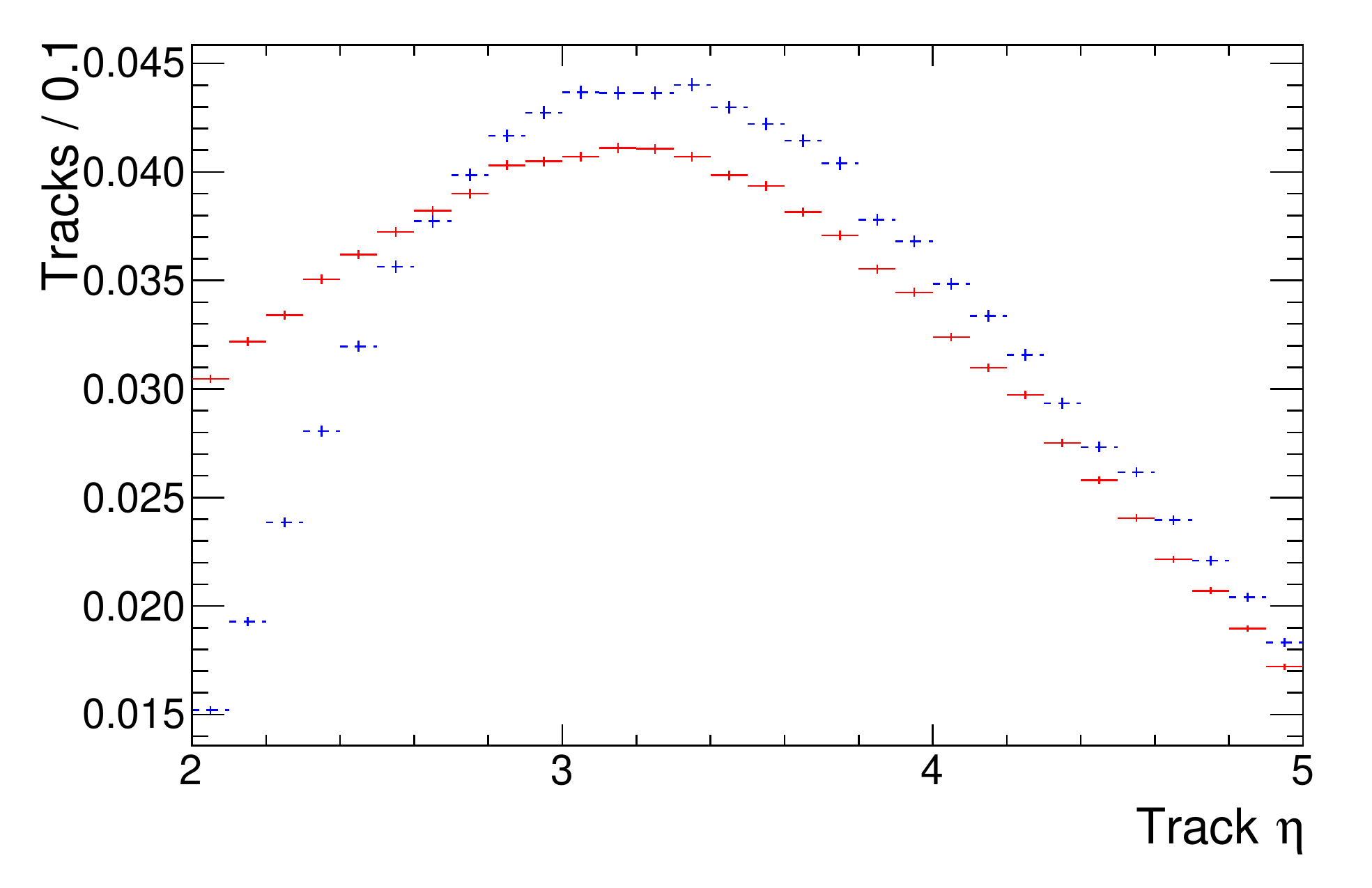}
\includegraphics[width=0.48\linewidth]{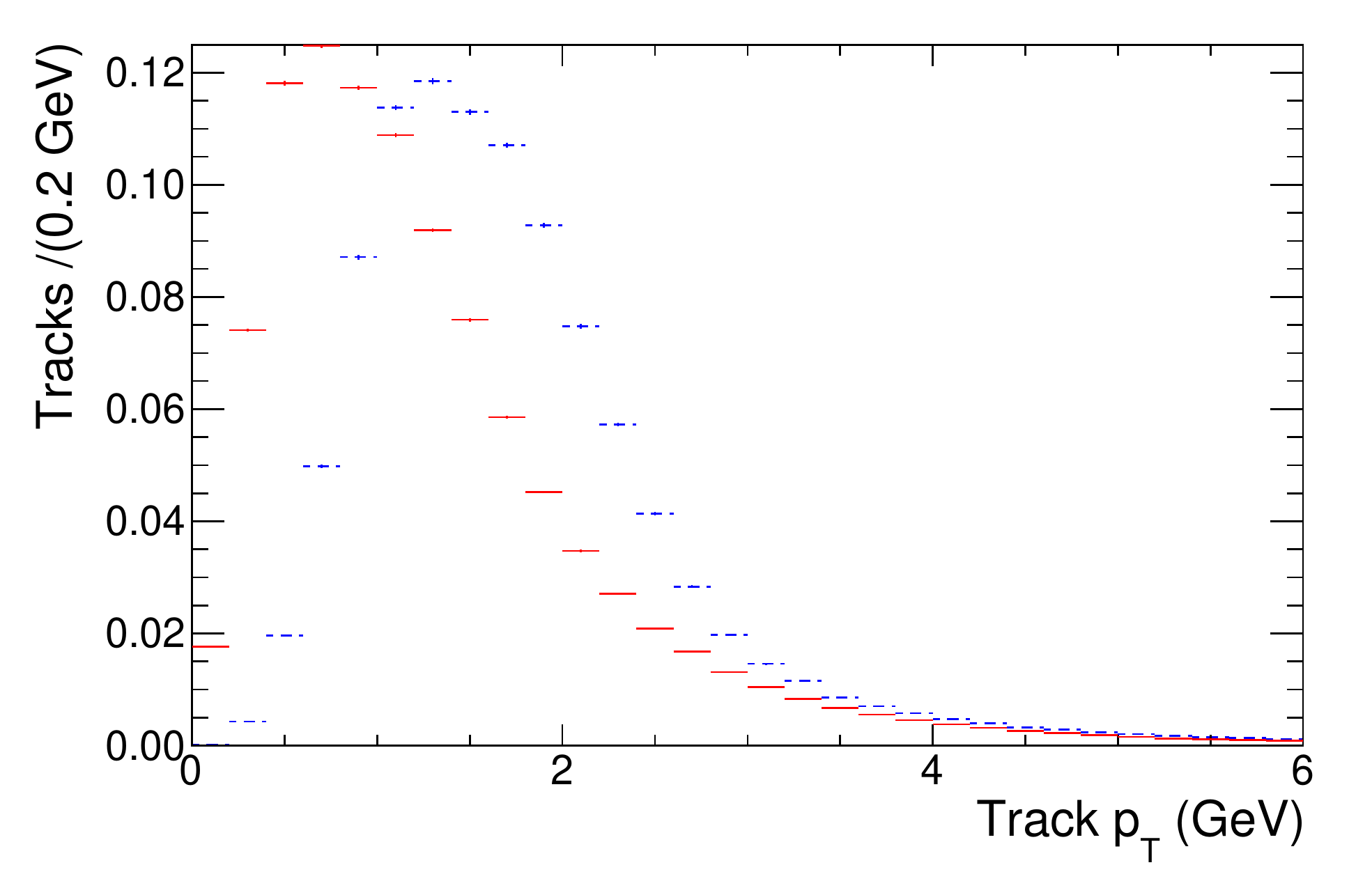}
\caption{The (left) pseudorapidity and (right) transverse momentum of the (solid red) signal and (dashed blue) calibration tracks.}
\label{fig:sigvscalkinematics}
\end{figure}

We again use toy samples simulated with RapidSim~\cite{Cowan:2016tnm} to illustrate the impact of this problem. A sample of simulated $J/\psi \to \mu\mu$
decays represents the tag-and-probe sample used to evaluate the tracking efficiencies, with a requirement that the tag muon has $p_{T} > 0.9$~GeV,
as in the muon trigger used for the early measurements data sample in 2015. The earlier simulated sample of $D^{*+}\to D^0 \pi^+$ decays is
reused here to represent the signal. The transverse momentum and rapidity distribution of the probe muon is compared to that of the kaon
and pion from the $D^0$ decay chain in Figure~\ref{fig:sigvscalkinematics}, where we can immediately see a substantial difference between them.

\begin{figure}
\centering
\includegraphics[width=0.7\linewidth]{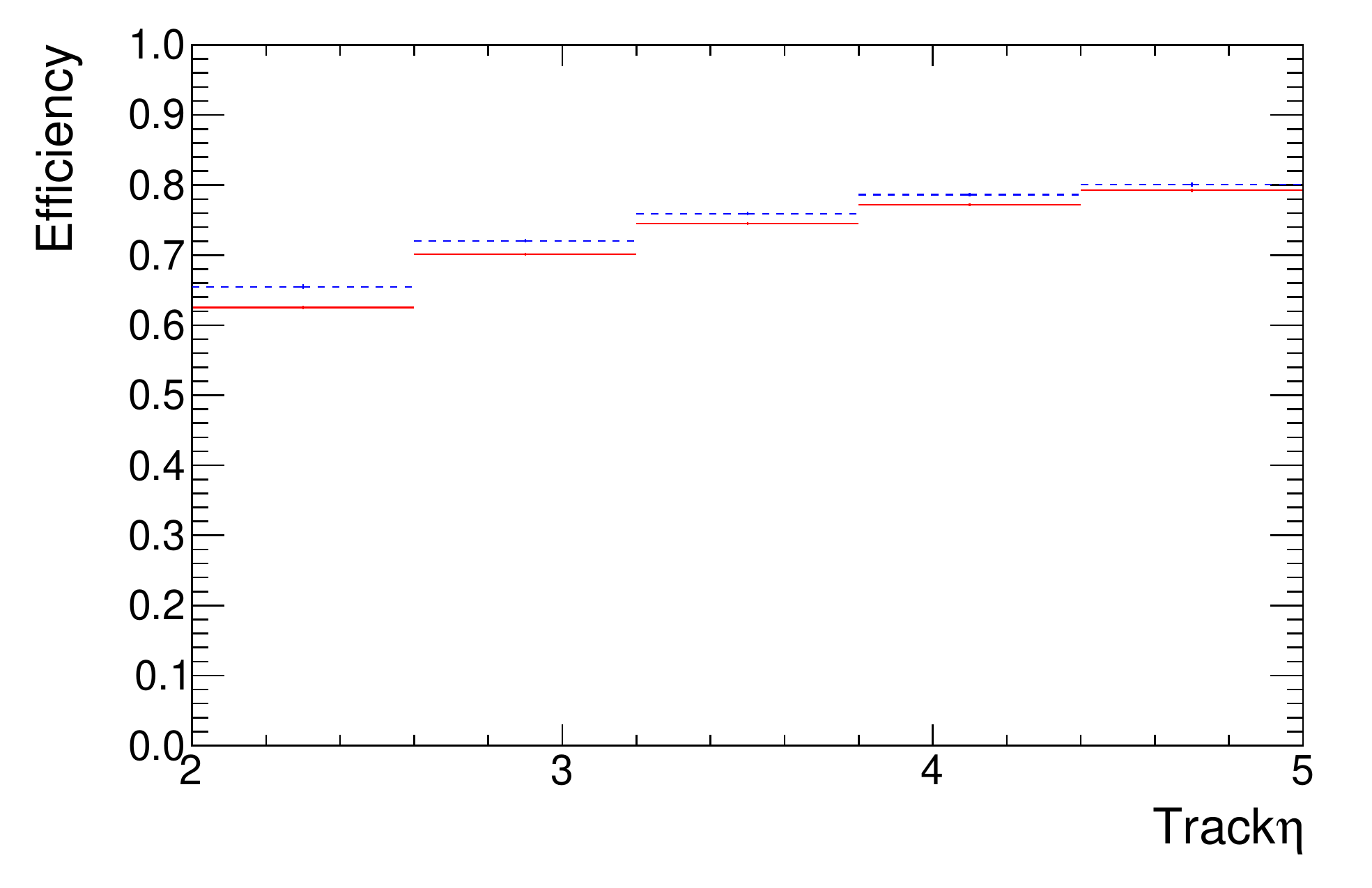}
\caption{The tracking efficiency measured on the (solid red) signal and (dashed blue) calibration samples as a function of track pseudorapidity.}
\label{fig:sigvscaleff}
\end{figure}

Next, we produce binned tracking efficiency plots for both samples. For simplicity, since the point here is to illustrate the effect
and not to reproduce the precise bias seen in the LHCb measurement, we simulate a true tracking efficiency with the following functional
form in transverse momentum and pseudorapidity:
\begin{equation*}
\epsilon = (0.5 + (\eta - 2)^{0.25} \cdot 0.35)\cdot (0.98-(6-p_{\textrm{T}})\cdot 0.03)
\end{equation*}
In Figure~\ref{fig:sigvscaleff} we compare the projection of this efficiency when binned in 8 bins of pseudorapidity for the signal and calibration samples
simulated earlier. We can clearly see that the efficiency measured from the calibration sample does not give a good estimate of the signal
efficiency in the relevant bin, because of the rapid variation of both the efficiency and the distribution of signal and calibration tracks 
inside each bin. This difference, which subsequently affects the overall measurement in a multiplicative way depending on the number of tracks produced by a given signal, 
is at the heart of what went wrong with the initial cross-section measurement, and indeed all the other LHCb measurements which
required errata in Run~2.

\section*{Lessons for the future}
Hindsight is a wonderful thing, but while it is natural to want to learn lessons from mistakes, this desire must be tempered by the fact
that stuff does, in fact, happen, and that no amount of procedures will ever eliminate all mistakes. For example, it would be very tempting
to preach about how code review would have caught the bugs which led to the first erratum, but in practice it is not at all obvious that
an external reviewer would have been able to see these mistakes any better than the analysis team itself. An independent set of analysts,
writing their own code, would have certainly reduced the probability of such a mistake happening. However, there is a conceptual contradiction between 
wanting to have independent analysis teams, with codebases which are sufficiently independent to reduce the risk of such mistakes, 
and at the same time wanting to standardize and professionalize code development in HEP, encourage
code reuse and minimize the extent to which everyone reinvents the same MINUIT fitting (or efficiency histogram filling) wheel.\footnote{There
is also the more mundane practical question of whether it is possible to duplicate every analysis in this way, which the field seems to give a rather clear answer to every time
the topic is brought up.} The latter in particular is a growing demand among the younger generations in HEP, who by and large prefer working in a team rather
than competitively.\footnote{This is also pushed from the management side by the idea that it better
prepares the majority of HEP researchers who will not have an academic career for life outside academia, although I note that my younger colleagues to whom I showed
a draft of this thesis dismissed that argument with the words ``if I wanted to work in industry I'd be working there already''.} Whether it will actually reduce the incidence
of errors in our analysis seems to me to be an open question.

The second erratum, on the other hand, is more instructive in every way. First of all, it taught LHCb the hard way that all our correction
tables for data-simulation differences relied on the fact that those differences were very small to begin with, or else that the
efficiencies or distributions we were correcting didn't rapidly vary with the signal geometry or kinematics. It is bitterly ironic that we in the charm-cross section analysis team
knew how to solve this problem before it happened and indeed took great care to optimize the binning of the PID calibration sample to avoid it,
but somehow simply failed to follow through with the logic for the tracking efficiencies. In any case, this has already led to
a significant reappraisal of the way analyses which correct for data-simulation differences are reviewed, and will certainly lead
to a more fine grained evaluation of data-simulation differences in the future. The significant expansion in the calibration samples
recorded by LHCb since 2015 will undoubtedly help with this. Secondly, this erratum was a salient reminder that we should not try and save
computing resources at all costs, nor assume that we can think our way out of all problems. The reality is that had we simply bruteforced the problem
by regenerating all the affected simulated samples, even at the cost of slowing down unaffected ongoing analyses, we would have saved ourselves at least
one and possibly several errata.\footnote{Of course we would likely have not learned about our hidden assumption of small data-simulation differences,
thus potentially leaving it to blow up in our faces later.} This is a lesson worth keeping in mind, particularly in light of the ongoing restrictions on the funding of HEP
computing resources, which will undoubtedly lead to more attempts at intelligent frugality in the future.

Perhaps the only aspect of the above errata which can be fairly laid at the door of real-time analysis, and which might have prevented the second erratum,
was our failure to cross check the production measurement of $D^0\to K^-\pi^+$ with $D^0\to K^-\pi^+\pi^+\pi^-$. Such a check, in which we would verify that both
methods gave the same answer, was planned in the run up to 2015, but was eventually dropped due to a lack of time, smaller signal yields and worse purities in the 
$D^0\to K^-\pi^+\pi^+\pi^-$ mode, and an evidently somewhat misplaced confidence that we knew what we were doing. 

\chapter{Staring at the sun: the future of real-time analysis}
\label{chpt:conclusion}
\pagestyle{myheadings}
\markboth{\bf Staring at the sun}{\bf Staring at the sun}

\begin{flushright}
\noindent {\it Al kad vidim pametare\linebreak 
\phantom{\textemdash} gde s' klanjaju \v{s}upljoj sili \linebreak
kad junake vidim stare \linebreak
\phantom{\textemdash} gde ih skotska naslad gnjili\linebreak
te sad ono pale, \v{z}are, \linebreak
\phantom{\textemdash} \v{c}em' su bili borci \v{c}ili;\linebreak
da zelucu nije smetnje, \linebreak
\phantom{\textemdash} ma\v{c} opa\v{s}u sa bedrine;\textemdash\linebreak
takve slike nisu vredne, \linebreak
\phantom{\textemdash} budaline Savedrine.\linebreak
}
\linebreak
-- Laza Kosti\'{c}, Don Kihotu, 1874
\end{flushright}

What, then, should we make of real-time analysis, which originated in a desire to exploit the full statistical power of the LHC,
not only for high energy New Physics searches or studies of the Electroweak sector but across the entire range of particles which its collisions produce.
I hope this document has convinced you that real-time analysis is a distinct and novel concept in HEP. Distinct because
the quantitative change in no longer storing raw detector data but only derived objects such as particles or
jets inevitably leads to significant qualitative changes in the nature of analysis\footnote{See \textit{The Encyclopedia Logic} 
by Hegel for a coherent account of how quantity transforms into quality in any natural system.}
, foremost among which is the requirement to fully map out the background and control samples needed before starting to collect data. And novel
because it is only recently that the software used to process HEP data has become sufficiently powerful and automatable
to allow such analysis without an unacceptable degradation in sensitivity. We have seen how these requirements translated into concrete
design choices when deployed within the LHCb experiment, and how LHCb's real-time analysis was commissioned and used to produce physics
results within weeks of datataking in 2015. We have also seen that those first physics results contained errors, although none of those
errors were problems caused by real-time analysis as such. Nevertheless the question remains whether real-time analysis is a passing fad, 
or whether it contains some seeds of a new and generally applicable approach to analysing data.

Interest in real-time analysis for HEP arises from, and depends on, a confluence of largely unrelated effects. Our experiments produce more interesting
data than can be indefinitely preserved in an affordable manner, while at the same time the computing power exists to analyse and dramatically
reduce this data volume in real-time without any significant impact on our ability to use this data in order to make inferences about the constants of nature which we
are interested in. In addition, modern software maintenance and versioning tools have made it possible to reliably execute thousands of real-time analyses,
largely written by students who often start with rather limited programming knowledge, and just as importantly to add new analyses over time without
breaking the existing ones. And of course the performance and long-term stability of our detector hardware, as well as the sophistication of modern
HEP simulation, makes it possible to keep our systematic uncertainties under sufficient control so that our data remains interesting with ever-increasing luminosities.

Most of these effects are unarguably here to stay. At the most basic level, real-time analysis is important in industry, and will only become more important as companies
race to ``automate'' transport, finance, medicine, manufacturing and a whole range of other aspects of life. We would be justified to
bet on real-time analysis for this reason if for no other. But even within HEP, the fundamental considerations speak less to a passing fad and more to a durable development. 
Our detectors have become more performant and more stable because of industry-driven advances in the underlying hardware
and because of the collective experience accumulated by high energy physicists from past experiments. Over the last decades every generation of HEP experiments has
delivered better and more stable basic performance than the preceeding one, and there is no reason to expect this trend to break down. The balance between storage
and processing cost might eventually tilt in the favour of storage, especially if the recent slowdown of Moore's law continues. However even if we ever got to the point where
this data were cheaper to store and distribute to analysts than to analyse in real-time, this would likely just lead to a shifting of the real-time analysis
software from the experimental areas in CERN to the distributed computing sites where the data are held. And the software maintenance tools which have been so central
to the success of real-time analysis are driven by the requirements of industrial software development and this is unlikely to stop in the foreseeable future; indeed
this is probably \textbf{the} software-related area in which we can most confidently expect to parasitically benefit from external developments. 

The genuine danger to real-time analysis in HEP lies less in the breakdown of any of the assumptions which led to its existence, and more in the ongoing
worldwide assault on the idea of, and consequently funding for, fundamental research. It is entirely possible that the next generation of HEP experiments will be shorn of the ability
to fully exploit the data which they collect, not by any intrinsic limitation in their granularity or resolution, but by a failure
to allocate sufficient resources to the processing of this data; the LHCb upgrade, which is expected to process, in real-time, around 100 times the data volume of the current LHCb
detector using a computing cluster budgeted at $1/2$ the cost of its current one, is a salient example. But LHCb is hardly alone in this: the computing resources
allocated to all the LHC experiments over the next decade are widely understood to be adequate only in the scenario where physicists manage to make their
reconstruction algorithms run around ten times faster. Some of this shortfall may be recovered using more intelligent and fine-grained cascade buffers,
some of it may be recovered because new reconstruction algorithms developed in industry finally begin to outperform the decades old duopoly of the Kalman filter and
Hough transform. It is generally a bad idea to bet against the collective intelligence of human beings, much less physicists, but it is also necessary to note the disparity
in the funds allocated for data processing in HEP and industry: LHCb's overall data processing turns over more than 10~Exabytes at a cost of around 10 million dollars per year,
while Facebook spends hundreds of millions of dollars per year analyzing a few hundred Petabytes of data. 
A failure because of these constraints won't therefore be a failure of real-time analysis as such, however, but rather a statement that we do not consider
the properties of nature which real-time analysis would have allowed us to measure worth measuring.

Even if the long term future holds potentially terminal challenges, the shift to real-time analysis in LHCb inspired a lot of other work and developments
which will not go away. In particular, the real-time alignment and calibration of the detector, and the significant expansion in the calibration samples
used to understand its performance, will benefit all of LHCb's analyses more and more with increasing luminosity. Another significant side-effect of real-time analysis
is an increasing acceptance across the collaboration that analyses should be implemented in publically accessible, automatically tested, and reproducible code. The development
of real-time analysis was not directly preconditioned on a move towards reproducible analysis, but it unquestionably contributed to a general shift in the collaboration's
culture which is currently underway, and which will lay the foundations for making LHCb's data openly accessible in a meaningful way. In this spirit, I want to end
as I began, by thanking all my LHCb colleagues for all the fun we had together making this seemingly impossible, and according to many people either trivial or undersirable, idea work. 
The circle, as Stephen King once wrote, has well and truly opened. As for the clicking bootheels, well, we'll see...

\cleardoublepage

\addcontentsline{toc}{chapter}{Bibliography}

\baselineskip=3ex

\newcommand{\BIBand}{\&}
\bibliographystyle{unsrt}

\pagestyle{myheadings}

\bibliography{references}

\end{document}